\documentclass[aps,prd,superscriptaddress,notitlepage,11pt,final,longbibliography]{revtex4-1}

\usepackage{graphicx}
\usepackage{hyperref}

\usepackage{amsmath,amssymb,mathrsfs}
\usepackage{subfigure}
\usepackage{epsf}
\usepackage{epsfig}
\usepackage[usenames,dvipsnames]{xcolor}
\usepackage{bbm}
\usepackage{color}
\usepackage{comment}
\usepackage{cleveref}

\usepackage{natbib}

\newcommand{\be}{\begin{equation}}
\newcommand{\ee}{\end{equation}}
\newcommand{\bea}{\begin{eqnarray}}
\newcommand{\eea}{\end{eqnarray}}

\newcommand{\eqn}{\begin{eqnarray}}
\newcommand{\eqnx}{\end{eqnarray}}

\usepackage{physics}
\numberwithin{equation}{section}

\begin{document}

\title{Compact Q-balls and Q-shells in a multi-component $\mathbb{C}P^N$ model}


\author{P. Klimas~}
\email{pawel.klimas@ufsc.br}
\affiliation{Departamento de F\'isica, Universidade Federal de Santa Catarina, Campus Trindade, 88040-900, Florian\'opolis-SC, Brazil}


\author{L.C. Kubaski~}
\email{luizckubaski@gmail.com}
\affiliation{Departamento de F\'isica, Universidade Federal de Santa Catarina, Campus Trindade, 88040-900, Florian\'opolis-SC, Brazil}

\author{N. Sawado~}
\email{sawadoph@rs.tus.ac.jp}
\affiliation{Department of Physics, Tokyo University of Science, Noda, Chiba 278-8510, Japan}

\author{S. Yanai}
\email{phyana0513@gmail.com}
\affiliation{Department of Physics, Tokyo University of Science, Noda, Chiba 278-8510, Japan}

\begin{abstract}
Coupled multi-component $\mathbb{C}P^N$ models with V-shaped potentials are analyzed. 
It is shown that the model has solutions being combinations of compact Q-balls and Q-shells. 
The compact nature of solutions permits the existence of novel harbor-type solutions having  the form of Q-balls sheltered by Q-shells.  
The relation between the energy $E$ and Noether charge $Q$ is discussed both analytically and numerically. 
The energy of the solutions behaves as $E\sim |Q|^\alpha,~\alpha<1$, {\it i.e.}, as for the standard Q-ball. 
Furthermore, the ratio $E/Q$ for various configurations in the multi-component model suggests that the solutions are at least classically stable.  
\end{abstract}

\maketitle

\section{Introduction}
\label{sec:intro}

Compactons are field configurations that exist on finite size supports.  The field takes its vacuum values  outside this support. 
For example, the signum-Gordon model, {\it i.e.}, the scalar field model with standard kinetic terms and V-shaped potential gives rise to such solutions~\cite{Arodz:2008jk,Arodz:2008nm}. 
	
A complex scalar field theory with some self-interactions has stationary
soliton solutions called Q-balls~\cite{Friedberg:1976me,Coleman:1985ki}. 
The U(1) invariance of the scalar field 
leads to the conserved charge $Q$ which, if the theory is coupled with the 
electromagnetism, is identified with the electric charge of the constituents.
Q-balls have attracted much attention in the studies of 
evolution of the early Universe~\cite{Friedberg:1986tq,Lee:1986ts}.
In supersymmetric extensions of the standard model, Q-balls appear as
the scalar superpartners of baryons or leptons forming coherent states with 
baryon or lepton number. They may survive as a major ingredient of dark matter
~\cite{Kusenko:1997zq,Kusenko:1997si,Kusenko:1997vp}.  

The Q-ball solutions of the $\mathbb{C}P^N$ nonlinear sigma model 
is given by the Lagrangian density~\cite{Klimas:2017eft}
\begin{align}
{\cal L}=-\frac{M^2}{2}\mathrm{Tr}(X^{-1}\partial_\mu X)^2-\mu^2V(X),
\label{lag0}
\end{align}
where the `V-shaped' potential
\begin{align}
V(X)=\frac{1}{2}[\mathrm{Tr}(\mathbb{I}-X)]^{1/2}
\end{align} 
is employed in order to obtain compact solutions, where $\mathbb{I}$ is an identity element. The behavior of fields at the outer border of compacton implies $X\to \mathbb{I}$. 
The coupling constants $M$ and $\mu$ have dimensions of $(\mathrm{length})^{-1}$ and $(\mathrm{length})^{-2}$, respectively. 
Taking into account coupling with gravity one gets that the model has several boson-star and boson-shell solutions~\cite{Klimas:2018ywv, Yanai:2019wpv, Sawado:2020ncc, Sawado:2021rsc}.
Among others the shell solutions are particularly interesting because they can harbor a localized body inside. 
The harbor solution consists of a massive body, such as the Schwarzschild-like black hole, localized in the center of the shell. The event horizon is by assumption located inside an empty region of the shell.  The field equations and Einstein equations are solved in the external region of the horizon. Such solutions are called the harbor~\cite{Kleihaus:2010ep}.
Harboring black holes in the solutions of the $\mathbb{C}P^N$ nonlinear sigma model 
were extensively studied in~\cite{Klimas:2018ywv,Sawado:2021rsc}. 

A theory implementing two or more scalar fields, {\it i.e.}, multi-component scalar field models, 
has attracted attention from both the theoretical and the experimental. 
On the theoretical side, it is known that some soliton solutions emerge in a two scalar field model. 
For instance, the vorton~\cite{Witten:1984eb} is a famous solution of this kind. The $U(1)\times\tilde{U}(1)$ gauge theory 
with an unbroken gauge symmetry $Q$ (electromagnetism) and a broken gauge symmetry $R$ supports the existence of stable vortex strings. 
The model implements a Higgs field $\varphi$ with $Q=0,~R=1$ 
and also a second static scalar field $\sigma$ with $Q=1,R=0$. The minimum of the potential employed in \cite{Witten:1984eb} 
was chosen at $\langle\sigma\rangle=0, \langle\varphi\rangle\neq 0$ which leads to the existence of vortex string, {\it i.e.}, the vorton. 
Note that the model possesses the soliton solutions even when the gauge coupling is 
ignored~\cite{Davis:1988jp,Lemperiere:2002en,Lemperiere:2003yt}. The stability or the condition of existence of the solutions is not so straightforward.  
The constraint conditions imposed on the potential parameters in order to guarantee the existence of the solutions were already proposed in \cite{Lemperiere:2002en}, but unfortunately they were not correct. 
The correct conditions can be obtained after  more careful study of the equations of motion of the model~\cite{Itaya:2014oza}.

Another example of a multi-component scalar field model is the oscillon. 
The oscillon was discovered in the seventies of the last century by Bogolyubsky and Makhankov \cite{Bogolyubsky:1976yu} and 
later revisited by \cite{Gleiser:1993pt,Copeland:1995fq}. 
Typically, the oscillon has a bell shape that oscillates sinusoidally in time \cite{Fodor:2008es,Fodor:2009xw}. 
There are several oscillon configurations with two interacting scalar fields, see \cite{Gleiser:2011xj, Achilleos:2013zpa, Correa:2019jaa}. 
Notable thing is that oscillons in the two-scalar model have lifetimes that are much larger than those associated with 
single-field oscillons. 
Many astrophysical observations suggest that the Universe is currently experiencing an
accelerated expansion. As usual, there are some possibilities of modelling such a behavior using scalar fields. Among others, a hybrid inflation model with two real scalar fields interacting quadratically~\cite{Gleiser:2011xj} is particularly interesting because it supports oscillon configurations. 
As shown in that work, such configurations persist for at least four cosmological
expansion times, accounting for up to 20\% of the total energy density of the early Universe. 
The multiple-scalar fields give us a good realistic explanation of the age of the Universe~\cite{Fadragas:2014mra}. 
Note that the long-term stability of the vortons or the oscillons always very much depend on the vacuum structure of the potentials. 

In this paper, we construct the Q-ball and Q-shell solutions of the multi-component field extension of the 
$\mathbb{C}P^N$ nonlinear sigma model. As shown in \cite{Klimas:2017eft,Klimas:2018ywv,Sawado:2020ncc}, 
for $\ell=0,1~(N:=2\ell +1)$, the solutions have the form of compact Q-balls, while for $\ell \geqq 2$ they become compact Q-shells.  
Hence, considering the multi-component field model, we are able to construct a new harbor solution involving the Q-balls or Q-shells inside the hollow space of a Q-shell. Furthermore, 
it leads to a funny extension: a compact Q-ball surrounded by a Q-shell surrounded by another Q-shell\ldots. 
This multiple-shell structure is surely possible to be models for some phenomena, such as 
nuclei, Bose-Einstein condensation, and star or galaxy formations, and so on. 
Recently, some multi-component Q-ball solutions were studied in a vorton-like model~\cite{Ishihara:2021iag} and the authors claimed at least classical their stability.

We employ multi-component extension of the V-shaped potentials for constructing the solutions. 
A first consideration of the V-shaped potentials in the field theoretical context has been presented in \cite{Arodz:2002yt,Arodz:2003mx} where the spontaneous symmetry breaking was considered as the mechanism of production of compact kinks. 
Such models were obtained as the continuous limit of systems containing coupled pendulums. 
It has been shown in \cite{Arodz:2002yt} that the size of the kink is directly related with the manner how the field approach its vacuum value. 
Dissipative solitons in complex Ginzburg-Landau equation can be obtained from a quite simple form of the V-shaped potential~\cite{Liu:2021}. 
The V-shaped potentials in systems with just few degrees of freedom  were already known in the literature in the context of passage 
to chaos known as grazing bifurcation \cite{PhysRevA.27.1741,PhysRevE.49.1073,PhysRevE.50.4427}. A new important context in which 
V-sheped potentials appear is discussed in \cite{Adam:2017pdh, Adam:2017srx} where the authors consider the decomposition of the 
Skyrme model on two coupled BPS submodels. 
The V-shaped potentials considered there are not imposed a priori but they rather appear as the result of symmetry reduction and the fact 
that the field values are limited to some interval. 
Consequently the trajectories in the phase space are not continuous (reflections) which unavoidably leads to V-shaped potentials.  
Some consequences of this symmetry reduction -- the presence of approximated oscillons and their properties were studied in \cite{Klimas:2018woi}. 
Another important point about the V-shaped potentials is the existence of threshold force at the minima of these potentials.  
It makes them very attractive for a description of the anchoring of topological defects. 
Such a phenomenon is observed in condensed matter systems \cite{nelson2002, PhysRevB.48.7030}. 
Finally, there are some other physical contexts in which the V-shaped potentials appear as for instance in plasma physics \cite{PhysRevLett.78.4761} 
or in description of the electron  gas in two dimensions \cite{Rodriguez_Coppola_1992}.

Notably, these configurations are thought to be stable against small perturbations. 
As it was pointed out in \cite{Arodz:2005gz}, there is no linear regime for the compacton in models with V-shaped potential. That is, for the linear perturbation of the equation one obtains a nonlinear differential equation even in the limit of small amplitude, which is generally not tractable. Hence the fact that any arbitrarily small perturbation is governed by a nonlinear equation indicates a lack of the harmonic oscillator paradigm as well as its applicability to minima of the potential. Moreover, the nonvanishing of the first derivative of the potential at the minimum gives rise to the threshold force effect which constitutes a serious obstacle on free expansion of compacton support or emission of small wave packages from the perturbed region.
Further, the resulting equation possesses the scaling symmetry which spoils the naive stability discussion.  
We just conclude that  the scale of  compactons in the $\mathbb{C}P^N$ model 
is frozen out from the phase transitions.

In this paper, we restrict solutions on the flat space-time but extensions into the gravitating solutions,{\it i.e.}, 
the boson shells are almost straightforward. 

The paper is organized as follows. In Section II we shall describe the model.
The ansatz which parametrizes the $\mathbb{C}P^N$ field and the definition of the Hamiltonian,  
the Noether change and their nontrivial relation are given in Section III. 	
In Section IV, several types of the solutions such as the ball-shell, shell-shell and the harbor are presented. 
Finally, we give some examples of the three field extensions in Section V. 
Conclusions and remarks are presented in the last Section.

\section{The model}\label{sec2}
We consider extension of the $\mathbb{C}P^N$ model~(\ref{lag0}) into the $n$ multi-component field models. The Lagrangian density has a simple form:
\be
{\cal L}=-\sum_{a=1}^n\frac{M_a^2}{2}{\rm Tr}(X_a^{-1}\partial_{\mu}X_a)^2-V(\{X_a\})
\label{lagr1}
\ee
where $X_a~(a=1,2,\cdots,n)$ are principal variables parametrizing the space $\mathbb{C}P^{N_1}\otimes \mathbb{C}P^{N_2}\otimes \cdots \otimes \mathbb{C}P^{N_n}$. 
Each $\mathbb{C}P^{N_a}$ space is a coset space $\mathbb{C}P^{N_a}=SU(N_a+1)/SU(N_a)\otimes U(1)$. There are many possible choices of the potential.  Formally, we can consider a quite general expression
\be
V(\{X_a\})=\sum_a\mu_a^2\;V_a(X_a)+\sum_{ab}\lambda_{ab}\; W_{ab}(X_a,X_b)+\sum_{abc}\omega_{abc}Z_{abc}(X_a,X_b,X_c)+\cdots
\label{formalpot}
\ee
which involves polynomials with arbitrary number of fields. $\sum_{ab},\sum_{abc}$ mean the summations over all possible terms. 
In particular, if the model contains only the first term with a single sum, we end up with $n$ decoupled $\mathbb{C}P^N-$type models with potentials. 

In what follows we consider the V-shaped (non differentiable at minima) form  of potentials $V_a$, which  support compact solutions in decoupled models. 
Namely, they are given by expressions
\be
V_a(X_a)\equiv V(X_a):=\frac{1}{2}\sqrt{{\rm Tr}(\mathbb{I}-X_a)},\qquad a=1,\cdots,n,\label{V12}
\ee
where $\mathbb{I}$ is an identity matrix ${N_a}\times {N_a}$. Although the V-shaped character is not directly visible from the expression \eqref{V12}, its name will be justified  in further parts of this section. 

Since we are interested in compact solutions of the coupled models, we choose the coupling terms of the potential in such a form that they allow for Q-balls/Q-shells supported by a single $\mathbb{C}P^N$ field component, while the other components take the vacuum value. In other words, when all the $\mathbb{C}P^N$ components, except one, take the vacuum value, then the non vanishing component decouples from the remaining fields even though the coupling constants have non zero values. For this reason we employ 
the following form of coupling terms
\begin{align}
&W_{ab}(X_a,X_b)\equiv W(X_a,X_b)
:=V(X_a)^{2\alpha}V(X_b)^{2\beta}
=\left(\frac{1}{4}{\rm Tr}({\mathbb I}-X_a)\right)^{\alpha}\left(\frac{1}{4}{\rm Tr}({\mathbb I}-X_b)\right)^{\beta},
\\
&Z_{abc}(X_a,X_b,X_c)\equiv Z(X_a,X_b,X_c)
:=V(X_a)^{2\alpha}V(X_b)^{2\beta}V(X_c)^{2\gamma}
\nonumber \\
&\hspace{5cm}=\left(\frac{1}{4}{\rm Tr}({\mathbb I}-X_a)\right)^{\alpha}\left(\frac{1}{4}{\rm Tr}({\mathbb I}-X_b)\right)^{\beta}
\left(\frac{1}{4}{\rm Tr}({\mathbb I}-X_c)\right)^{\gamma}
\end{align}
and so on. Here $\alpha,\beta,\gamma,\cdots \geqq 1$. In addition, we assume that there are no terms with repetition of indices, 
{\it i.e.} $a\neq b$, $a\neq c$ etc. It avoids  modifications of the V-shaped potential $V_a(X_a)$.
The coefficients of (\ref{formalpot}) $\lambda_{ab},\omega_{abc},\cdots$ are then symmetric 
with interchange of the indices $a,b,c,\cdots$. 
For example, in the case of three field model, we can define two parameters $\lambda,\omega$ such that
\begin{align}
\lambda:=\lambda_{12}=\lambda_{21}=\lambda_{23}=\lambda_{32}=\lambda_{31}=\lambda_{13},~~~~
\omega:=\omega_{123}~~\textrm{and it's all permutations}
\end{align}
and all  others coefficients are zero. 
This simple prescription enables us to realize a direct extension for a model with an arbitrary number of  fields. 
In order to avoid unnecessary complications in construction of solutions  we concentrate in this paper  on a formulation with $n\leqq 3$, {\it i.e.}, on implementation of maximally three different principal variables $X_a$.

The principal variables $X_a(g_a)$ have the form 
\be
X_a(g_a):=g_a\sigma_a(g_a)^{-1},~~~~a=1,\cdots,n,
\ee 
and $\sigma_a(k_a)=k_a$ are involutive automorphisms $\sigma_a(\sigma_a(g_a))=g_a$.
Each principal variable parametrizes a coset space, nalmely $X_a(g_ak_a)=X_a(g_a)$ and $\sigma_a(k_a)=k_a$ where $k_a\in SU(N_a)\otimes U(1)$. 
In matrix representation the group elements $g_a$ are parametrized by a set of the scalar fields 
\[
u^{(a)}\equiv\begin{pmatrix}
u^{(a)}_1\\ \vdots\\  u^{(a)}_{N_a}\end{pmatrix},\qquad a=1,2\ldots,n.
\]
The group elements $g_a\in SU(N_a+1)$ are parametrized as follows
\begin{eqnarray}
&g_a\equiv\dfrac{1}{ \sqrt{1+u^{(a)\dagger} \cdot u^{(a)}}}\left(\begin{array}{cc}
\Delta(u^{(a)})&iu^{(a)}\\
iu^{(a)\dagger}&1
\end{array}\right),
\\
&\Delta_{ij}(u^{(a)})\equiv  \sqrt{1+u^{(a)\dagger} \cdot u^{(a)}}\,\delta_{ij}
-\dfrac{u^{(a)}_iu_j^{(a)*}}{1+ \sqrt{1+u^{(a)\dagger} \cdot u^{(a)}}}.
\label{parametryzacja}
\end{eqnarray}
The principal variables read
\begin{eqnarray}
X_a(g_a)=g_a^2=
\left(\begin{array}{cc}
{\mathbb{I}}_{N_a\times N_a} & 0 \nonumber \\
0 & -1 \nonumber 
\end{array}\right)
+
\frac{2}{1+u^{(a)\dagger}\cdot u^{(a)}}\left(\begin{array}{cc}
-u^{(a)}\otimes u^{(a)\dagger} & iu^{(a)} \nonumber \\
iu^{(a)\dagger} & 1  
\end{array}\right).
\end{eqnarray}

The Lagrangian density \eqref{lagr1} has the form
\begin{align}
&{\cal L}=\sum_a{\cal L}_a+{\cal L}_{\rm int}\label{lagr2}
\\
&{\cal L}_a:={\cal L}_{{\rm kin},a}-\mu_a^2\;V_a(u^{(a)\dagger},u^{(a)}),~~
\quad {\cal L}_{{\rm kin},a}:=-M_a^2\eta_{\mu\nu}\tau_a^{\nu\mu}
\\
&{\cal L}_{\rm int}:=-\lambda \sum_{a\neq b}\;W(u^{(a)\dagger},u^{(a)}, u^{(b)\dagger},u^{(b)})
-\omega Z(u^{(1)\dagger},u^{(1)}, u^{(2)\dagger},u^{(2)}, u^{(3)\dagger},u^{(3)})
\end{align}
with $\eta_{\mu\nu}={\rm diag}(1,-1,-1,-1)$ being components of the metric tensor in Minkowski spacetime.
The summation $\sum_{a\neq b}$ is taken for all possible pairs $(a,b)$, {\it i.e.}, $(a,b)=(1,2),(2,1),(1,3),(3,1),(2,3),(3,2)$. 
The symbols $\tau_a^{\nu\mu}$  are defined as
\be
\tau_a^{\nu\mu}=-4\frac{\partial^{\mu} u^{(a)\dagger}\cdot\Delta^2(u^{(a)})\cdot\partial^{\nu} u^{(a)}}{(1+u^{(a)\dagger}\cdot u^{(a)})^2}
\label{tau}
\ee
and $\Delta^2$ is a square of $\Delta$ matrix, $\Delta^2_{ij}\equiv\Delta_{ik}\Delta_{kj}$, which is of the form
\be
\Delta^2_{ij}(u^{(a)})=\Big(1+u^{(a)\dagger}\cdot u^{(a)}\Big)\delta_{ij}-u_i^{(a)}u^{(a)*}_j.
\ee

The Euler-Lagrange equations are obtained taking variation of the action $S=\int d^4x {\cal L}$ with respect 
to fields $u^{(a)}_i$ as well as its complex conjugates (which became redundant). 
Here we present equations obtained from $\delta_{u^{(a)*}_k}{\cal L}=0$. They have the form
\begin{align}
\frac{\delta{\cal L}_{{\rm kin},a}}{\delta u^{(a)*}_k}-\partial_{\alpha}\frac{\delta {\cal L}_{{\rm kin},a}}{\delta(\partial_{\alpha}u_k^{(a)*})}
-\mu_a^2\frac{\delta V_a}{\delta u_k^{(a)*}}-\lambda\sum_{b\neq c}\frac{\delta W}{\delta u_k^{(a)*}}-\frac{\delta Z}{\delta u_k^{(a)*}}=0
\label{eq01}
\end{align}
where $a$ has fixed value for each equation and  $W\equiv W(u^{(b)\dagger},u^{(b)}, u^{(c)\dagger},u^{(c)})$.
Multiplying \eqref{eq01} by $-\frac{1}{4}\big(1+u^{(a)\dagger}\cdot u^{(a)}\big)^2$  one gets a set of equations 
\begin{align}
\Delta^2_{kl}(u^{(a)})K^{(a)}_l(u^{(a)})+\frac{1}{4}(1+u^{(a)\dagger}\cdot u^{(a)})^2\left[\mu_a^2\frac{\delta V_a}{\delta u_k^{(a)*}}
+\lambda\sum_{b\neq c}\frac{\delta W}{\delta u_k^{(a)*}}+\omega\frac{\delta Z}{\delta u_k^{(a)*}}\right]=0,
\label{eqresres}
\end{align}
where
\be
K_l^{(a)}(u^{(a)})\equiv M_a^2\eta_{\mu\nu}\left[\partial^{\mu}\partial^{\nu}u^{(a)}_l
-\frac{(u^{(a)\dagger}\cdot\partial^{\mu}u^{(a)})\partial^{\nu}u_l+(u^{(a)\dagger}\cdot\partial^{\nu}u^{(a)})
\partial^{\mu}u_l^{(a)}}{1+u^{(a)\dagger}\cdot u^{(a)}}\right].
\ee
The matrix $\Delta^2_{kl}(u^{(a)})$ is invertible. 
Multiplying the resulting equations \eqref{eqresres}  by its inverse which is of the form $\Delta^{-2}_{kl}=\frac{1}{1+u^{\dagger}\cdot u}(\delta_{kl}+u_ku^*_l)$ 
one can decouple the terms containing second order derivatives. The resulting equations read
\begin{align}
K^{(a)}_l(u^{(a)})+\frac{1}{4}(1+u^{(a)\dagger}\cdot u^{(a)})^2\Delta^{-2}_{lk}(u^{(a)})\left[\mu_a^2\frac{\delta V_a}{\delta u_k^{(a)*}}
+\lambda\sum_{b\neq c}\frac{\delta W}{\delta u_k^{(a)*}}+\omega\frac{\delta Z}{\delta u_k^{(a)*}}\right]=0.
\label{eq1a}
\end{align}
Note that the $\mathbb{C}P^N$ equations of motion correspond with $V=0$ and they read $K^{(a)}_l(u^{(a)})=0$. 

In the next two sections, we present the results for $n=2$, {\it i.e.}, the model containing two $\mathbb{C}P^N$ fields  (two principal variables). 
In such a model $Z\to 0$ and thus 
only the potential $W(X_1,X_2)$ is responsible for the coupling between the fields.  
For convenience, we write $u^{(1)}\to u, u^{(2)}\to v$. 
In this paper we are interested in compact finite-energy solutions. 
Thus we shall consider potentials $V_1(u^{\dagger}, u)$ and $V_2(v^{\dagger},v)$ in the form given by \eqref{V12} which have visibly V-shaped character at their minima
\be
V_1=\left(\frac{u^{\dagger}\cdot u}{1+u^{\dagger}\cdot u}\right)^{\frac{1}{2}},\qquad V_2=\left(\frac{v^{\dagger}\cdot v}{1+v^{\dagger}\cdot v}\right)^{\frac{1}{2}}.\label{V1e2}
\ee
The potentials \eqref{V1e2}   have exactly the same form as in our previous work. 
Thus in the case $\lambda=0$ the model contains two docoupled $\mathbb{C}P^N-$type models with potentials supporting compactons. 
The coupling term $W(X_1,X_2)$ reads
\begin{align}
W(X_1,X_2)=
\left(\frac{u^{\dagger}\cdot u}{1+u^{\dagger}\cdot u}\right)^{\alpha}\left(\frac{v^{\dagger}\cdot v}{1+v^{\dagger}\cdot v}\right)^{\beta}
\end{align}
where, $\alpha,\beta\ge 1$. Plugging the explicit expression for the potential into \eqref{eq1} and \eqref{eq2} one gets
\begin{align}
&K^{(1)}_l(u)+\frac{\mu_1^2}{8}\sqrt{1+u^{\dagger}\cdot u}\frac{u_l}{\sqrt{u^{\dagger}\cdot u}}+\frac{\lambda\alpha}{4}\left(\frac{u^{\dagger}\cdot u}{1+u^{\dagger}\cdot u}\right)^{\alpha-1}\left(\frac{v^{\dagger}\cdot v}{1+v^{\dagger}\cdot v}\right)^{\beta}u_l=0,\label{eq1}\\
&K^{(2)}_r(v)+\frac{\mu_2^2}{8}\sqrt{1+v^{\dagger}\cdot v}\frac{v_r}{\sqrt{v^{\dagger}\cdot v}}+\frac{\lambda\beta}{4}\left(\frac{u^{\dagger}\cdot u}{1+u^{\dagger}\cdot u}\right)^{\alpha}\left(\frac{v^{\dagger}\cdot v}{1+v^{\dagger}\cdot v}\right)^{\beta-1}v_r=0 \label{eq2}
\end{align}
where $l=1,2,\ldots,N_1$ and $r=1,2,\ldots,N_2$.

In the last step we  write the field equations in dimensionless form. First, we observe that the variables $X_a$ and  the fields $u_l$ and $v_r$ are dimensionless. On the other hand, the spatial coordinates $x^{\mu}$ have dimension of length. The characteristic length scale can be given in terms of  dimensional constants  $M_a^2$ , $\mu^2_a$  and $\lambda$ (or their combinations), where $[M_a^2]=L^{-2}$, $[\mu_a^2]=[\lambda]=L^{-4}$. Let $r_0$ be such a characteristic length scale $[r_0]=L$. We replace the dimensional coordinates by their dimensionless counterparts putting the length scale explicitly, $x^{\mu}\rightarrow r_0 x^{\mu}$, $\partial_{\mu}\rightarrow r_0^{-1}\partial_{\mu}$. It leads to dimensionless expressions 
\[
\widetilde K_l^{(1)}:=M_1^{-2}r_0^2\,K_l^{(1)},\qquad \widetilde K_r^{(2)}:=M_2^{-2}r_0^2\,K_r^{(2)}.
\]
Next we multiply \eqref{eq1} by $M_1^{-2}r_0^2$ and \eqref{eq2} by $M_2^{-2}r_0^2$ and define dimensionless coupling constants
\[
\widetilde \mu_1^2:=M_1^{-2}r_0^2\mu_1^2,\qquad \widetilde \mu_2^2:=M_2^{-2}r_0^2\mu_2^2,\qquad \lambda_1:= M_1^{-2}r_0^2\,\lambda,\qquad \lambda_2:= M_2^{-2}r_0^2\,\lambda.
\]
Thus dimensionless field equations have the form of equations \eqref{eq1} and \eqref{eq2} with $K_l^{(1)}\rightarrow \widetilde K_l^{(1)}$, $K_r^{(2)}\rightarrow \widetilde K_r^{(2)}$, $\mu_1^2\rightarrow \widetilde \mu^2_1$, $\mu_2^2\rightarrow \widetilde \mu^2_2$ and  $\lambda\rightarrow \lambda_1$ in \eqref{eq1} and   $\lambda\rightarrow \lambda_2$ in \eqref{eq2}. 

\section{The Q-ball/Q-shell ansatz }\label{sec3}

In similarity to our previous works, we consider dimensional reduction (in spherical coordinates) combined with the Q-ball ansatz. Thus, we consider the case with $N_1$ and $N_2$ being two odd numbers.
Taking $N_1=2l_1+1$ and $N_2=2l_2+1$ we introduce the fields $u_{m_1}$ and $v_{m_2}$ which are proportional to spherical harmonics
\begin{align}
u_{m_1}(t,r,\theta,\varphi)&=\sqrt{\frac{4\pi}{2l_1+1}}\,f(r)\,Y_{l_1,m_1}(\theta,\varphi)e^{i\omega_1 t},\label{ans1}\\
v_{m_2}(t,r,\theta,\varphi)&=\sqrt{\frac{4\pi}{2l_2+1}}\,g(r)\,Y_{l_2,m_2}(\theta,\varphi)e^{i\omega_2 t},\label{ans2}
\end{align}
where coordinates $t,r,\theta,\varphi$ are dimensionless and $m_1=0,\pm 1,\ldots,\pm l_1$ and $m_2=0,\pm 1,\ldots,\pm l_2$.  The parameters $\omega_1$ and $\omega_2$ are real numbers and they represent rotation in the internal space of fields.

The ansatz reduces the field equations to the set of two coupled ordinary equations which depend on radial variable $r$
\begin{align}
&\Sigma^{(1)}(r)=\frac{\widetilde \mu_1^2}{8}\sqrt{1+f^2}\;{\rm sgn}(f)+\lambda_1\frac{\alpha}{4}\left(\frac{f^2}{1+f^2}\right)^{\alpha-1}\left(\frac{g^2}{1+g^2}\right)^{\beta}f,\label{eq1f}\\
&\Sigma^{(2)}(r)=\frac{\widetilde \mu_2^2}{8}\sqrt{1+g^2}\;{\rm sgn}(g)+\lambda_2\frac{\beta}{4}\left(\frac{f^2}{1+f^2}\right)^{\alpha}\left(\frac{g^2}{1+g^2}\right)^{\beta-1}g\label{eq1g}
\end{align}
where left hand side expressions that originate in kinetic and gradient terms  read
\begin{align}
&\Sigma^{(1)}(r)\equiv f''+\frac{2}{r}f'+\omega_1^2\frac{1-f^2}{1+f^2}f-2\frac{ff'^2}{1+f^2}-\frac{l_1(l_1+1)}{r^2}f,\label{Sigma-f}\\
&\Sigma^{(2)}(r)\equiv g''+\frac{2}{r}g'+\omega_2^2\frac{1-g^2}{1+g^2}g-2\frac{gg'^2}{1+g^2}-\frac{l_2(l_2+1)}{r^2}g.\label{Sigma-g}
\end{align}
Here by definition ${\rm sgn}(0)=0$.

Now we would like to comment on the origin of expressions ${\rm sgn}(f)$ and ${\rm sgn}(g)$ in equations \eqref{eq1f} and \eqref{eq1g}. 
A distinguishable property of models with V-shaped potentials is the fact that their dynamics is not completely described by  Euler-Lagrange equations. The equations derived from a variational principle do not ``see" vacuum solutions which represent some physical configurations. For instance, this is a very characteristic property of the model with the Lagrangian density  ${\cal L}=\frac{1}{2} \eta^{\mu\nu}{\partial_{\mu}}\phi{\partial_{\nu}}\phi-|\phi|$,  where $\phi\in{\mathbb R}$. Its energy is given by $E=\int d^3x\left[\frac{1}{2}(\partial_0\phi)^2+\frac{1}{2}\nabla^2\phi+|\phi|\right]$ whereas the Euler-Lagrange equations have the form $\partial_{\mu}\partial^{\mu}\phi\pm 1=0$. The physical configuration  which minimizes the energy, $\phi=0$, is not a solution of the Euler-Lagrange equations! On the other hand, this solution plays an important role in  the field dynamics.  The vacuum solution can be formally included in the set of solutions of the field equation providing that the term $\pm 1$ is replaced by the ${\rm sgn}(\phi)$ term such that ${\rm sgn}(0):=0$. The equation of motion became $\partial_{\mu}\partial^{\mu}\phi+{\rm sgn}(\phi)=0$  (signum-Gordon model). When a scalar field is complex valued $\phi=fe^{i\chi}$ and $f\ge 0$ the signum term is restricted to two values ${\rm sgn}(f)=0$ and  ${\rm sgn}(f)=+1$. This is exactly the case of fields considered in the current work. By assumption the functions $f(r)$ in \eqref{ans1} and $g(r)$ in \eqref{ans2} are non-negative. Our definition of the signum function makes the vacuum configuration $f=0$ and $g=0$ an  explicit  solution of \eqref{eq1f} and \eqref{eq1g}.

\subsection{\label{hdensity}The Hamiltonian density}
The Hamiltonian density associated with \eqref{lagr2} is of the form
\begin{align}
{\cal H}&=\frac{\delta{\cal L}_1}{\delta(\partial_0u_i)}\partial_0u_i+\frac{\delta{\cal L}_1}{\delta(\partial_0u^*_i)}\partial_0u^*_i-{\cal L}_1+\nonumber\\
&+\frac{\delta{\cal L}_2}{\delta(\partial_0v_j)}\partial_0v_j+\frac{\delta{\cal L}_2}{\delta(\partial_0v^*_j)}\partial_0v^*_j-{\cal L}_2-{\cal L}_{int}
\end{align}
and it can be cast in the form
\be
{\cal H}=\frac{M_1^2}{r_0^2}{\cal H}_1+\frac{M_2^2}{r_0^2}{\cal H}_2+\lambda W
\ee
where $W$ and
\be
{\cal H}_a:=-\tau^{a}_{00}+\sum_{k=1}^3\,\tau^{a}_{kk}+\widetilde \mu^2_aV_a,\qquad a=1,2\label{hamilt1}
\ee
 are dimensionless functions. The definition of expressions $\tau^a_{\mu\nu}$ is given in \eqref{tau}. The only difference is that expressions $\tau^a_{\mu\nu}$ considered in \eqref{hamilt1} contain derivatives with respect to dimensionless coordinates. Providing that  $\lambda\neq 0$ one can cast the Hamiltonian density in the form 
 \be
 {\cal H}=\lambda\left(\frac{1}{\lambda_1}{\cal H}_1+\frac{1}{\lambda_2}{\cal H}_2+W\right).\label{hamiltfin}
 \ee
 The expressions ${\cal H}_a$ and $W$ considered for Q-ball/Q-shell ansatz read
 \begin{align}
 {\cal H}_1&=\frac{4}{(1+f^2)^2}\left[f'^2+\omega_1^2 f^2+\frac{l_1(l_1+1)}{r^2}(1+f^2)f^2\right]+\widetilde \mu^2_1\frac{f}{\sqrt{1+f^2}},\\
 {\cal H}_2&=\frac{4}{(1+g^2)^2}\left[g'^2+\omega_2^2 g^2+\frac{l_2(l_2+1)}{r^2}(1+g^2)g^2\right]+\widetilde \mu^2_2\frac{g}{\sqrt{1+g^2}},\\
 W&=\left(\frac{f^2}{1+f^2}\right)^{\alpha}\left(\frac{g^2}{1+g^2}\right)^{\beta}, \qquad \alpha,\beta \ge 1.
 \end{align}
 The vacuum configuration $f(r)=0$ and $g(r)=0$ has zero total energy.
 
 \subsection{Noether charges}
 
 In terms of the parametrization (\ref{parametryzacja}), the symmetry $SU(N_a+1)$ reduces to $SU(N_a)\otimes U(1)$
	and the Lagrangian (\ref{lagr2}) possesses the symmetry $U(1)^{N_1}\otimes U(1)^{N_2}$ under the transformation
	\begin{align}
	&u_i\to e^{i\sigma_i}u_i,~~i=1,\cdots,N_1,
	\label{ui} \\
	&v_i\to e^{i\tau_i}v_i,~~i=1,\cdots,N_2
	\label{vi}
	\end{align}
	where $\sigma_i,\tau_i$ are some global parameters. 
	This global symmetry and the Noether charge have an important role in stabilization of solutions. 
	The Noether currents corresponding to (\ref{ui}) are
	\begin{align}
	J_\mu^{(i)}=-\frac{4i}{(1+u^{\dagger} \cdot u)^2}\sum_{j=1}^N[u_i^*\Delta_{ij}(u)^2\partial_\mu u_j-\partial_\mu u_j^*\Delta_{ji}(u)^2u_i]\,.
	\end{align}
	Considerning that (\ref{vi}) leads to  similar expressions as (\ref{ui}) we show 
	the results for (\ref{ui}). The results for (\ref{vi}) can be obtained replacing $u\leftrightarrow v$ and $f\leftrightarrow g$.
	In terms of parametrization (\ref{ans1}), the currents have the form
	\begin{align}
	&J^{(m)}_0=8\omega\frac{(n-m)!}{(n+m)!}\frac{f^2}{(1+f^2)^2}(P_n^m(\cos\theta))^2\,,
	\label{currentt}\\
	&J^{(m)}_\varphi=8m\frac{(n-m)!}{(n+m)!}\frac{f^2}{1+f^2}(P_n^m(\cos\theta))^2.
	\label{currentp}
	\end{align}
	The other two components read $J_r^{(m)}=J_\theta^{(m)}=0$, for $m=-n,-n+1,\cdots,n-1,n$. 
	It can be seen directly that the current is conserved because the components (\ref{currentt}),(\ref{currentp}) depend only on 
	$r,\theta$. Hence
	\begin{align}
	\frac{1}{\sqrt{-g}}\partial_\mu(\sqrt{-g}g^{\mu\nu} J_\nu^{(m)})
	=\partial_0J_0^{(m)}+\frac{1}{r^2\sin^2\theta}\partial_\varphi J_\varphi^{(m)}=0\,.
	\label{current_cons}
	\end{align}
	Integrating the continuity equation (\ref{current_cons}) involving the four-current  we get
	\begin{align}
	\int dt\int d^3x \sqrt{-g}\biggl(
	\partial_0J_0^{(m)}+\frac{1}{r^2\sin^2\theta}\partial_\varphi J_\varphi^{(m)}
	\biggr)=0\,.
	\end{align}
	Assuming that the spatial component $J_\varphi^{(m)}$  decreases sufficiently quickly at the spatial 	boundary, we obtain the conserved Noether charge 
	\begin{align}
	Q_1^{(m)}&=\frac{1}{2}\int d^3 x\sqrt{-g}J_0^{(m)}(x)
	=\frac{16\pi\omega}{2n+1}\int^\infty_0r^2dr\frac{f^2}{(1+f^2)^2}\,.
	\end{align}
	The spatial components of the Noether currents do not contribute to the charges; however,
	they may be useful to define the integrals
	\begin{align}
	q_1^{(m)}&:=\frac{3}{2}\int d^3x\sqrt{-g}\frac{J^{(m)}_\varphi (x)}{r^2} 
	=\frac{48\pi m}{2n+1}\int^\infty_0dr\frac{f^2}{1+f^2}\,.
	\label{qintegral}
	\end{align}
	As it was already pointed out, the Noether charge associated with the second model $Q_2^{(m)}$ as well as  the integral $q_2^{(m)}$ corresponding to (\ref{qintegral})
	for (\ref{vi}) can be obtained just replacing $f\to g$. 
	The total energy $E=\int d^3 x\;{\cal H}$, where $\cal H$ is given by \eqref{hamiltfin}, can be expressed in terms of $Q_i^{(m)}$ and $q_i^{(m)}$ 
	\begin{align}
	&E=\sum_{m=-l_1}^{m=l_1}\Big(\omega_1 Q_1^{(m)}+m q_1^{(m)}\Big)+\sum_{m=-l_2}^{m=l_2}\Big(\omega_2 Q_2^{(m)}+m q_2^{(m)}\Big)
	\nonumber \\
	&+4\pi\int r^2 dr\biggl[\frac{4f'^2}{(1+f^2)^2}+\frac{4g'^2}{(1+g^2)^2}+\tilde{\mu}_1^2\frac{f}{\sqrt{1+f^2}}
	+\tilde{\mu}_2^2\frac{g}{\sqrt{1+g^2}}
	+\biggl(\frac{f^2}{1+f^2}\biggr)^\alpha\biggl(\frac{g^2}{1+g^2}\biggr)^\beta\biggr]
	\end{align}
	where for simplicity we set $\lambda=\lambda_1=\lambda_2=1$.
	For the full understanding, we need to know $E$ as a function of $Q_1,Q_2$. In particular, it is useful to know 
	the relation between $E$ and the sum of the charges $Q=Q_1+Q_2$. 
	The analytical study might be possible only in limited cases such as a thin-wall approximation of the model.  
	However, for the general case we have to rely on the numerical analysis, which will be discussed in  the subsequent section.

\section{Solutions}\label{secondmodel}

First we have looked at the most generic case of two overlapping compactons. 
There are some qualitatively different cases: Q-ball--Q-ball (BB), Q-ball--Q-shell (BS) and Q-shell--Q-shell (SS). The numerical examples presented in this paper were obtained for $\alpha=\beta=1$. We have checked that for higher values of $\alpha$ and $\beta$ there is no qualitative change in the form of profile functions.

\subsection{Q-ball--Q-ball}

\begin{figure}[t]
\centering
\subfigure[]{\includegraphics[width=0.45\textwidth,height=0.27\textwidth, angle =0]{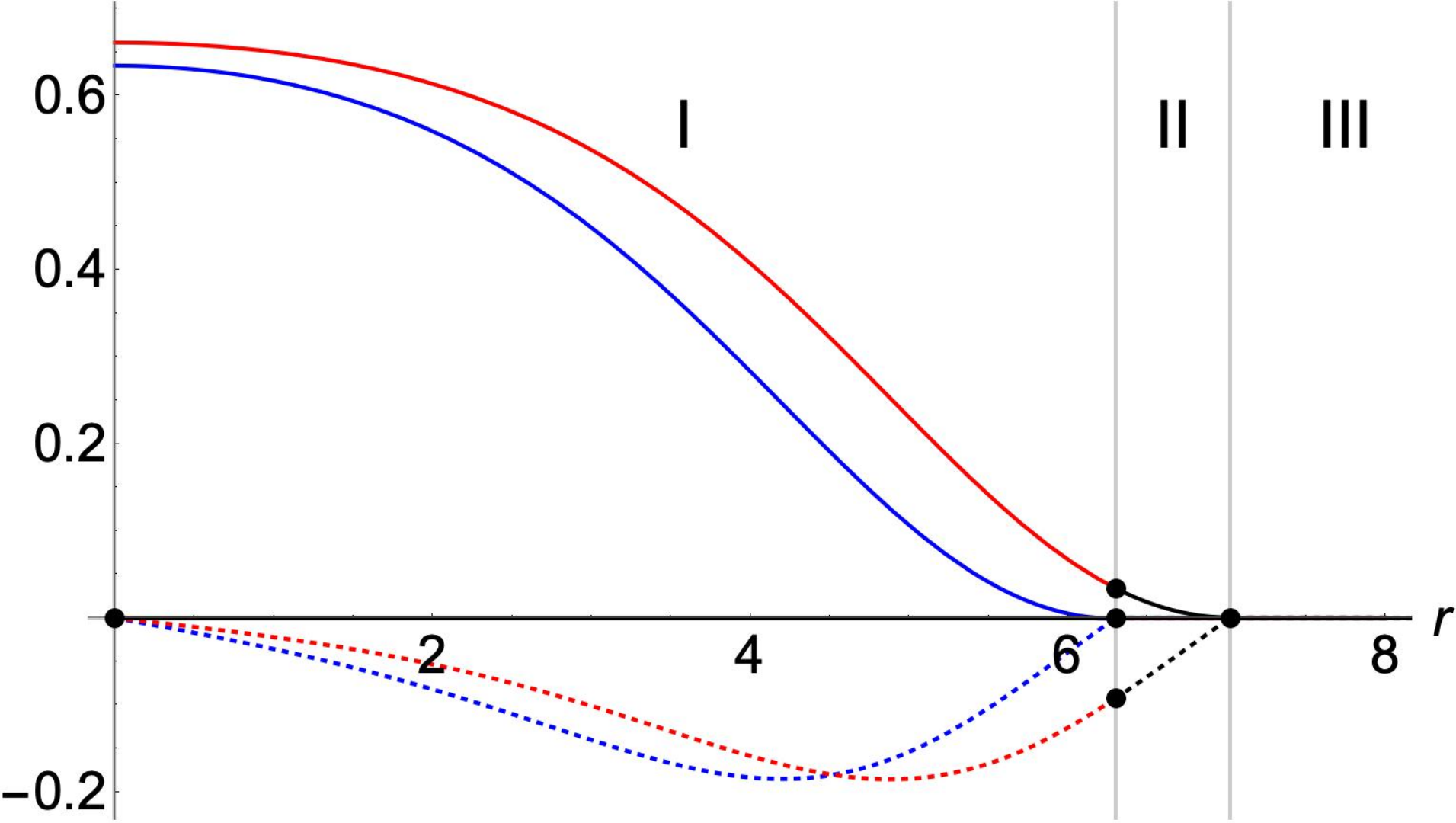}}\hspace{0.5cm}
\subfigure[]{\includegraphics[width=0.45\textwidth,height=0.27\textwidth, angle =0]{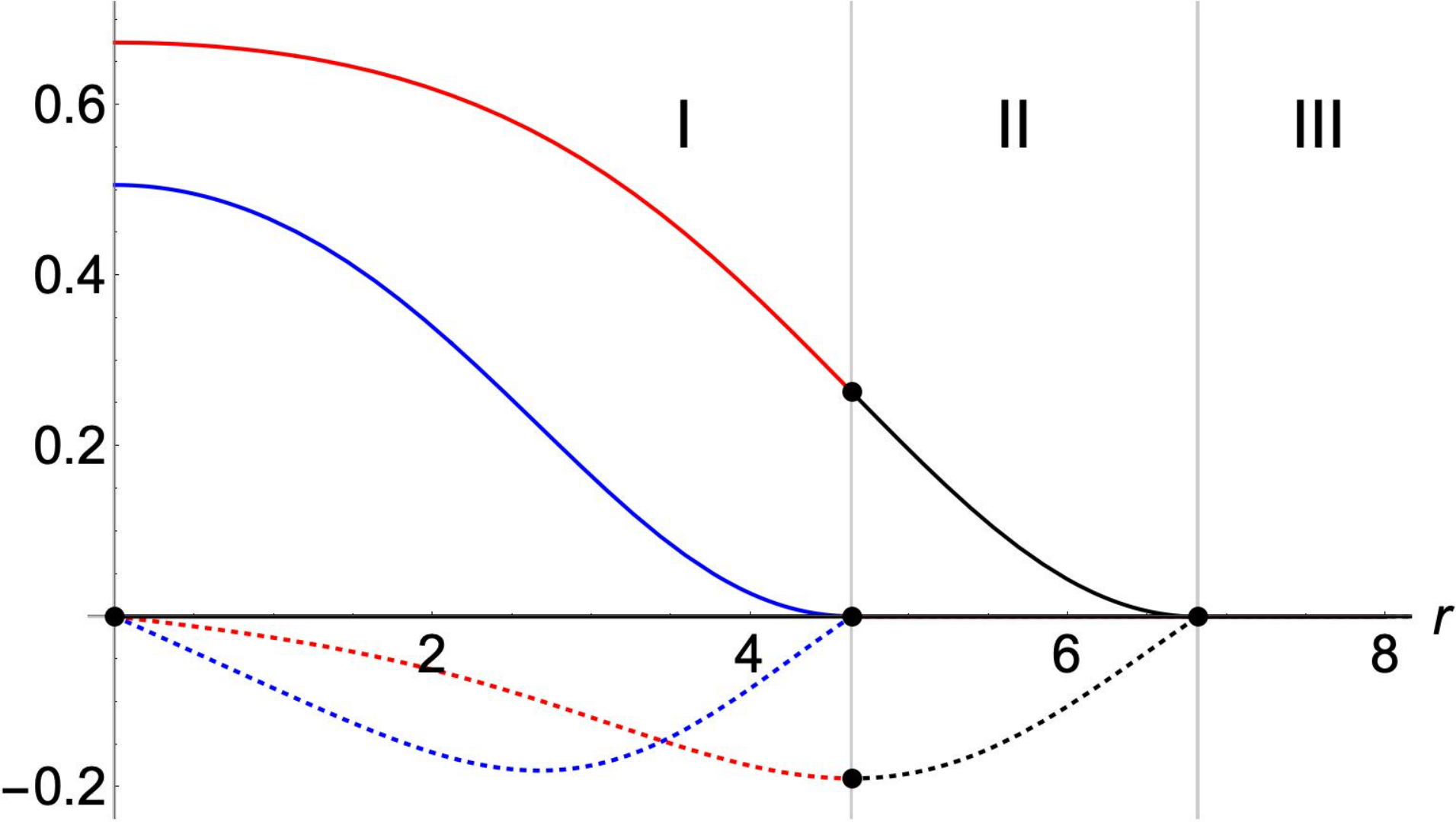}}
\caption{\label{fig:fig1}The $\mathbb{C}P^1-\mathbb{C}P^1$ Q-balls for 
(a) $(\omega_1,\omega_2)=(1.05,1.0)$ and (b) $(\omega_1,\omega_2)=(1.2,1.0)$. 
The other constants have values $\widetilde \mu_1^2=\widetilde \mu_1^2=1.0$, $\lambda_1=\lambda_2=1.0$. 
The radii of Q-balls have values (a) $(R_1,R_2)=(6.30,7.02)$ and (b) $(R_1,R_2)=(4.64,6.82)$. 
The profile function $f(r)$ has lower amplitude than function $g(r)$ (solid curves). 
The derivatives $f'(r)$ and $g'(r)$ are represented by dashed curves.}
\end{figure}

The simplest BB solution contains two overlapping $\mathbb{C}P^1$ Q-balls. 
The profile functions $f(r),g(r)$ and their derivatives are sketched in Fig.\ref{fig:fig1}. 
We consider  $\omega_1>\omega_2$ because for $\omega_1=\omega_2$ 
and  $\widetilde \mu_1^2=\widetilde \mu_1^2$, $\lambda_1=\lambda_2$ the profile functions are equal 
$f(r)=g(r)$ and the model reduces to a single $\mathbb{C}P^1$ case with analytical deformation of the V-shaped potential.
The bigger difference between $\omega_1$ and $\omega_2$ is  the less alike the radial curves are (the discrepancy between  $R_1$ and  $R_2$ as well as between $f(0)$ and $g(0)$ grows when $|\omega_1-\omega_2|$ increases).

\begin{figure}[ht]
\centering
\subfigure[]{\includegraphics[width=0.45\textwidth,height=0.27\textwidth, angle =0]{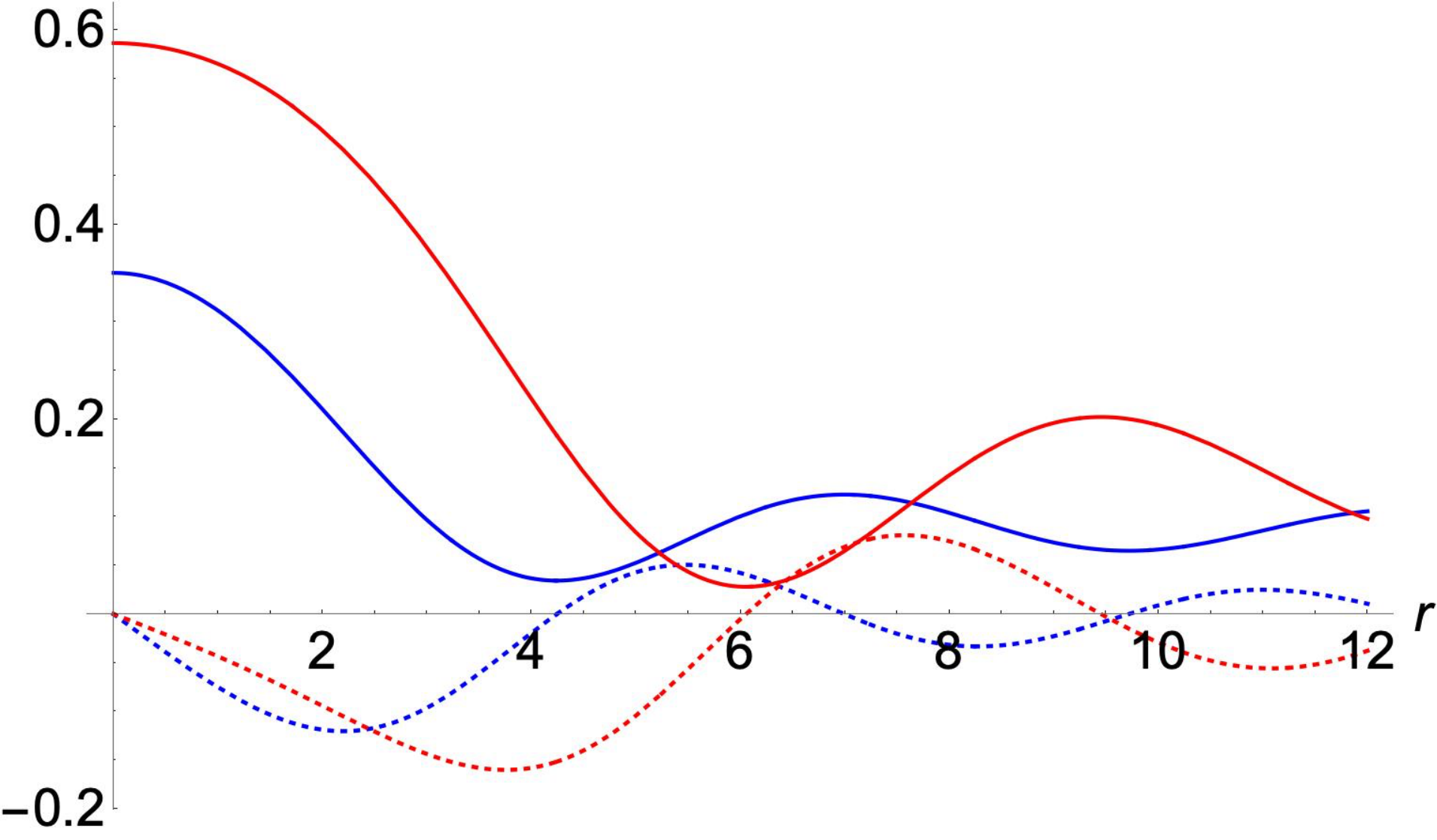}}\hspace{0.5cm}
\subfigure[]{\includegraphics[width=0.45\textwidth,height=0.27\textwidth, angle =0]{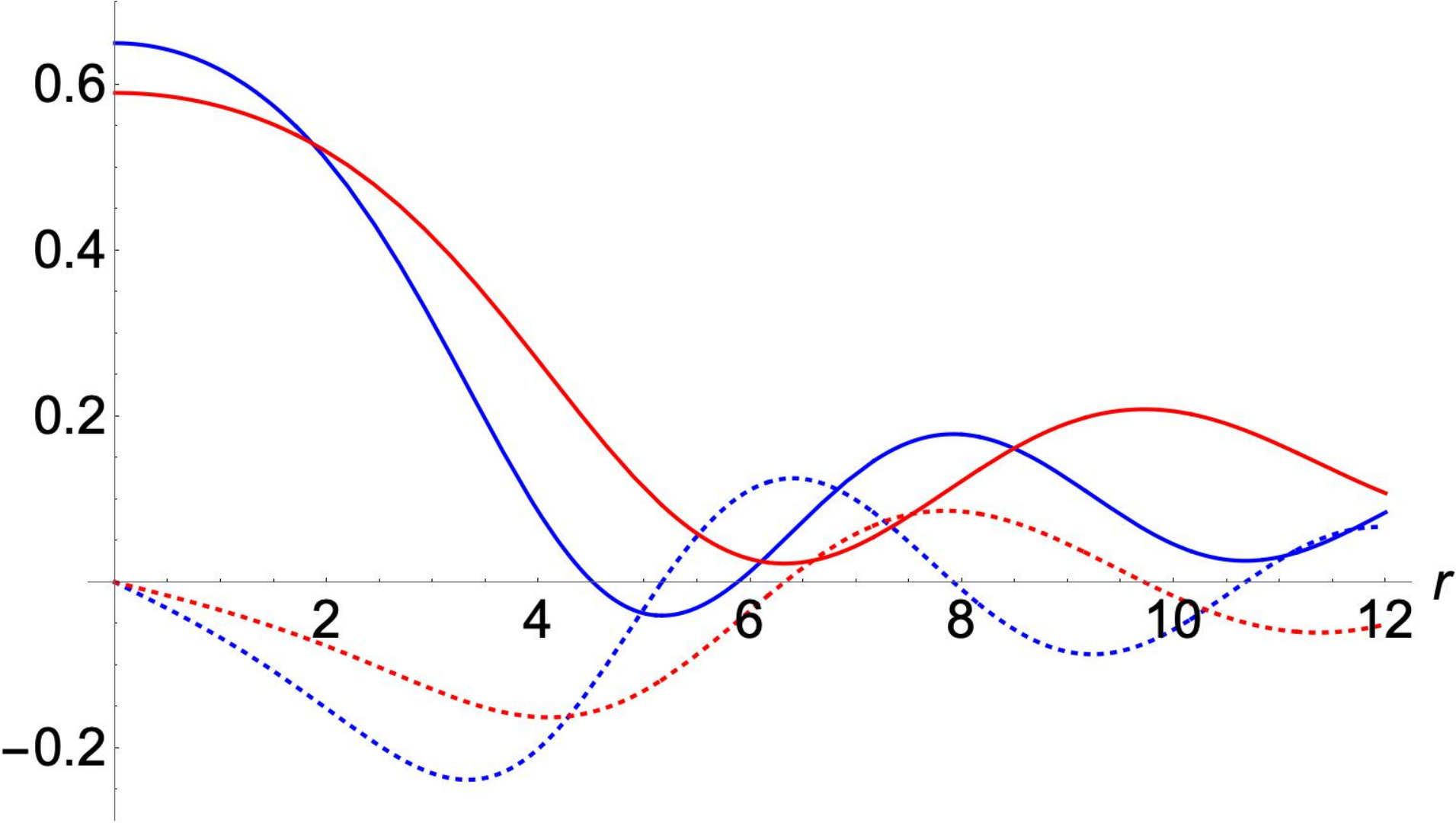}}
\caption{Wrong shooting curves for the $\mathbb{C}P^1-\mathbb{C}P^1$ case. (a) Both shooting parameters $a_0$ and 
$b_0$ are too small and numerical solution never reaches the vacuum value, (b) $a_0$ is too big and $b_0$ too small -- the function $f(r)$ changes its sign.}\label{fig:fig2}
\end{figure}

\begin{figure}[t]
\centering
\subfigure{\includegraphics[width=0.7\textwidth,height=0.4\textwidth, angle =0]{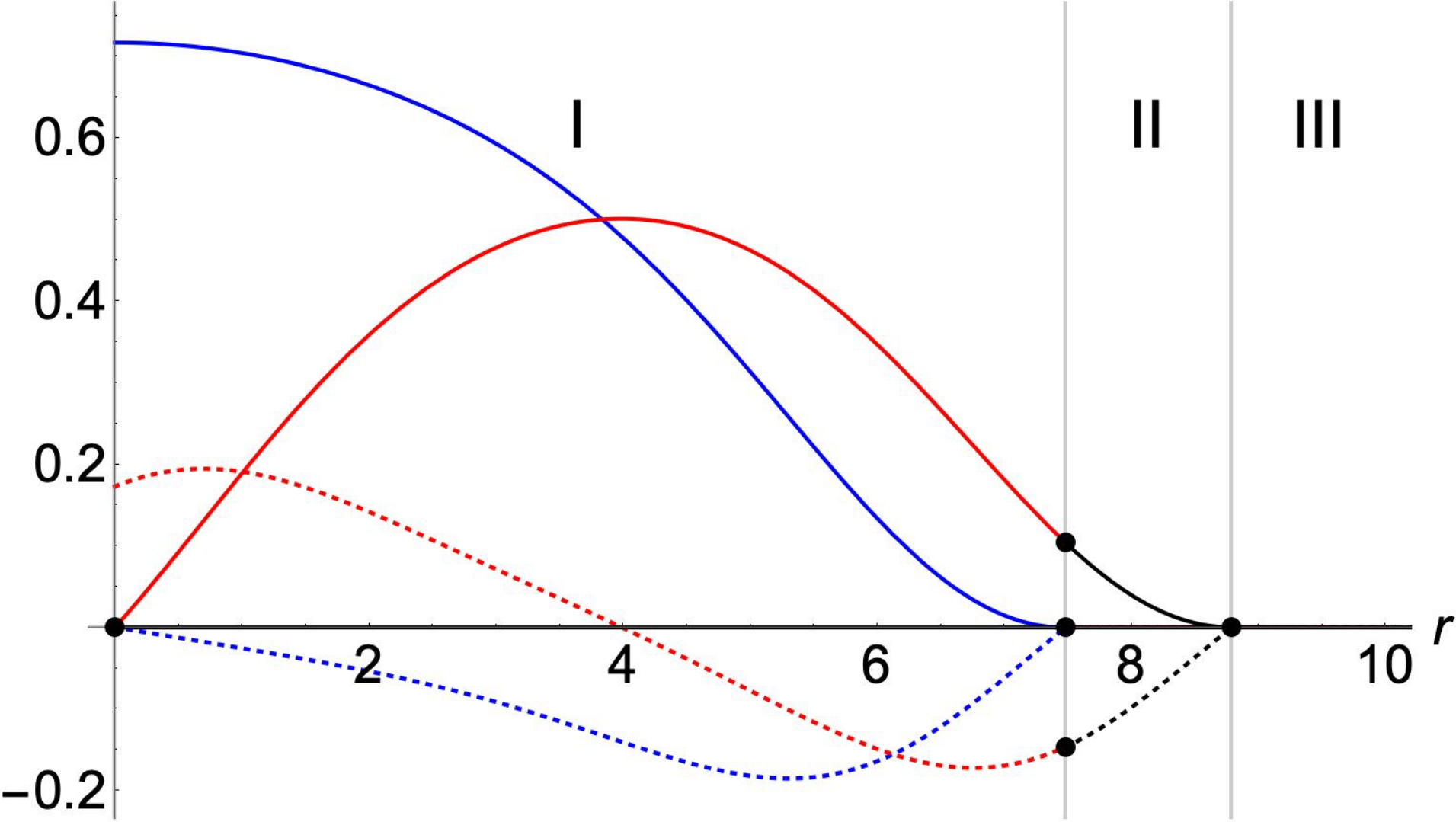}}
\caption{The BB solution for the $\mathbb{C}P^1-\mathbb{C}P^3$ case. 
The values of free parameters $\omega_1=\omega_2=1.0$, $\lambda_1=\lambda_2=1.0$, $\widetilde \mu_1^2=\widetilde \mu_1^2=1.0$. 
The shooting parameters have values $(a_0,b_1)=(0.717, 0.173)$. Both profile functions are nontrivial in I, $0<r<R_1$, the function $f(r)=0$ whereas $g(r)>0$ in II  ($R_1<r<R_2$) and $f(r)=g(r)=0$ in III ($r>R_2$). The dashed curves represent derivatives of profile functions with respect to $r$.}\label{fig:fig3}
\end{figure}

\begin{figure}[t]
\centering
\subfigure[]{\includegraphics[width=0.6\textwidth,height=0.5\textwidth, angle =0]{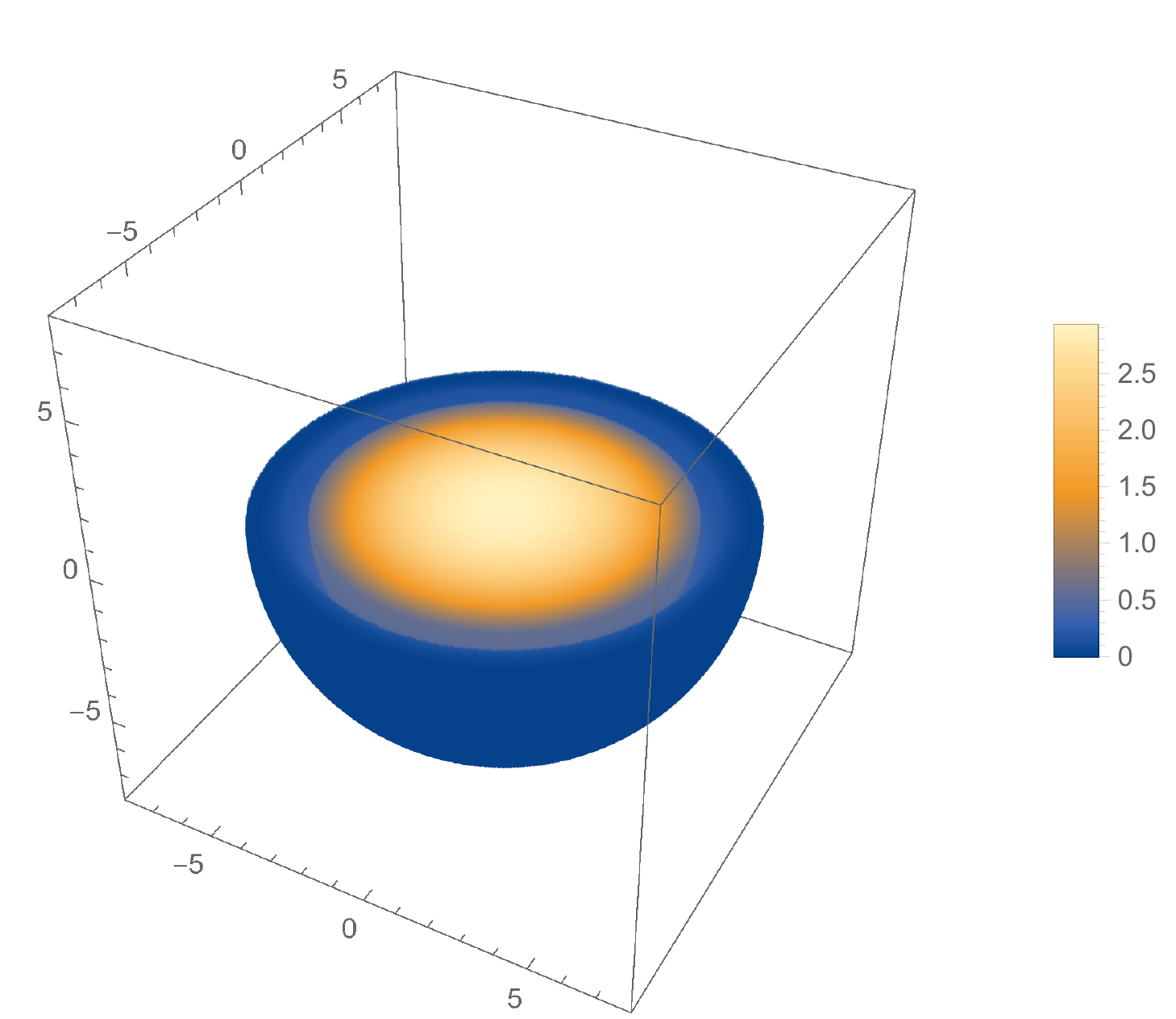}}\hspace{0.5cm}
\subfigure[]{\includegraphics[width=0.3\textwidth,height=0.2\textwidth, angle =0]{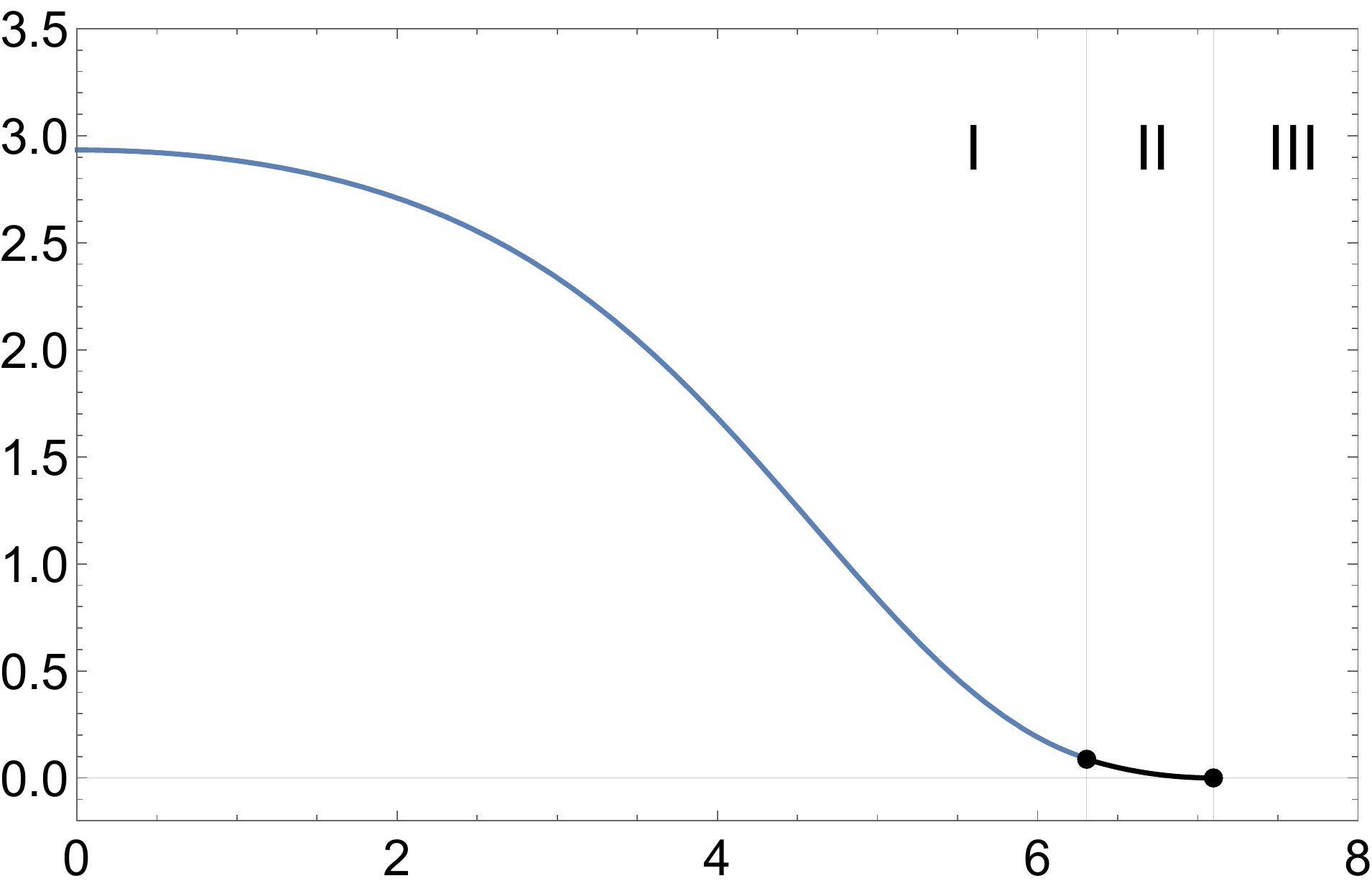}}
\caption{\label{hdensityBB}The Hamiltonian density of the BB solution corresponding to Fig.\ref{fig:fig1}(a) 
for the  $\mathbb{C}P^1-\mathbb{C}P^1$ model. Figure (a) the 3D plot and (b) 
the radial density associated with (a). The Hamiltonian density is the monotonically decreasing function.}
\end{figure}

The solution is obtained by numerical integration of radial equations using the shooting method. The shooting parameters  $a_0\equiv f(0)$ and $b_0\equiv g(0)$ are two free parameters of series expansion at $r=0$
\be
f(r)=\sum_{k=0}^{\infty} a_kr^k,\qquad g(r)=\sum_{k=0}^{\infty} b_kr^k.
\ee
Plugging these formulas into equations \eqref{eq1f} and \eqref{eq1g} one can determine recursively coefficients with $k=1,2,\ldots$. 
The coefficients with $k=1,3,5,\ldots$ vanish. First two even coefficients read
\begin{align}
a_2&=\frac{1}{6}\left[\frac{a_0b_0^2}{4(1+b_0^2)}\lambda_1+\frac{\widetilde\mu_1^2}{8}\sqrt{1+a_0^2}-\frac{a_0(1-a_0^2)}{1+a_0^2}\omega_1^2\right],\\
b_2&=\frac{1}{6}\left[\frac{b_0a_0^2}{4(1+a_0^2)}\lambda_2+\frac{\widetilde\mu_1^2}{8}\sqrt{1+b_0^2}-\frac{b_0(1-b_0^2)}{1+b_0^2}\omega_2^2\right].
\end{align}
It shows that the second derivative of the profile function $f(r)$ at $r=0$ depends on the value of $g(0)$ and {\it vice versa}. 
In the region I ($0<r<R_1$) both functions $f(r)$ and $g(r)$ are nontrivial. 
In order to get a compact solution, one has to choose properly both shooting parameters.  
If $a_0$ and $b_0$ are too small the region I extends to spatial infinity and the numerical solution never reaches vacuum value. 
Such field configuration has infinite energy and thus it cannot be acceptable, see Fig.\ref{fig:fig2}(a). 
On the other hand, when the shooting parameter is too big the function changes its sign. 
As we look for solutions with a non-negative profile function this numerical solution also must be rejected. 
An example of such a solution is shown in Fig.\ref{fig:fig2}(b). The behavior of numerical curves shows 
that there exists a certain point in the space $(a_0,b_0)$ such that profile $f(R_1)=0$ and $f'(R_1)=0$ and $g(R_2)=0$ and $g'(R_2)=0$. 
The profile function $f(r)$ vanishes on the segment II, $R_1<r<R_2$, whereas $g(r)$ is nontrivial. Finally, on the third segment $r>R_2$ both profile functions $f(r)$ and $g(r)$ take vacuum value. 

The behavior of profile functions at $R_1$ 
\begin{align}
f(r)&=\frac{\widetilde\mu_1^2}{16}(r-R_1)^2-\frac{\widetilde\mu_1^2}{24 R_1}(r-R_1)^3+\ldots,\nonumber\\
g(r)&=B_0+B_1(r-R_1)+\ldots
\end{align}
where $B_0\equiv g(R_1)$ and $B_1\equiv g'(R_1)$ are determined by numerical integration of coupled equations. Similarly, in vicinity of $R_2$
\[
f(r)=0,\qquad{\rm and}\qquad g(r)=\frac{\widetilde\mu_2^2}{16}(r-R_2)^2-\frac{\widetilde\mu_2^2}{24 R_2}(r-R_2)^3+\ldots
\]
This expansion shows that both fields have a parabolic approach to  vacuum.

Another BB compacton is obtained in the $\mathbb{C}P^1-\mathbb{C}P^3$ model. In this case the radial function $g(r)$ has zero at the center $r=0$. The expansion has two free coefficients $a_0$ and $b_1$.  The other coefficients are determined in terms of these two parameters. The leading terms of expansion at $r=0$ have the form
\begin{align}
f(r)&=a_0+\left[\frac{\widetilde\mu_1^2}{48}\sqrt{1+a_0^2}-\frac{a_0}{6}\frac{1-a_0^2}{1+a_0^2}\omega_1^2\right]r^2+\ldots,\\
g(r)&=b_1 r+\frac{\widetilde\mu_2^2}{32}r^2+\frac{b_1}{40}\left[8b_1^2+\frac{\lambda_2 a_0^2}{1+a_0^2}-4\omega_2^2\right]r^3+\ldots
\end{align}
Looking at higher terms of expansion we get $a_3=0$ and observe that $a_4$ depends on $b_1$. Note that term $b_2$ is fixed by the value of constant $\widetilde \mu_2^2$. Looking at Fig.\ref{fig:fig3} we see that the function $f(r)$ associated with the $\mathbb{C}P^1$ model reaches the vacuum value at $r=R_1$ {\it i.e.} before the function $g(r)$ changes its sign. For  $r>R_1$ $f(r)=0$ and $g(r)>0$ until it reaches zero at $r=R_2$. The free parameters $a_0$ and $b_1$ are chosen in order to get local minima of two profile functions at the $r$ axis. In the considered case $(a_0,b_1)=(0.717, 0.173)$ and $(R_1,R_2)=(7.48,8.79)$.

In Fig.\ref{hdensityBB}, we present the 3D plot of the Hamitlonian density discussed in \ref{hdensity} corresponding to the BB solution (plotted in Fig.\ref{fig:fig1}) for the $\mathbb{C}P^1-\mathbb{C}P^1$ model.  
The radial function of density decreases monotonically in similarity to radial functions describing fields.

\begin{figure}[t]
\centering
\subfigure[]{\includegraphics[width=0.45\textwidth,height=0.27\textwidth, angle =0]{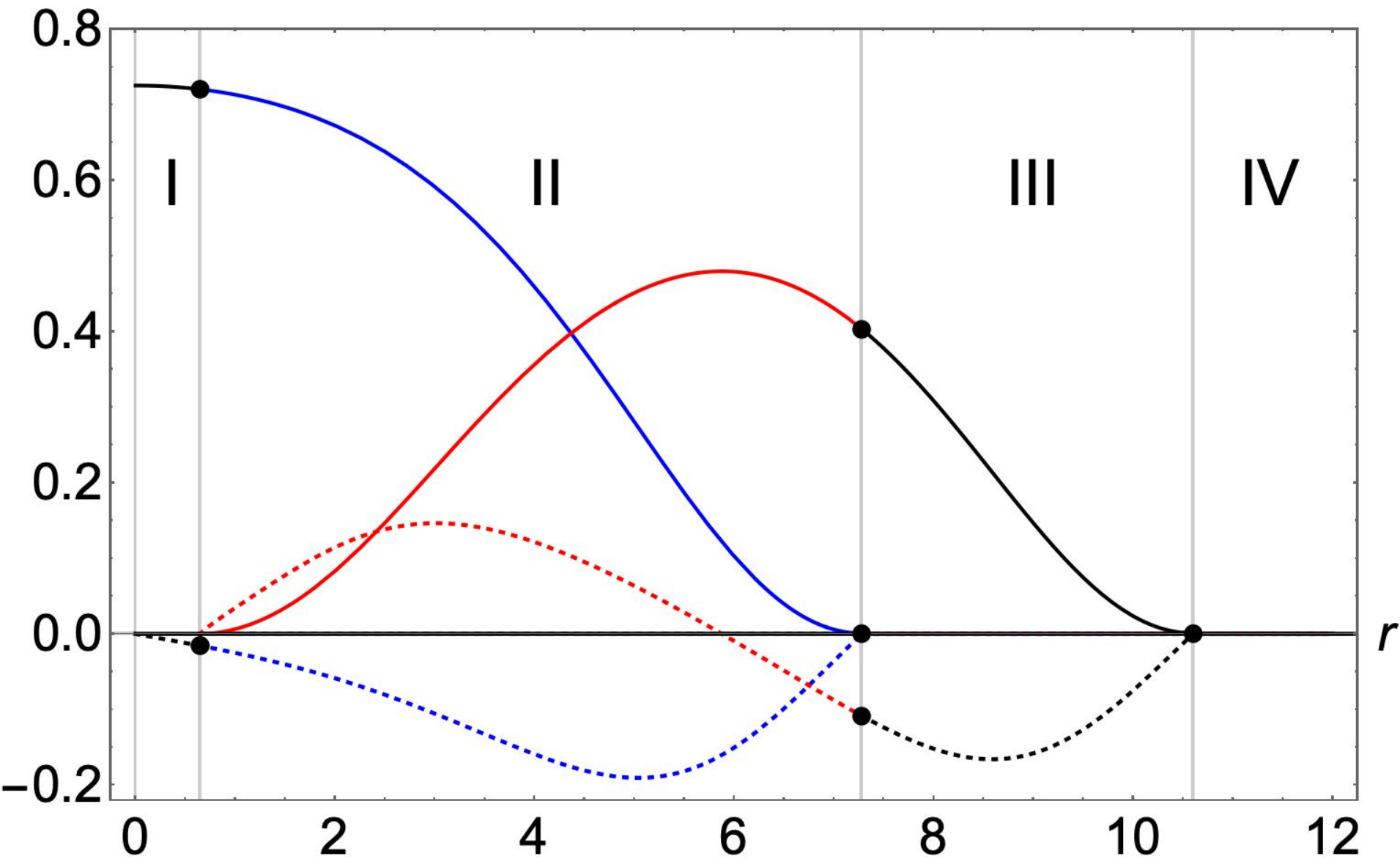}}\hspace{0.5cm}
\subfigure[]{\includegraphics[width=0.45\textwidth,height=0.27\textwidth, angle =0]{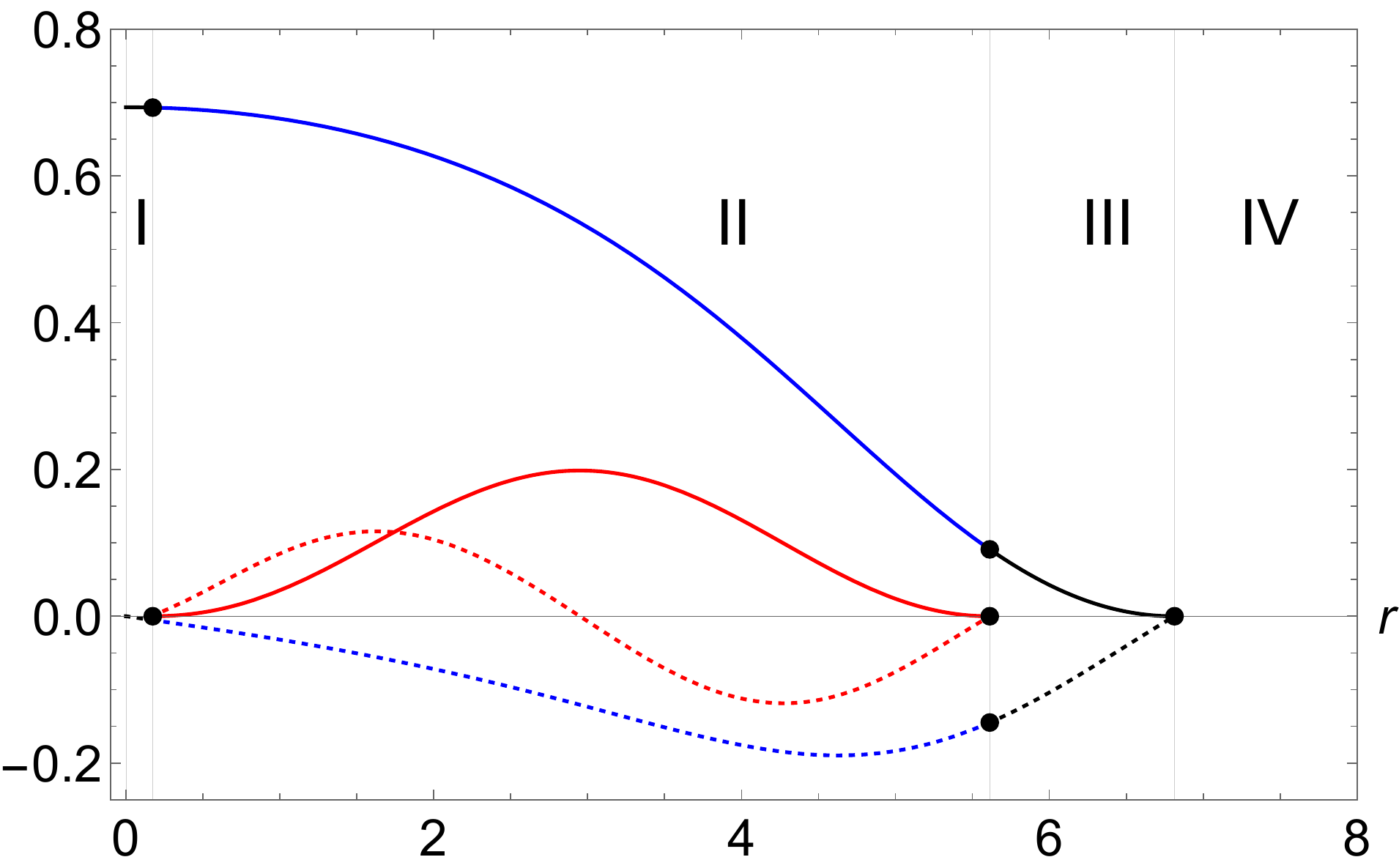}} 
\subfigure[]{\includegraphics[width=0.45\textwidth,height=0.27\textwidth, angle =0]{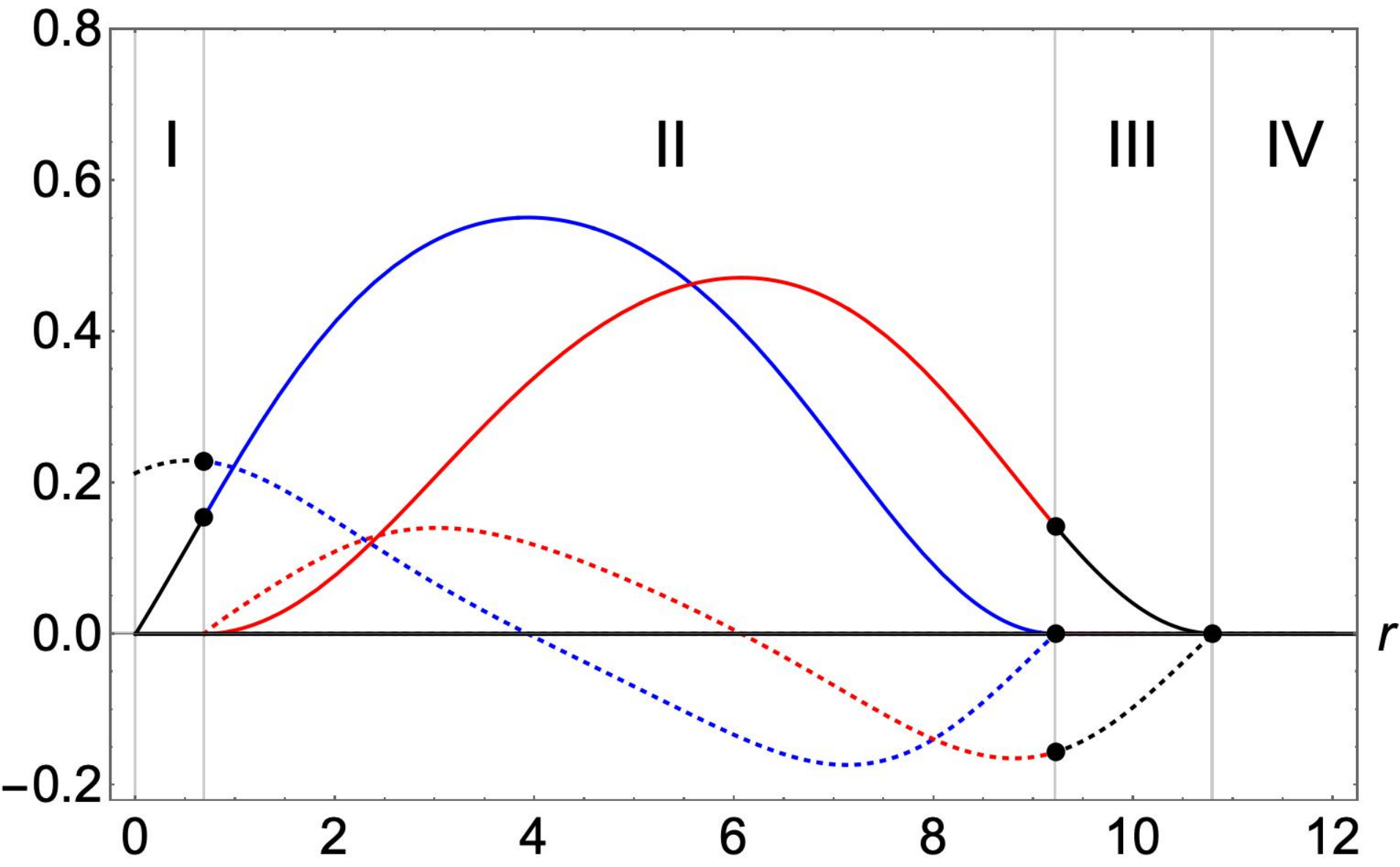}}\hspace{0.5cm}
\subfigure[]{\includegraphics[width=0.45\textwidth,height=0.27\textwidth, angle =0]{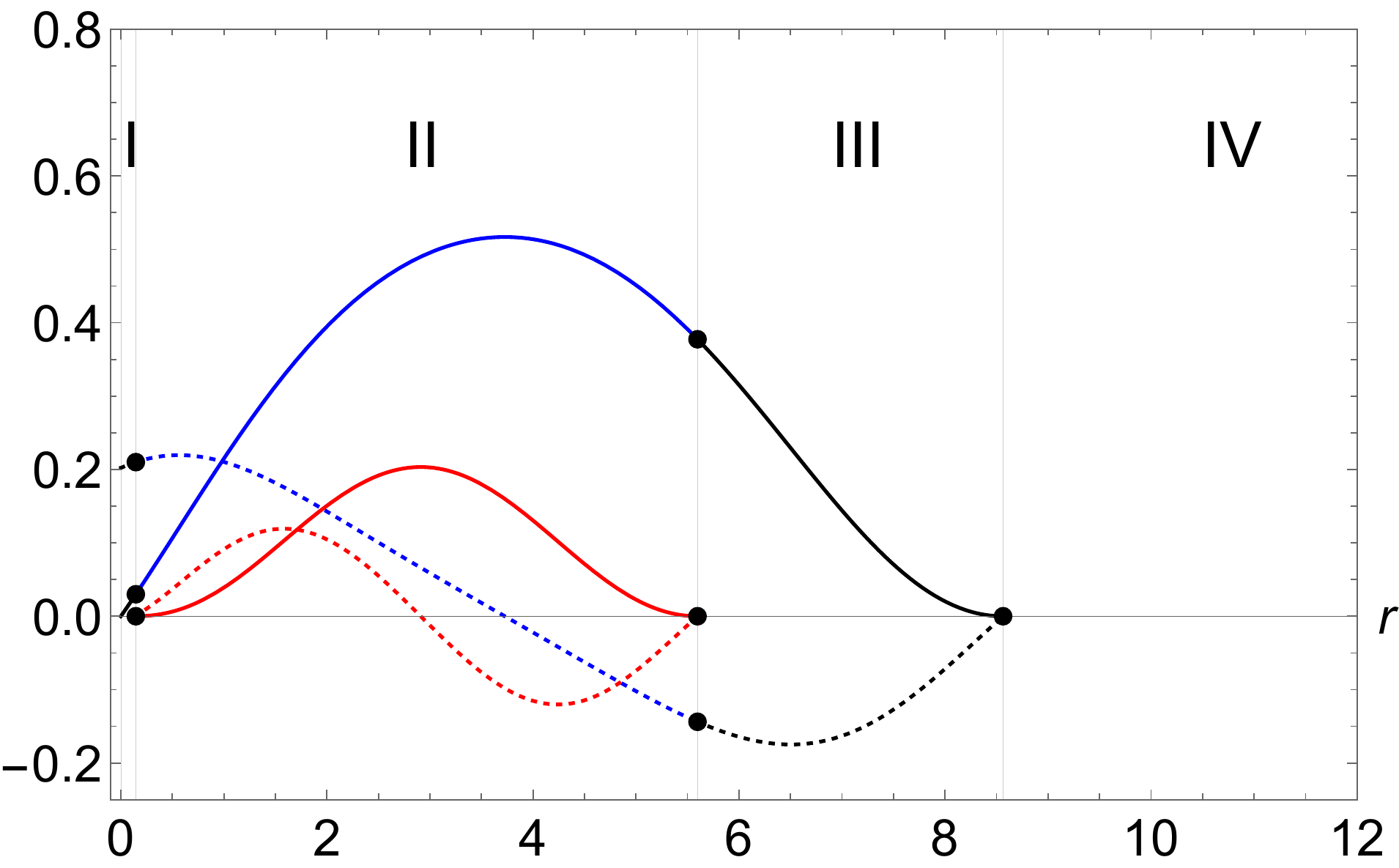}}

\caption{\label{fig:fig5}The BS solution for $(\lambda_1,\lambda_2)=(1.0,1.0)$, $(\widetilde \mu_1^2,\widetilde \mu_2^2)=(1.0,1.0)$.  
(a)  The $\mathbb{C}P^1-\mathbb{C}P^5$ model with $(\omega_1, \omega_2) = (1.0,1.0)$, shooting parameters $(a_0,R_1)=(0.725,0.643)$. 
The outer radii of Q-ball and Q-shell are given by $(R_2,R_3)=(7.28,10.60)$.  
(b)  $\mathbb{C}P^1-\mathbb{C}P^5$ model with $(\omega_1, \omega_2) = (1.0,1.5)$, shooting parameters $(a_0,R_1)=(0.694,0.174)$. 
The outer radii of Q-ball and Q-shell are given by $(R_2,R_3)=(6.81,5.61)$. 
(c) The $\mathbb{C}P^3-\mathbb{C}P^5$ model with $(\omega_1 , \omega_2) =(1.0,1.0)$, shooting parameters $(a_1,R_1)=(0.212,0.683)$. 
The outer radii of Q-ball and Q-shell are given by $(R_2,R_3)=(9.22,10.81)$.  
(d) The $\mathbb{C}P^3-\mathbb{C}P^5$ model with $(\omega_1 , \omega_2) =(1.0,1.5)$, shooting parameters $(a_1,R_1)=(0.202,0.145)$. The outer radii of Q-ball and Q-shell are given by $(R_2,R_3)=(8.56,5.60)$.}
 \end{figure}

\begin{figure}[t]
\centering
\subfigure[]{\includegraphics[width=0.6\textwidth,height=0.5\textwidth, angle =0]{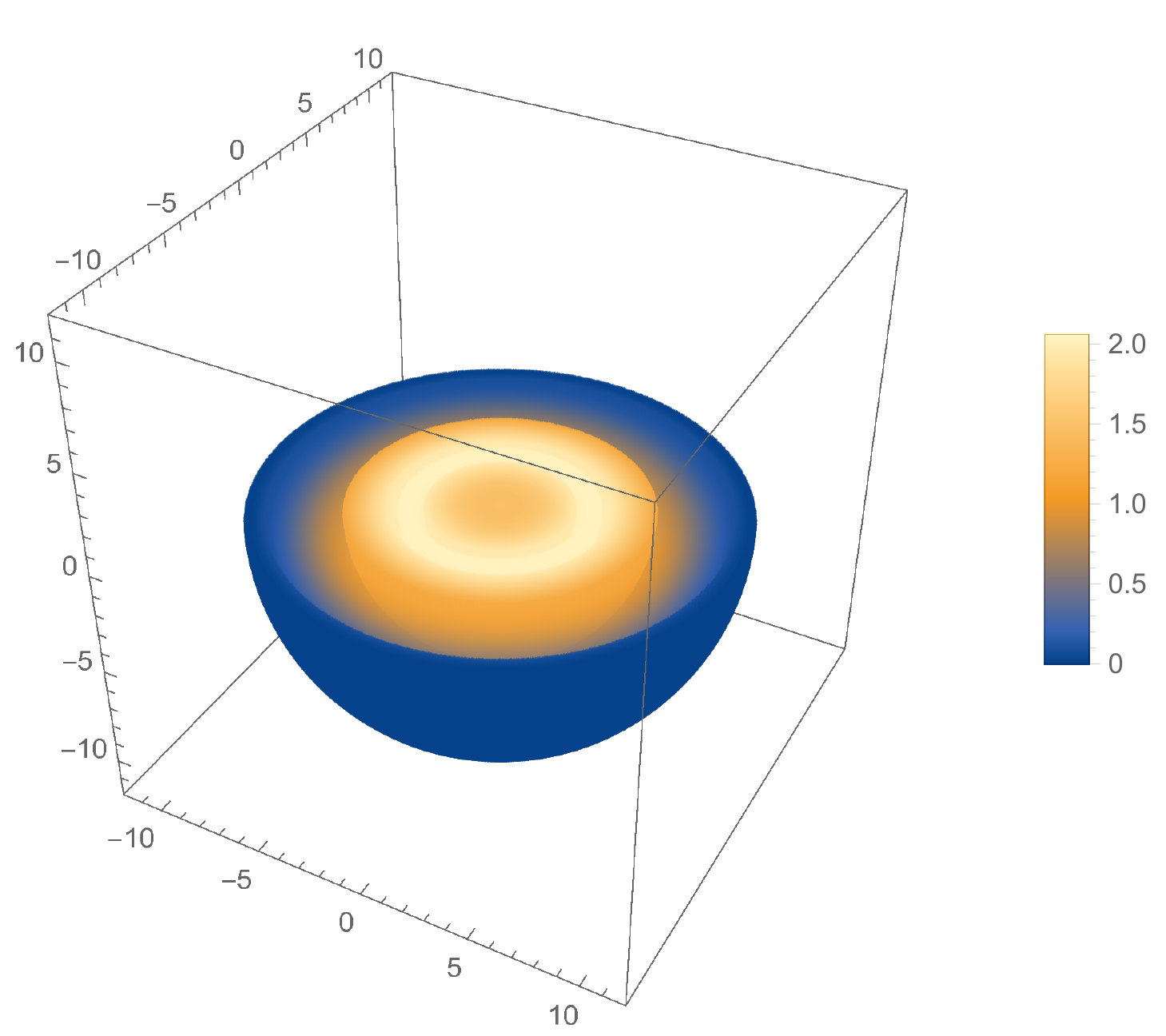}}\hspace{0.5cm}
\subfigure[]{\includegraphics[width=0.3\textwidth,height=0.2\textwidth, angle =0]{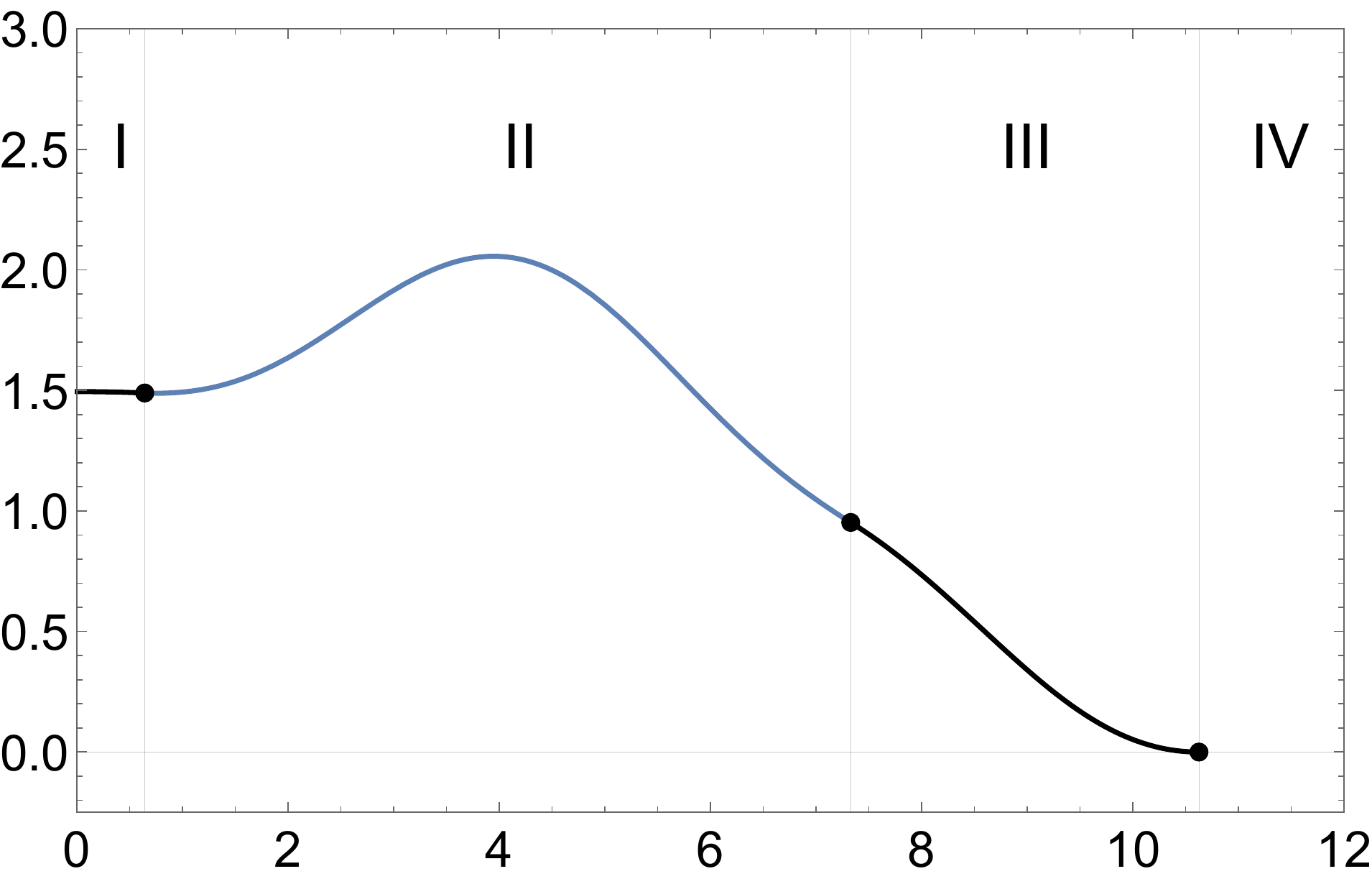}}
\caption{\label{hdensityBS}The Hamiltonian density of the BS solution corresponding to Fig.\ref{fig:fig5}(a)
for the $\mathbb{C}P^1-\mathbb{C}P^5$ model. Figure (a) the 3D plot and (b)
the radial density associated with (a). The Hamiltonian density is still a ball like shape, however, it contains a slight peak 
localized around $r\sim 4$.}
\end{figure}

\subsection{Q-ball--Q-shell}

The BS solution is obtained when one of the coupled models is $\mathbb{C}P^1$ or $\mathbb{C}P^3$ and the other one is $\mathbb{C}P^N$ with $N\ge 5$. Even for a single $\mathbb{C}P^N$ model the Q-shell solution appears because expansion at the origin $r=0$ does not lead to any nontrivial behaviour (all $b_k$ coefficients vanish). The same is true for two coupled models. Only the profile function  associated with $N=1$ or $N=3$ can have nontrivial behavior in the vicinity of $r=0$. Thus, expansion at $r=0$ involves only $f(r)$ and $g(r)=0$ in the vicinity of the center.

The solution consists of four partial solutions matched at $r=R_1$, $r=R_2$ and $r=R_3$. The profile function $f(r)$ of the first solution has the following behavior at the vicinity of $r=0$
\begin{align}
\mathbb{C}P^1:&\quad f(r)=a_0^2+\left[\frac{\widetilde\mu_1^2}{48}\sqrt{1+a_0^2}-\frac{a_0}{6}\frac{1-a_0^2}{1+a_0^2}\omega_1^2\right]r^2+{\cal O}(r^4)\\
\mathbb{C}P^3:&\quad f(r)=a_1 r+\frac{\widetilde\mu_1^2}{32}r^2+\frac{a_1}{10}(2a_1^2-\omega_1^2)r^3+{\cal O}(r^4)
\end{align}
whereas $g(r)=0$. This partial solution exists in the ball $r<R_1$. For the second partial solution $R_1<r<R_2$ the profile function is nontrivial $g(r)>0$. It has the following expansion at $r=R_1$ 
\begin{align}
g(r)=\frac{\widetilde\mu_2^2}{16}(r-R_1)^2-\frac{\widetilde\mu_2^2}{24R_1}(r-R_1)^3+{\cal O}\big((r-R_1)^4\big)
\end{align}
for both the $\mathbb{C}P^1-\mathbb{C}P^5$ and $\mathbb{C}P^3-\mathbb{C}P^5$ models. The numerical results are shown in 
Fig.\ref{fig:fig5}.
The inner radius of the shell compacton $R_1$ is not known a priori and thus it can be considered as a free parameter. 
The higher-order coefficients  depend on $f(R_1)$ and $f'(R_1)$ as well as the model constants. In this region II both 
profile functions $f(r)$ and $g(r)$ depend on each other, see Fig.\ref{fig:fig5}. 
At $r=R_2$ the Q-ball component reaches vacuum value $f=0$. 
In region III ($R_2<r<R_3$) the function $f(r)=0$ is constant and $g(r)$ decreases reaching vacuum value at $r$ (parabolic approach). 
The shooting parameters are $(a_0,R_1)$ for the $\mathbb{C}P^1-\mathbb{C}P^5$ model and $(a_1,R_1)$ for the $\mathbb{C}P^3-\mathbb{C}P^5$ model. 
Similar with BB solutions, as increasing the difference of $\omega_1,\omega_2$, the deviation of the profiles are more explicit. 

In Fig.\ref{hdensityBS}, we present the 3D plot of the Hamiltonian density that corresponds to the BS solution (plotted in Fig.\ref{fig:fig5})
for the $\mathbb{C}P^1-\mathbb{C}P^5$ model.

\subsection{Q-shell--Q-shell}
\begin{figure}[t]
\centering
\subfigure{\includegraphics[width=0.7\textwidth,height=0.4\textwidth, angle =0]{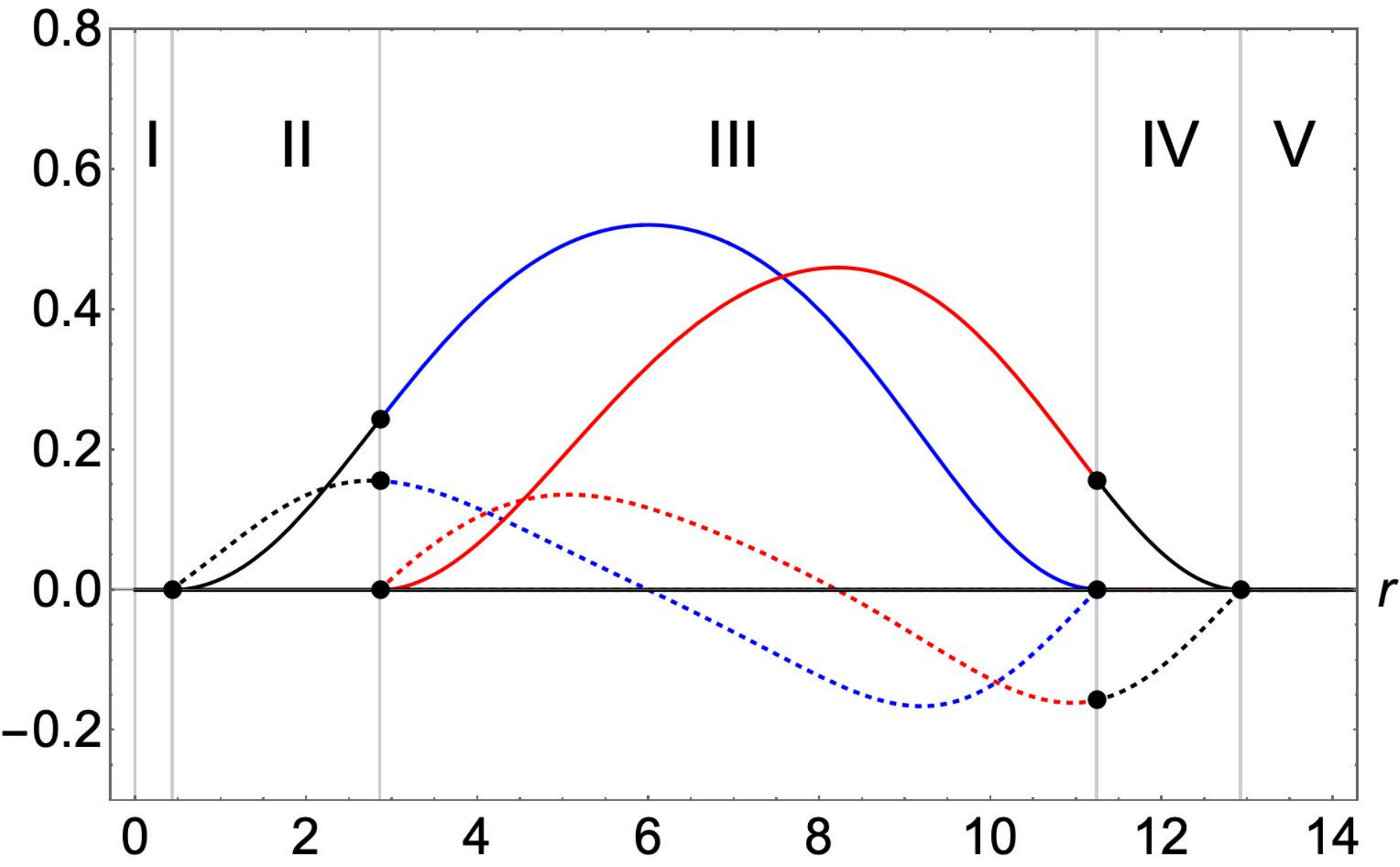}}
\caption{The SS solution for the $\mathbb{C}P^5-\mathbb{C}P^7$ model for $(\omega_1,\omega_2)=(1.0,1.0)$, 
$(\widetilde \mu^2_1,\widetilde \mu^2_2)=(1.0,1.0)$ and $(\lambda_1,\lambda_2)=(1.0,1.0)$.  
The compact SS solution is obtained for $(R_1,R_2)=(0.43,2.86)$. 
The outer radii have values $(R_3,R_4)=(11.24,12.92)$.}\label{fig:fig7}
\end{figure}

As an example of  SS compactons we take two overlapping Q-shells in the $\mathbb{C}P^5-\mathbb{C}P^7$ model. Equivalently one can consider  $\mathbb{C}P^5-\mathbb{C}P^5$  model with parameters $\omega_1\neq\omega_2$. Here we  consider  $\omega_1=\omega_2=1.0$.  
The solution consists of five regions, see Fig.\ref{fig:fig7}. 
There are four characteristic radii $R_1$, $R_2$, $R_3$ and $R_4$. 
In region I, given by $0<r<R_1$, as well as in region V, $R>R_4$, both profile functions take vacuum value. 
In region II the profile function $f(r)>0$ whereas the second function $g(r)$ vanishes. 
Similarly, in region IV $f(r)=0$ and $g(r)>0$. Both profile functions obey the set of coupled equations only in region III (shells overlap). 
 
\begin{figure}[t]
\centering
\subfigure[]{\includegraphics[width=0.6\textwidth,height=0.5\textwidth, angle =0]{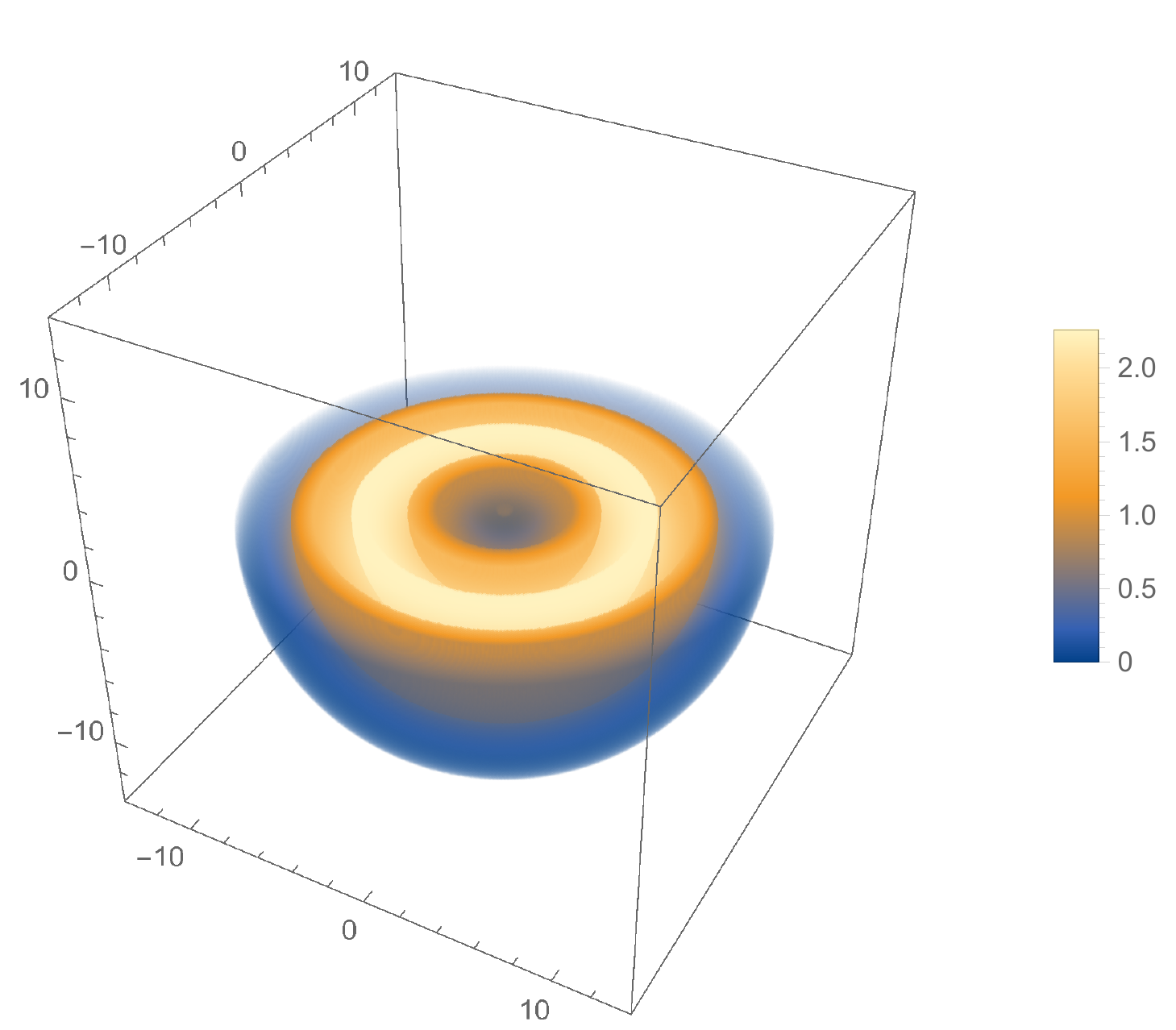}}\hspace{0.5cm}
\subfigure[]{\includegraphics[width=0.3\textwidth,height=0.2\textwidth, angle =0]{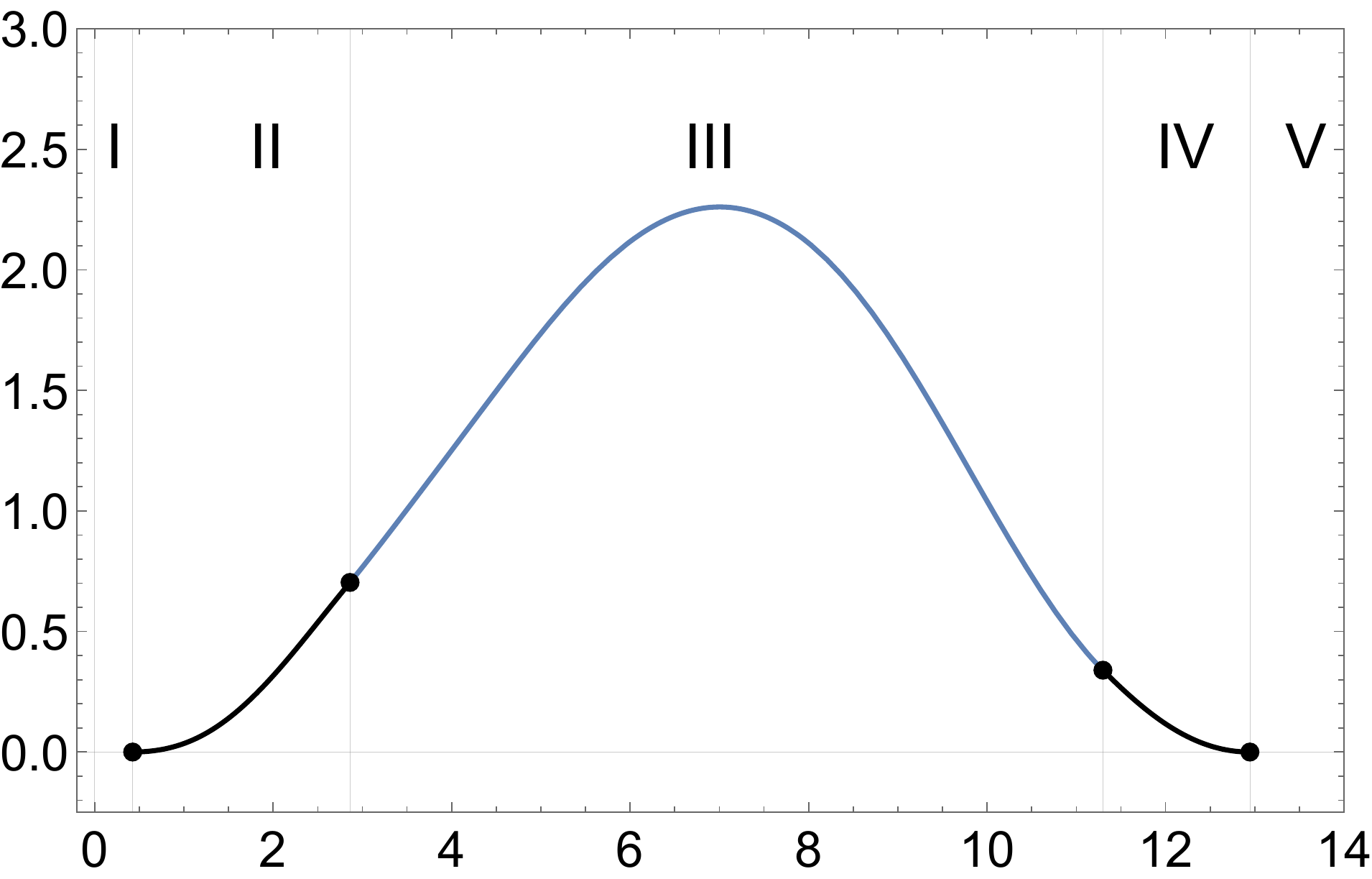}}
\caption{\label{hdensitySS}The Hamiltonian density of the SS solution corresponding to Fig.\ref{fig:fig7}(a)
for the $\mathbb{C}P^5-\mathbb{C}P^7$ model. Figure (a) the 3D plot and (b)
the radial density associated with (a). The shell structure becomes apparent in this solution. }
\end{figure}

The radii $R_1$ and $R_2$ are considered as shooting parameters. 
The profile functions at $R_1$  have expansions
\begin{align}
f(r)=\frac{\widetilde\mu_1^2}{16}(r-R_1)^2-\frac{\widetilde\mu_1^2}{24R_1}(r-R_1)^3+\ldots,\qquad g(r)=0. 
\end{align}
At $r=R_2$ the function $f(r)$ (determined by numerical integration) can be represented by series $f(r)=A_0+A_1(r-R_2)+A_2(r-R_2)^2+\ldots$ and function $g(r)$ is given by
\be
g(r)=\frac{\widetilde\mu_2^2}{16}(r-R_2)^2-\frac{\widetilde\mu_2^2}{24 R_2}(r-R_2)^3+\ldots
\ee
In order to get the solution we assume certain $R_1$ and $R_2$  and  integrate the system of equations. 
Next we determine the first local minimum $r=R_3$ of $f(r)$. Then changing  values of shooting parameters we get $f(R_3)=0$. 
The correct solution is obtained if for $R_4$ such that $g'(R_4)=0$ and $g(R_4)=0$ hold. 
Clearly, it requires simultaneous determination of two shooting parameters. 

In Fig.\ref{hdensitySS}, we present the 3D plot of the Hamitlonian density that corresponds with
the BS solution (plotted in Fig.\ref{fig:fig7}) for the $\mathbb{C}P^5-\mathbb{C}P^7$ model.

\subsection{Harbored compactons}\label{secondmodelh}

The harbored compactons are BS and SS structures such that $f(r)>0$ overlap only with $g(r)=0$ and $g(r)>0$ overlap only with $f(r)=0$. 
It means that the coupling terms proportional to $\lambda_1$ and $\lambda_2$ does not really matter since one of the fields is in its vacuum state. 
This is possible exclusively due to the compact nature of solutions. 
The possibility of having a compacton surrounded by the other compacton is quite intriguing, especially,  
in the context of application of the model to boson stars. The harbored solutions containing a black hole in the center are discussed in 
\cite{Kleihaus:2009kr,Kleihaus:2010ep,Klimas:2018ywv,Sawado:2021rsc}.
 
\begin{figure}[t]
\centering
\subfigure[]{\includegraphics[width=0.45\textwidth,height=0.27\textwidth, angle =0]{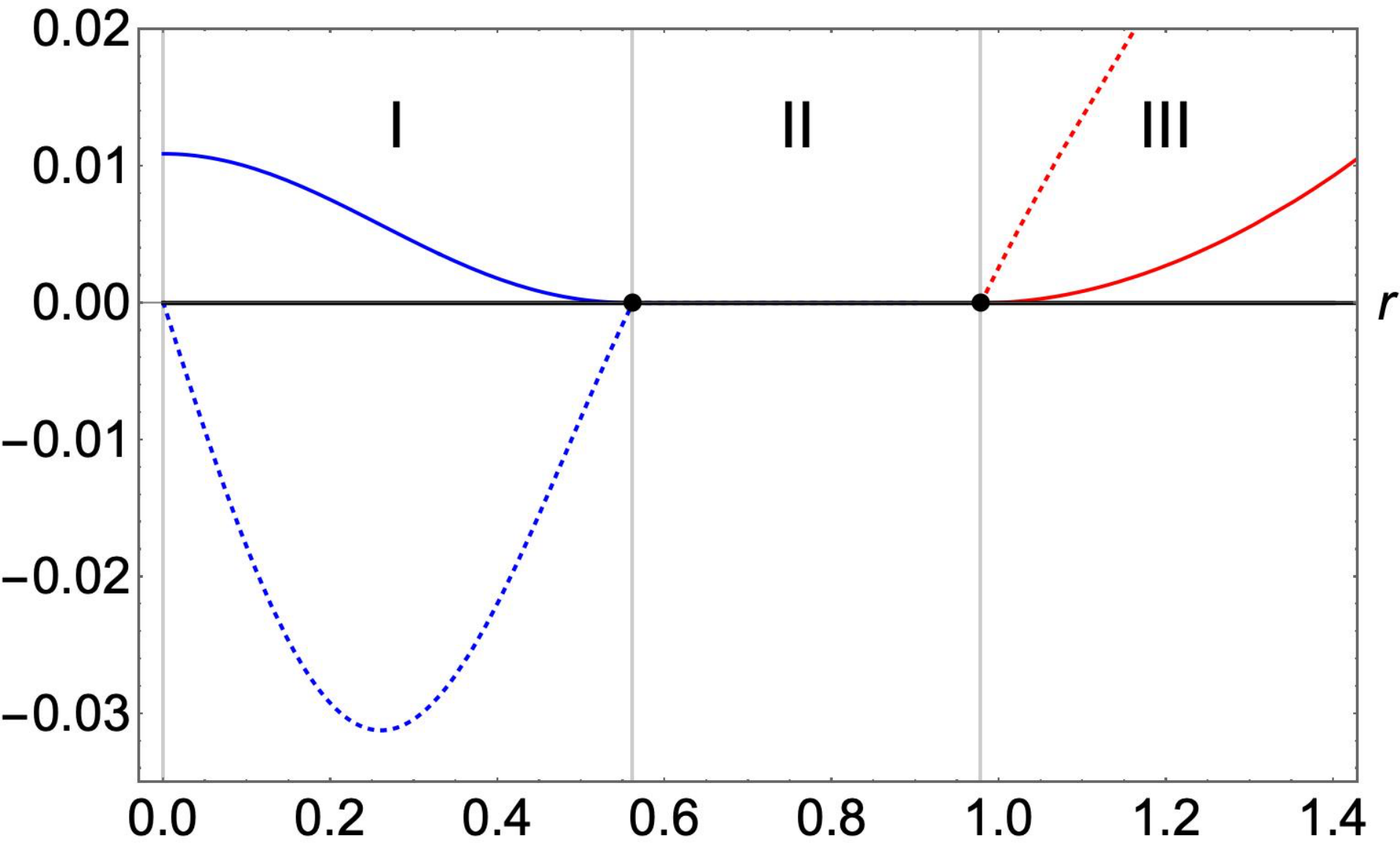}}\hspace{0.5cm}
\subfigure[]{\includegraphics[width=0.45\textwidth,height=0.27\textwidth, angle =0]{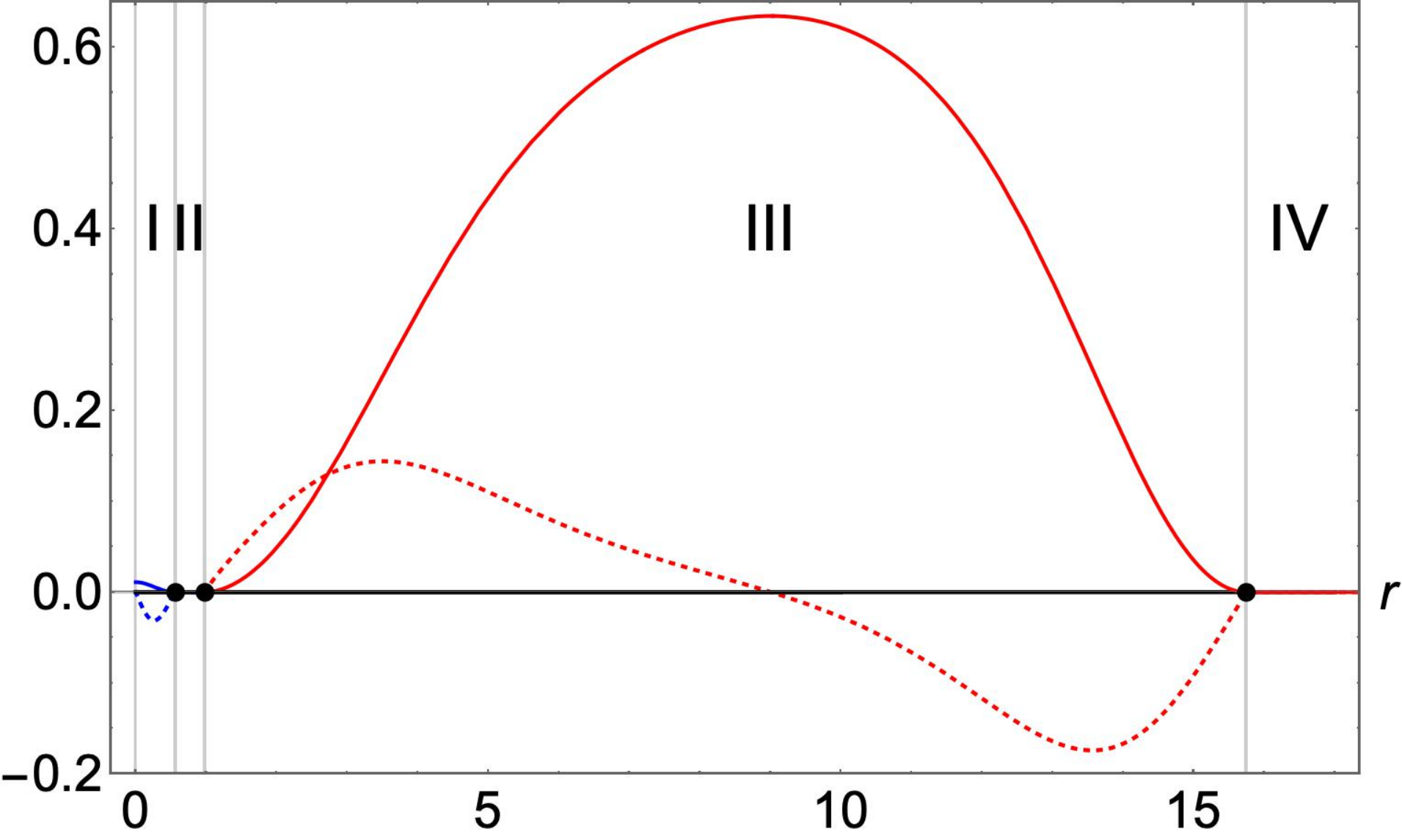}}
\caption{The $\mathbb{C}P^1$ Q-ball which is surrounded by the $\mathbb{C}P^5$ Q-shell. 
The Q-ball outer radius has the value $R_0=0.56$ whereas the Q-shell radii reads $R_1=0.97$ and $R_2=15.75$. 
The solutions were obtained for $\omega_1=8.0$ and $\omega_2=0.9$. 
Figure (a) shows the blow up of the central region where the Q-ball is localized. 
In  II and IV there exists vacuum solution $f=g=0$. In I $g=0$ and $f$ is nontrivial whereas in III $f=0$ and $g$ is nontrivial.}\label{fig:fig9}
\end{figure}

In Fig.\ref{fig:fig9} we show an example of a solution that consists of a $\mathbb{C}P^1$ Q-ball surrounded by the $\mathbb{C}P^5$ Q-shell. 
The supports of these two compact configurations do not overlap {\it i.e.} when $f>0$, $g=0$ and vice versa. 
For the harbored compactons the Q-ball radius $R_0$ is smaller or equal to the inner Q-shell radius $R_1$.

\begin{figure}[t]
\centering
\subfigure[]{\includegraphics[width=0.35\textwidth,height=0.2\textwidth, angle =0]{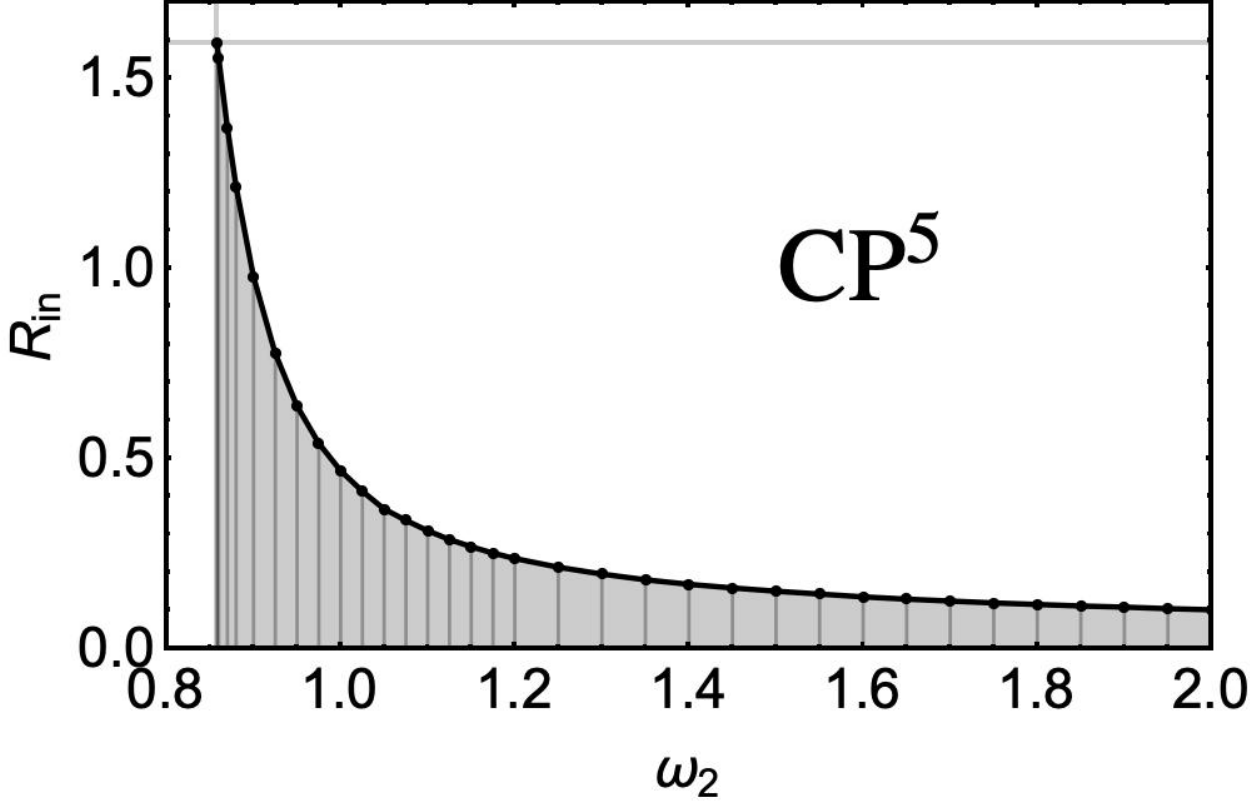}}\hspace{1.0cm}
\subfigure[]{\includegraphics[width=0.35\textwidth,height=0.2\textwidth, angle =0]{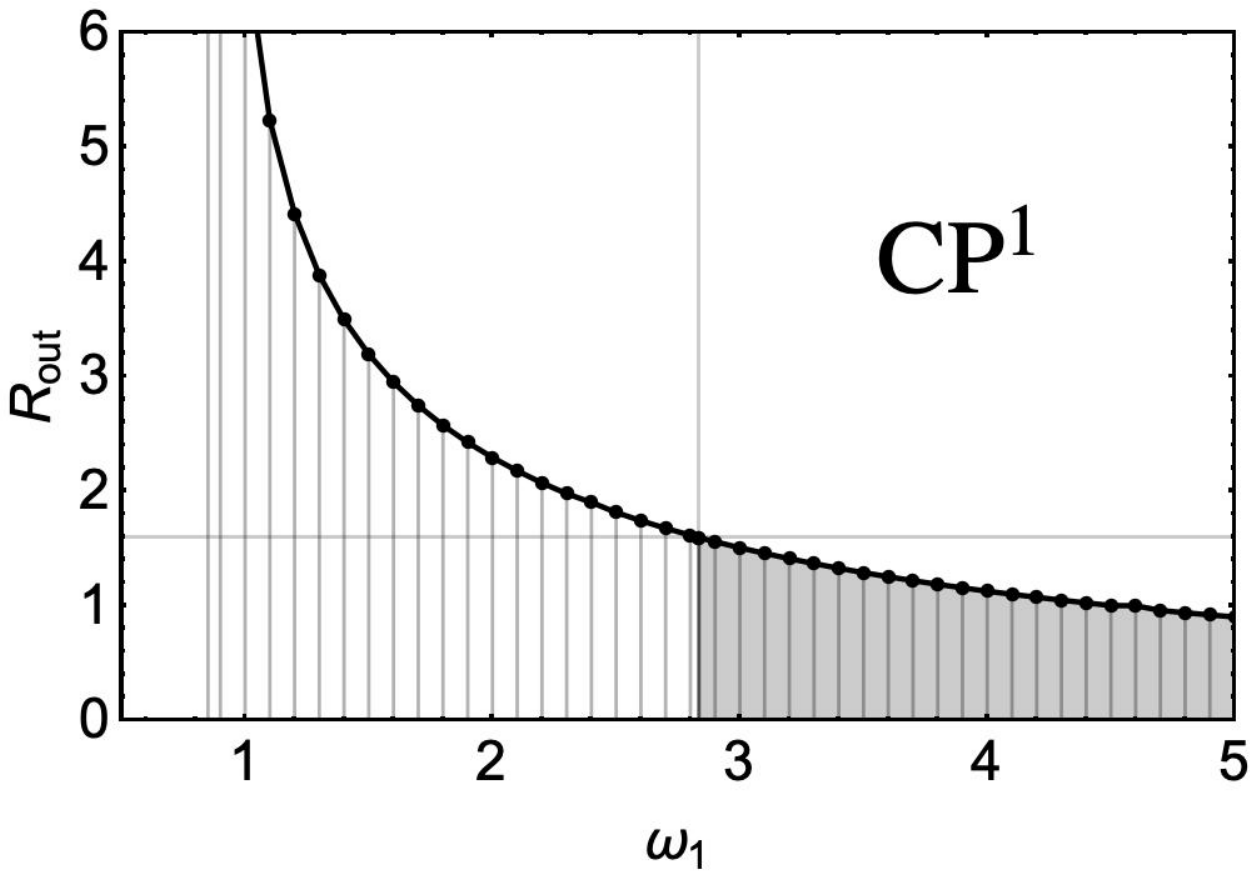}}
\subfigure[]{\includegraphics[width=0.35\textwidth,height=0.2\textwidth, angle =0]{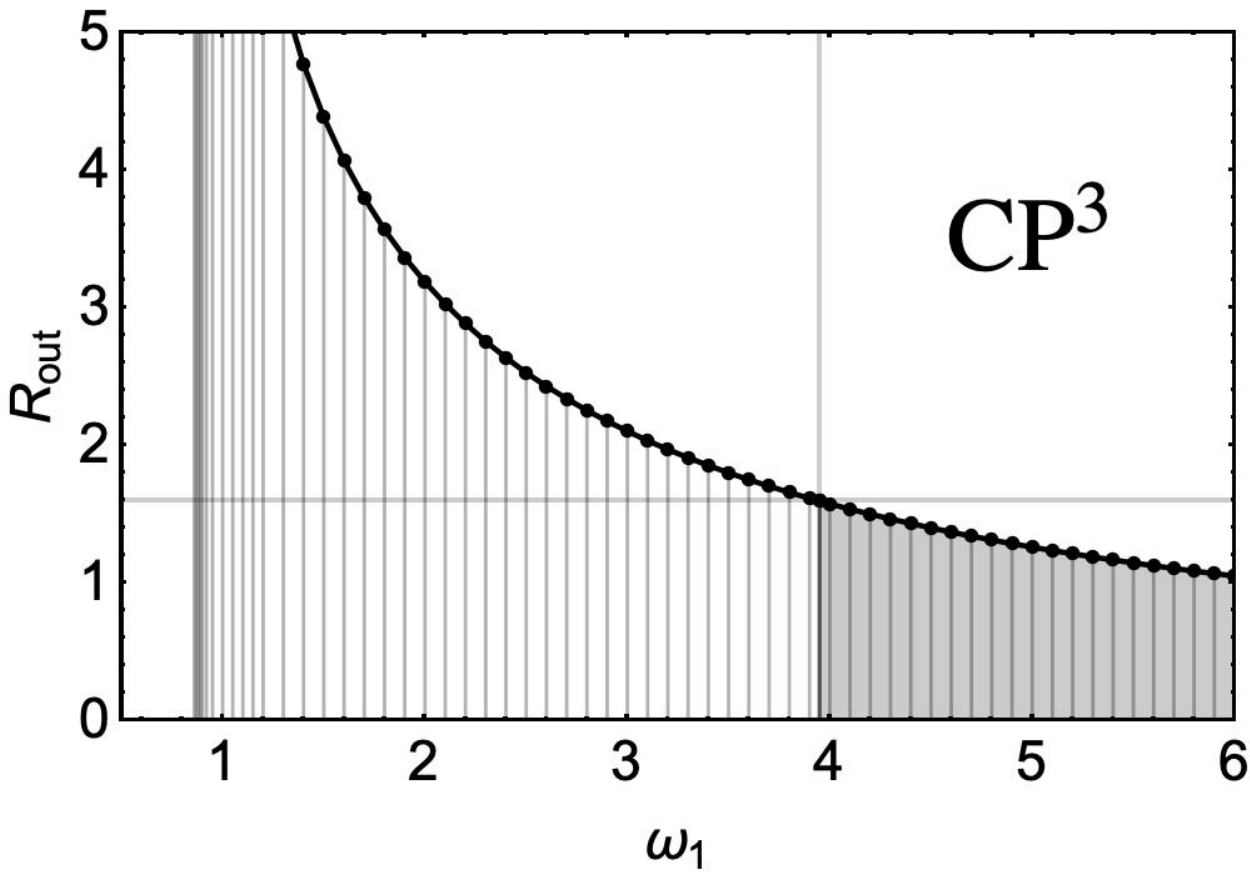}}\hspace{1.0cm}
\subfigure[]{\includegraphics[width=0.35\textwidth,height=0.2\textwidth, angle =0]{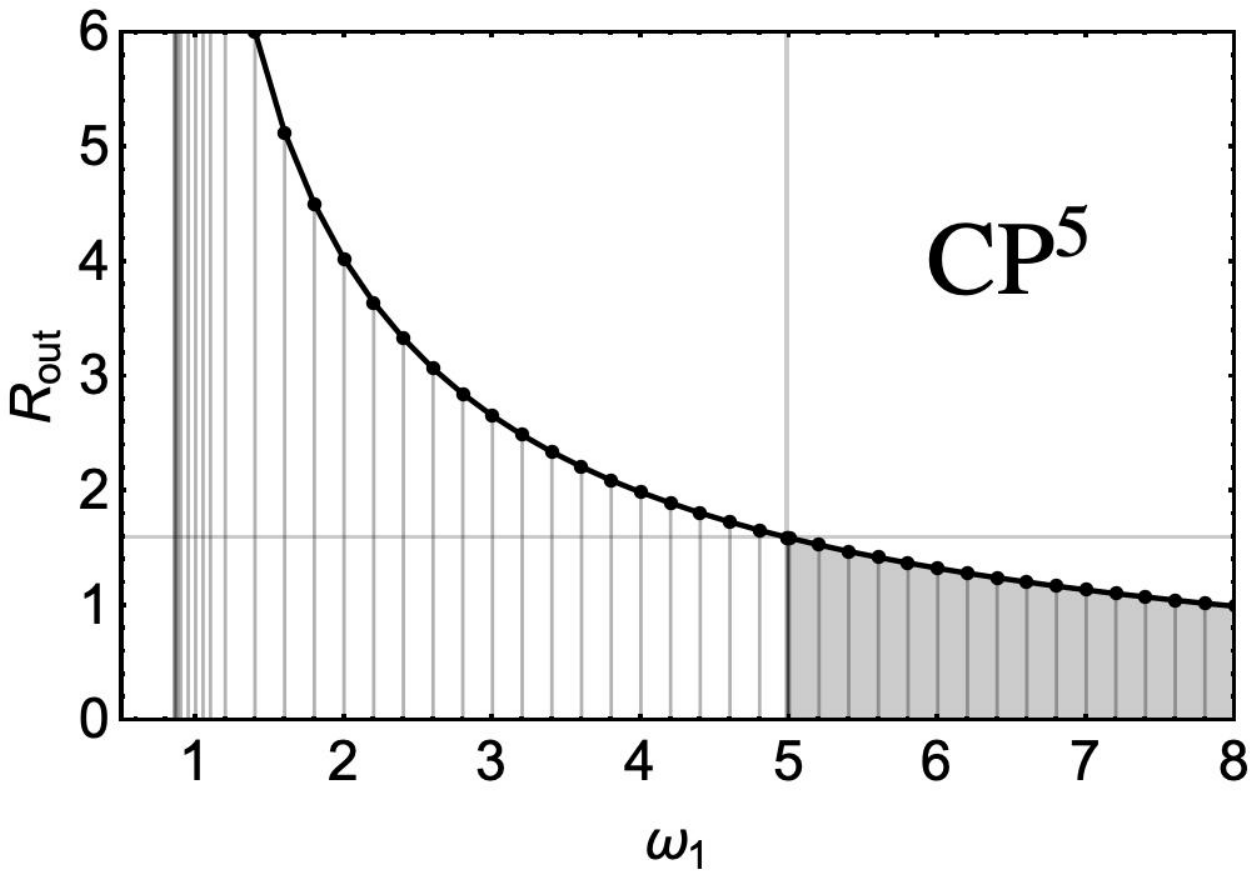}}
\caption{(a) Inner radius of the $\mathbb{C}P^5$ Q-shell. It is maximal $R_{\rm max}\approx 1.592$ for $\omega_{\rm min}\approx 0.858$. 
(b) The $\mathbb{C}P^1$ Q-ball radius is smaller than $R_{\rm max}$ for $\omega_1 \gtrapprox 2.835$.
 (c) The $\mathbb{C}P^3$ Q-ball radius is smaller than $R_{\rm max}$ for $\omega_1 \gtrapprox 3.95$. 
(d) The outer radius of the $\mathbb{C}P^5$ Q-shell. It is smaller than $R_{\rm max}$ for $\omega_1\gtrapprox4.99$.}\label{fig:fig10}
\end{figure}

The inner  Q-shell radius $R_1\equiv R_{\rm in}$ is limited from above. In the case of the $\mathbb{C}P^5$ compacton the maximum value of 
this radius is $R_1=R_{\rm max}\approx 1.592$ for $\omega_2=\omega_{\rm min}\approx 0.858$. 
This radius decreases as $\omega_2$ increases, see Fig.\ref{fig:fig10}(a). 
Consequently, the adequate Q-ball/shell which is parametrized by $\omega_1$ must have (outer) radius  $R_{\rm out}=\{R_0,R_2\}$ 
which value does not exceed the value of the inner radius of the shell parametrized by $\omega_2$. 
In Fig.\ref{fig:fig10}(b) we plot the outer radius of the $\mathbb{C}P^1$ Q-ball in dependence on $\omega_1$. 
The shadowed region ($\omega_1\gtrapprox2.835$) contains radii which are smaller than $R_{\rm max}$ of the $\mathbb{C}P^5$ Q-shell. 
The radius of the $\mathbb{C}P^3$ Q-shell is plotted in Fig.\ref{fig:fig10}(c). It is smaller than $R_{\rm max}$ for $\omega_1\gtrapprox3.95$. 
Another possibility for harboring is a double Q-shell solution. 
Two Q-shells do not overlap if the outer radius of a smaller shell is no bigger than the inner radius of a bigger Q-shell. 
The case of the $\mathbb{C}P^5-\mathbb{C}P^5$ compactons is shown in Fig.\ref{fig:fig10}(d), 
where outer radius of the $\mathbb{C}P^5$ shell is smaller than $R_{\rm max}\approx 1.592$ for $\omega_1\gtrapprox4.99$.

\subsection{The energy-charge scaling}\label{EQ}

\begin{figure}[t]
\centering
\subfigure[]{\includegraphics[width=0.65\textwidth,height=0.45\textwidth, angle =0]{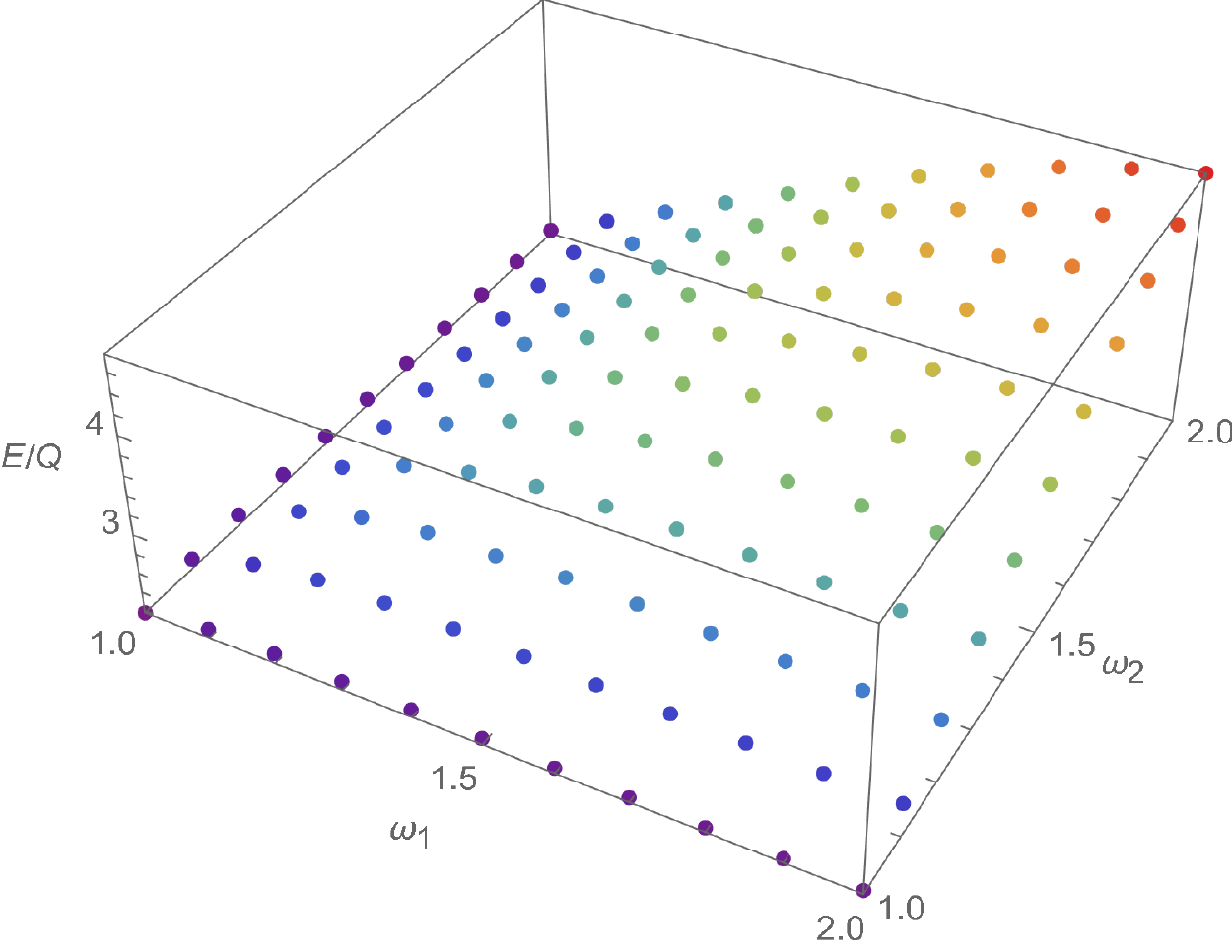}} \\
\subfigure[]{\includegraphics[width=0.65\textwidth,height=0.45\textwidth, angle =0]{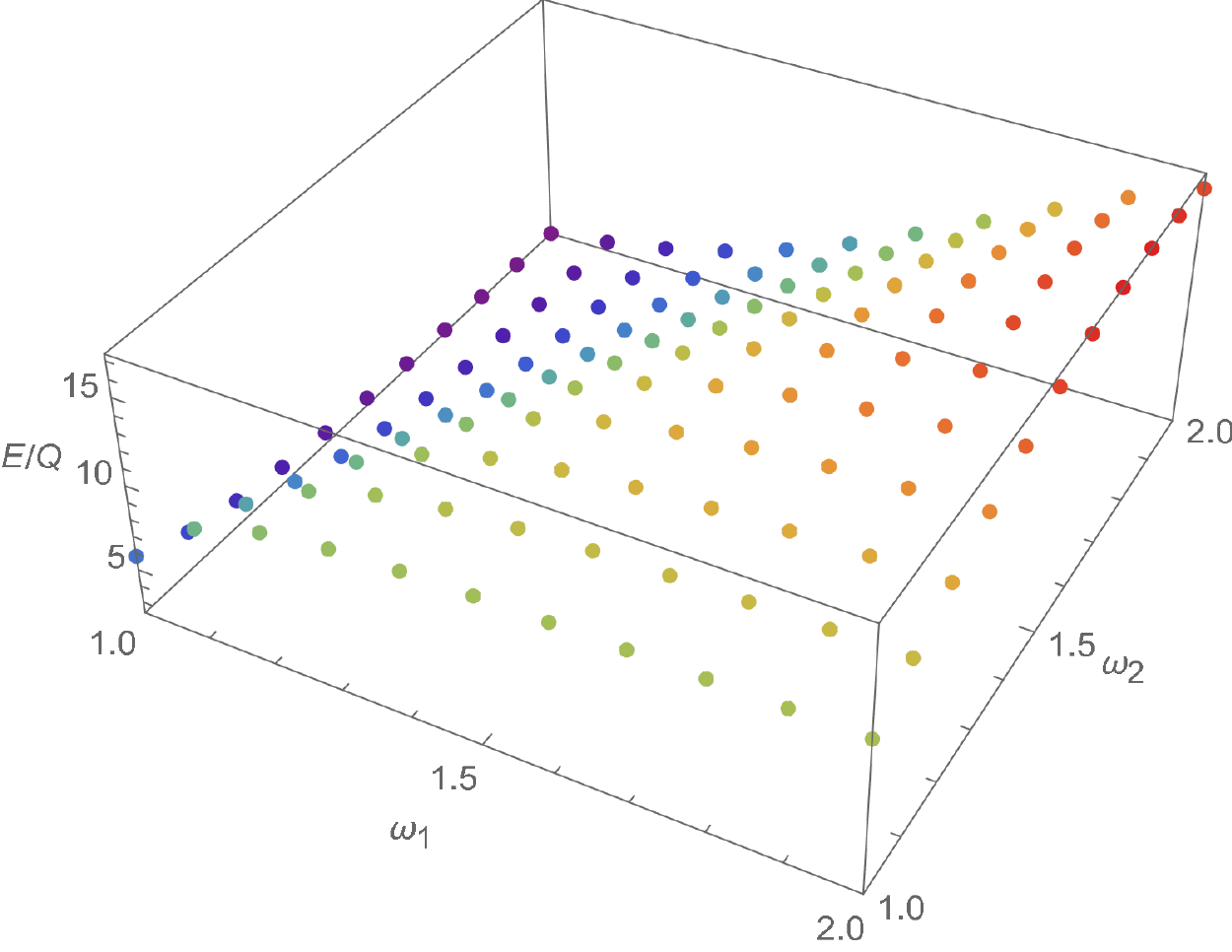}}
\caption{\label{Edensity}The ratio of the energy versus charge $E/Q$ for (a)~$\mathbb{C}P^1-\mathbb{C}P^1$: the BB solution
 and for (b)~$\mathbb{C}P^1-\mathbb{C}P^5$: the BS solution.}
\end{figure}

\begin{table}[t]
\caption{The power $\alpha$ of the energy charge scaling $E \sim Q^{\alpha}$ of the BB:~$\mathbb{C}P^1-\mathbb{C}P^1$, 
the BS:~$\mathbb{C}P^1-\mathbb{C}P^5$ and the SS:~$\mathbb{C}P^5-\mathbb{C}P^7$ solutions. 
In the numerical fitting, we fix the $\omega_1$ and evaluate the energy $E$ and the charge $Q$ with changing $\omega_2$. }
  \label{table1}
    \centering
\begin{tabular}{lccc}\hline\hline
$\omega_1 $ & ~~~~~~BB~~~~~~ & ~~~~~~BS~~~~~~ & ~~~~~~SS~~~~~~\\ \hline
1.0 & 0.689472 & 1.177404 & 0.560152 \\
1.1 & 0.781990 & 0.829030 & 0.735320 \\
1.2 & 0.824313 & 0.803957 & 0.776537 \\
1.3 & 0.848673 & 0.827690 & 0.801405 \\
1.4 & 0.863741 & 0.844339 & 0.821611 \\
1.5 & 0.872798 & 0.854584 & 0.836811 \\
1.6 & 0.878805 & 0.861000 & 0.847649 \\ 
1.7 & 0.882801 & 0.865147 & 0.855240 \\
1.8 & 0.885367 & 0.867902 & 0.860562 \\
1.9 & 0.887303 & 0.869777 & 0.864316 \\
2.0 & 0.888564 & 0.871078 & 0.866988 \\ 
5.0 & 0.892214 & 0.874687 & 0.873244 \\ 
10.0 & 0.892229 & 0.874702 & 0.873263 \\ \hline\hline
\end{tabular}
\end{table}

\begin{table}[t]
  \caption{The power $\alpha$ of the energy charge scaling $E \sim Q^{\alpha}$ of the BB:~$\mathbb{C}P^1-\mathbb{C}P^1$, 
the BS:~$\mathbb{C}P^1-\mathbb{C}P^5$ and the SS:~$\mathbb{C}P^5-\mathbb{C}P^7$ solutions. 
In the numerical fitting, we fix the $\omega_2$ and evaluate the energy $E$ and the charge $Q$ in dependence  on  $\omega_1$. }
  \label{table2}
  \centering
\begin{tabular}{lccc}\hline\hline
$\omega_2 $ & ~~~~~~BB~~~~~~ & ~~~~~~BS~~~~~~ & ~~~~~~SS~~~~~~\\ \hline
1.0 & 0.689472 & 0.649519 & 0.582622 \\
1.1 & 0.781990 & 0.814268 & 0.753623 \\
1.2 & 0.824313 & 0.852421 & 0.795216 \\
1.3 & 0.848673 & 0.865346 & 0.821697 \\
1.4 & 0.863741 & 0.871563 & 0.838833 \\
1.5 & 0.872798 & 0.875649 & 0.849806 \\
1.6 & 0.878805 & 0.878831 & 0.856950 \\ 
1.7 & 0.882801 & 0.881463 & 0.861704 \\
1.8 & 0.885367 & 0.883623 & 0.864936 \\
1.9 & 0.887303 & 0.885386 & 0.867173 \\
2.0 & 0.888564 & 0.886790 & 0.868747 \\ 
5.0 & 0.892214 & 0.892200 & 0.874671 \\ 
10.0 & 0.892229 & 0.892229 & 0.874701 \\ \hline\hline
\end{tabular}
\end{table}

 \begin{figure}[t]
\centering
\subfigure[]{\includegraphics[width=0.45\textwidth,height=0.27\textwidth, angle =0]{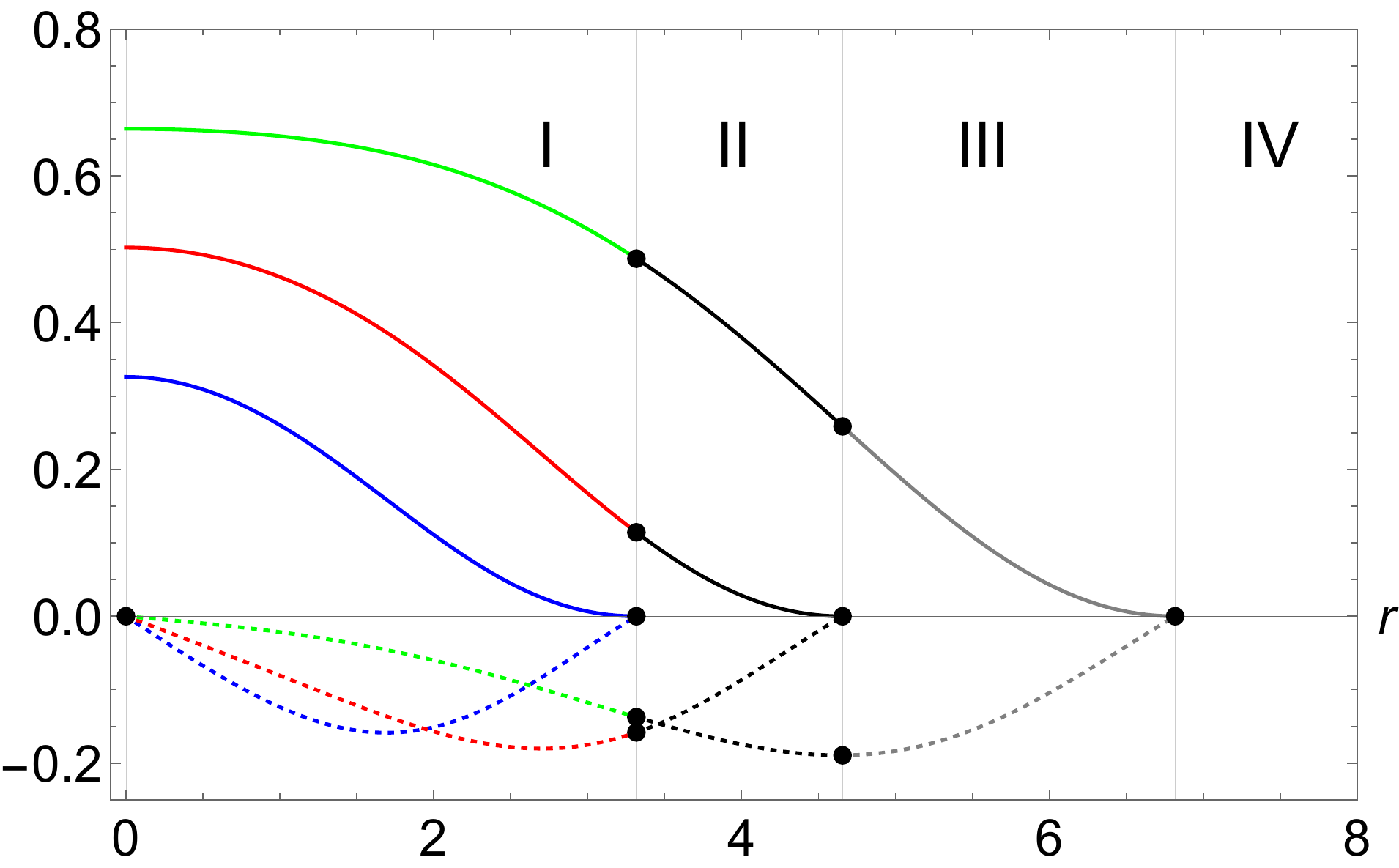}}\hspace{1.0cm}
\subfigure[]{\includegraphics[width=0.45\textwidth,height=0.27\textwidth, angle =0]{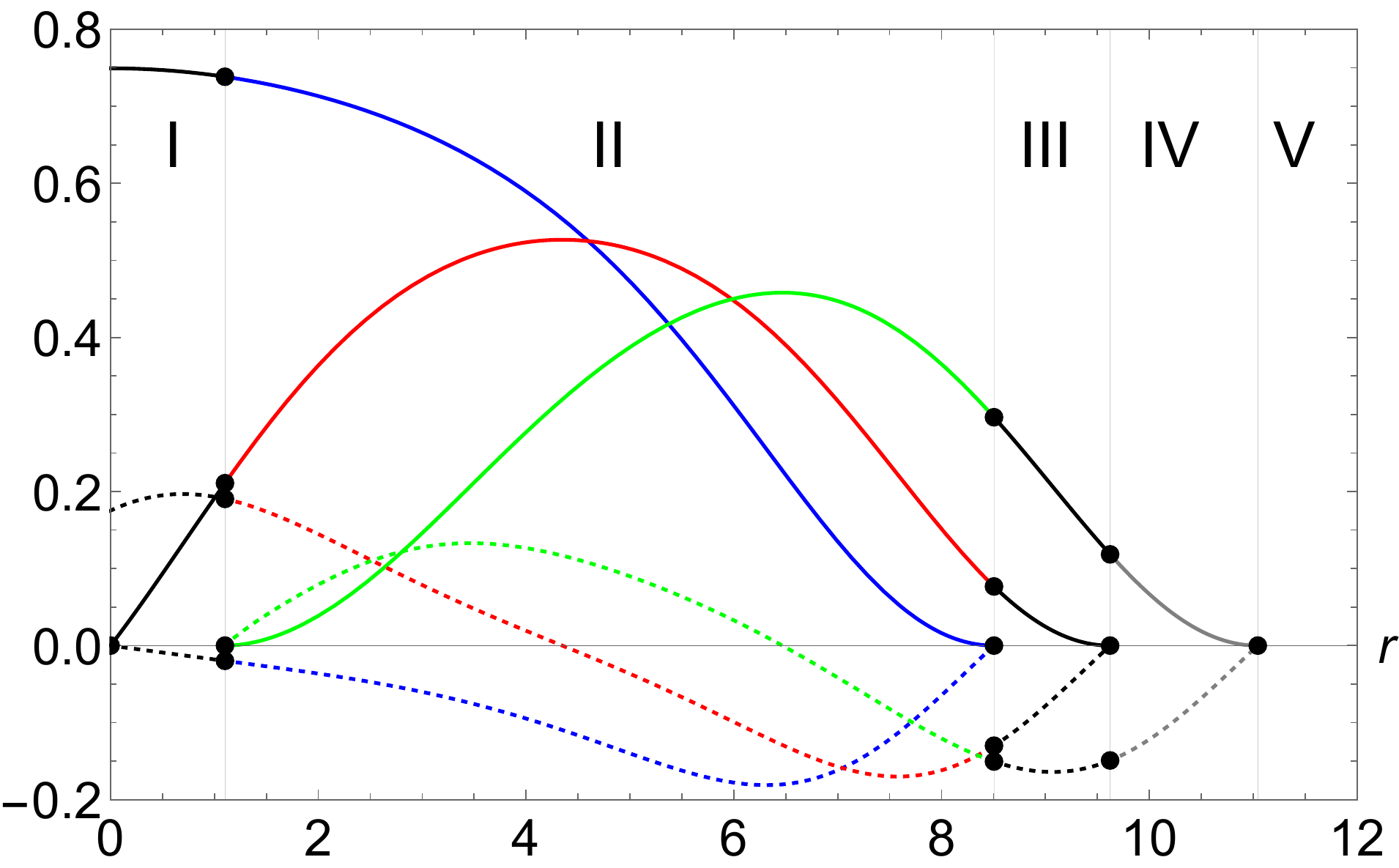}}
\caption{(a) The $\mathbb{C}P^1-\mathbb{C}P^1-\mathbb{C}P^1$ Q-balls for $(\omega_1,\omega_2, \omega_3)=(1.5,1.2,1.0)$. 
The other constants have values $\widetilde \mu_1^2=\widetilde \mu_2^2=\widetilde \mu_3^2=1.0$, $\lambda_1=\lambda_2=\lambda_3=1.0$. 
The shooting parameters for $f(r),g(r)$ and $h(r)$ are $(0.3261812,0.5025038,0.6642960)$. 
The radii of Q-balls have values $(R_1,R_2,R_3)=(3.3167,4.65616,6.81789)$.
The derivatives $f'(r),g'(r)$ and $h'(r)$ are represented by dashed curves.
(b) The $\mathbb{C}P^1-\mathbb{C}P^3-\mathbb{C}P^5$ Q-balls and Q-shell for $(\omega_1,\omega_2, \omega_3)=(1.0,1.0,1.0)$. 
The other constants have values $\widetilde \mu_1^2=\widetilde \mu_2^2=\widetilde \mu_3^2=1.0$, $\lambda_1=\lambda_2=\lambda_3=1.0$. 
The shooting parameters have values $(a_0, b_1,R_1) = (0.74928874, 0.175212,1.1012037)$. 
The outer radii of Q-balls and Q-shell are given by $(R_2,R_3,R_4)=(8.50442,9.62203,11.0439)$. 
The derivatives $f'(r),g'(r)$ and $h'(r)$ are represented by dashed curves.
}\label{fig:fig12}
\end{figure}


We have succeeded to find several multi-shell solutions such as a ball-shell or shell-shell ones 
in our multi-component model. A natural question arises: do such solutions exist without any dissipation or collapse?
The plausible answer for the compacton is not found so far.  A perturbative analysis might be a possible (and a best) 
approach but, as it was pointed out in \cite{Arodz:2005gz}, there are no linear regime for compacton solutions in models with V-shaped potential. 
Any arbitrarily small perturbation in such models is governed by a nonlinear equation that contains the signum function. 
It clearly indicates a lack of the harmonic oscillator paradigm as well as  its applicability in linearization of the 
field equation in the vicinity of minima of the potential. 
Moreover, nonvanishing of the first derivative of the potential at the minimum gives rise to the threshold force effect 
which constitutes a serious obstacle on free expansion of the compacton support or emission of small wave packages from the perturbed region.
We just conclude that the scale of compactons in the $\mathbb{C}P^N$ model is frozen out from the phase transitions.

The analysis of the energy for various geometrical shapes of the solutions may give us an insight  
into the question of the existence of  field configurations presented in this paper. 
Since the properties of solutions such as their size,  location and the height depends on the value of the frequencies 
$\omega_1,\omega_2$, then we examine the ratio of total energy of the solution over the total Noether charge 
in dependence on $\omega_1,\omega_2$. Taking into account that the Noether charge associated with a  Q-ball 
can be identified with the number of constituents,  the energy over the charge estimates the mass of the constituent. 
The smaller the value of this parameter, the more stable the system is.  
Fig.\ref{Edensity}(a) shows the results for the $\mathbb{C}P^1-\mathbb{C}P^1$ model in the space of parameters $\omega_1$ and $\omega_2$.
The expression ($E/Q$) takes the highest value for  
$\omega_1=\omega_2$, {\it i.e.}, when both radial profiles  
coincide with each other (total overlapping of compactons). This function decreases when the 
values of $\omega_1,\omega_2$ deviate from each other. It corresponds with diminishing of the grade of overlapping of compact profile functions, see Fig.\ref{fig:fig1}. 
This result clearly suggests  that the multi-shell or the multi-nodal configurations are energetically more favorable than the lump-shaped solutions. In Fig.\ref{Edensity}(b) we plot the results for the $\mathbb{C}P^1-\mathbb{C}P^5$ case. 
The behavior of the function $(E/Q)$ is quite similar, however, a notable difference is that the ratio decreases for small values $\omega_1$ 
and high values of $\omega_2$.  In such a case
the support of the $\mathbb{C}P^5$ shell is entirely located inside the support of the $\mathbb{C}P^1$ ball, see Fig.\ref{fig:fig9}(a)(b). 

Another important point in the study of stability of the solutions is their energy-charge scaling relation. 
In the single $\mathbb{C}P^N$ field model, both in the flat and in the curved space-time, 
the solutions exhibit $E\sim Q^{5/6}$ scaling which means that they are classically stable against splitting into 
fragments with lower charges~\cite{Klimas:2017eft,Klimas:2018ywv}. 
In Tables \ref{table1},\ref{table2}, we present the power $\alpha$ of the scaling $E\sim Q^\alpha$ of the BB, BS, SS solutions 
as the function of $\omega_2$ with fixed $\omega_1$ (Table \ref{table1}) or as the function of $\omega_1$ with fixed $\omega_2$ (Table \ref{table2}). 
The parameter $\alpha$ is almost always (except a single value) $\alpha<1$ and relaxes into $\sim 9/10$ for larger $\omega_1$ or $\omega_2$. 
This clearly indicates that the solutions became more stable with  growth of the charge. 

\subsection{Comment on analytical solutions}
It has been shown in \cite{Klimas:2017eft} that in the limit of small amplitude fields the model tends to a  modified signum-Gordon model. The profile function satisfy a linearized equation containing the signum function (which is the only term that cannot be linearized). Consequently, in this limit there are exact solutions of the radial equation for $f(r)$.  One would wonder if a similar approach can be used in the case of coupled models analyzed in this paper. First of all we see that \eqref{Sigma-f} and \eqref{Sigma-g} take the form
\begin{align}
&\Sigma^{(1)}(r)\equiv f''+\frac{2}{r}f'+\left(\omega_1^2-\frac{l_1(l_1+1)}{r^2}\right)f,\nonumber\\
&\Sigma^{(2)}(r)\equiv g''+\frac{2}{r}g'+\left(\omega_2^2-\frac{l_2(l_2+1)}{r^2}\right)g.\nonumber
\end{align}
The radial field equations \eqref{eq1f} and \eqref{eq1g} are then of the form
\begin{align}
&\Sigma^{(1)}(r)-\frac{\mu_1^2}{8}\;{\rm sgn}(f)=\frac{\lambda_1\alpha}{4}g^{2\beta}f^{2\alpha-1},\label{eq1lin}\\
&\Sigma^{(2)}(r)-\frac{\mu_2^2}{8}\;{\rm sgn}(f)=\frac{\lambda_2\beta}{4}f^{2\alpha}g^{2\beta-1}.\label{eq2lin}
\end{align}
For $\alpha=\beta=1$ the r.h.s of the first equation \eqref{eq1lin} is proportional to $g^2(r)f(r)$ and r.h.s of the second equation \eqref{eq2lin} is proportional to $f^2(r)g(r)$. It is clear that the coupling terms are higher order nonlinear terms and they should be omitted  in the linear approximation. When skipping these terms we end up with two non interacting $\mathbb{C}P^N$ models. It shows that in the limit of small amplitudes of fields the compactons do not interact in the linear regime. The  analytical solutions of each model are just solutions presented in \cite{Klimas:2017eft}. On the other hand, if one keeps the coupling terms then the nonlinearity of equations would be a serious obstacle in getting their exact solutions. Hence in this situation a numerical approach would  necessary imply that there are no benefits from linearization of the equations.  

\section{The three-component solutions}\label{sec4}
 
\begin{figure}[ht]
\centering
\subfigure[]{\includegraphics[width=0.6\textwidth,height=0.5\textwidth, angle =0]{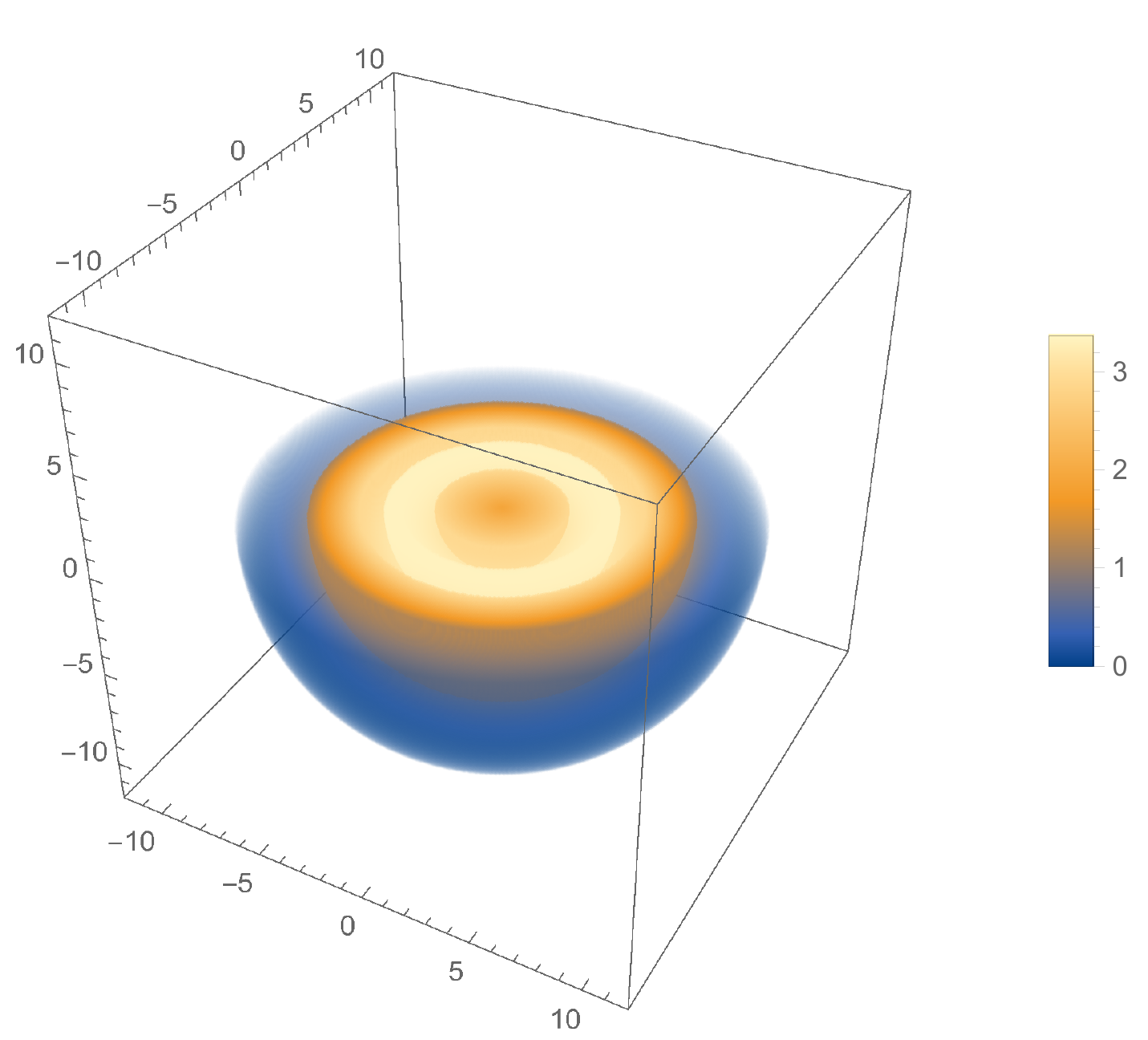}}\hspace{0.5cm}
\subfigure[]{\includegraphics[width=0.3\textwidth,height=0.2\textwidth, angle =0]{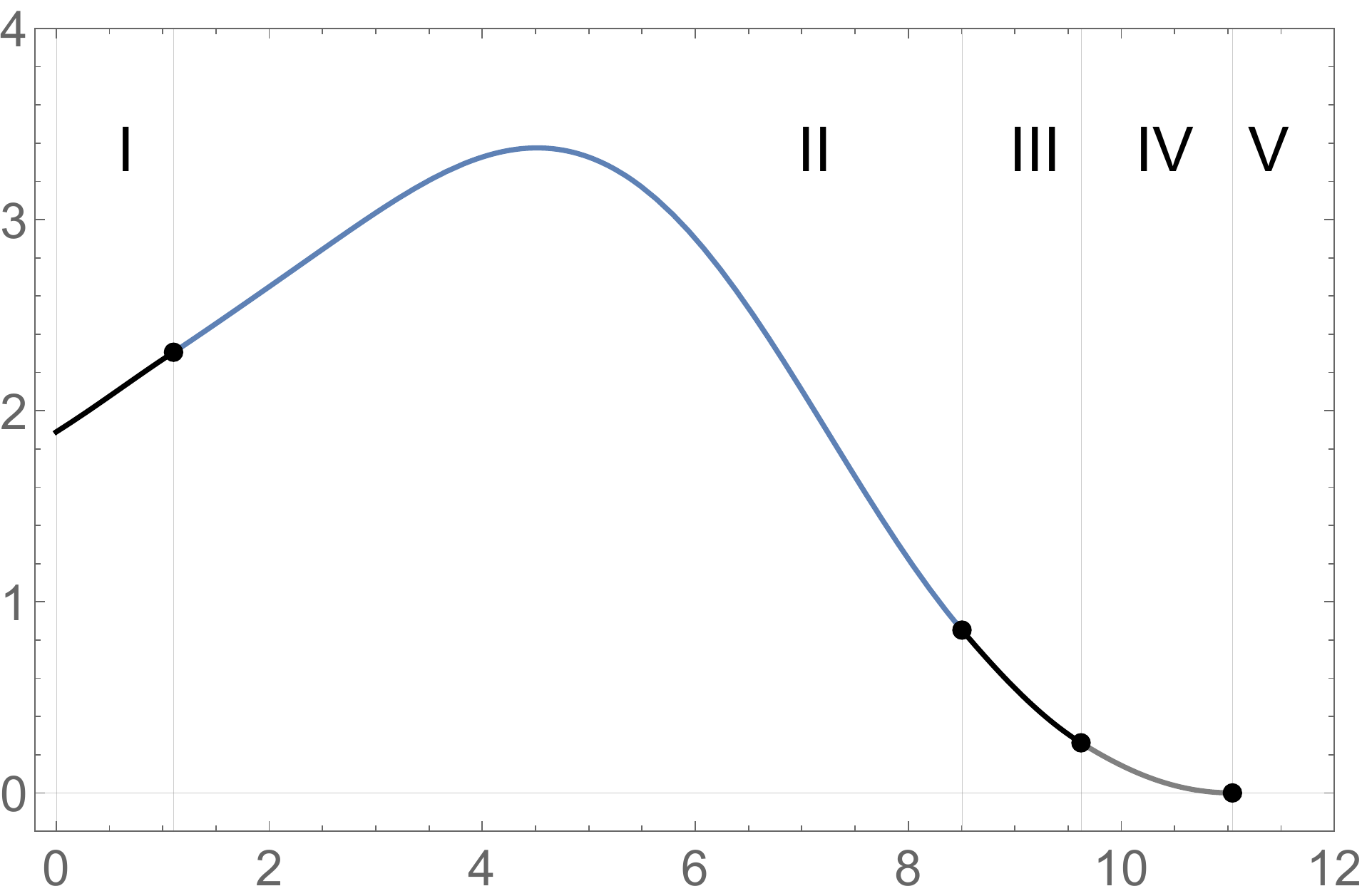}}
\caption{\label{hdensityBBS}The Hamiltonian density corresponding to Fig.\ref{fig:fig12}(b) 
for the $\mathbb{C}P^1-\mathbb{C}P^3-\mathbb{C}P^5$ model. Figure (a) the 3D plot and (b)
the radial density function. The structure is resembles the function presented for the BS solution. }
\end{figure}
A formal generalization of the presented approach by increasing a number of coupled models is straightforward, however, it leads to growth of technical difficulties in obtaining numerical solutions. Here we show some results obtained studying the three-component field model. We look at two examples, one containing a triple Q-ball and the other one containing the mixture of Q-balls and a Q-shell. 
Fig.\ref{fig:fig12}(a) shows the results for three Q-balls (BBB) for $(\omega_1,\omega_2, \omega_3)=(1.5,1.2,1.0)$. 
The figure shows that for frequencies $\omega_1>\omega_2>\omega_3$ the radii of Q-balls satisfy the relation $R_1<R_2<R_3$. 
The Q-balls-Q-shell solution (BBS) for $(\omega_1,\omega_2, \omega_3)=(1.0,1.0,1.0)$ is presented in 
Fig.\ref{fig:fig12}(b). In the central region $0\le r\le R_1$ there are only $\mathbb{C}P^1-\mathbb{C}P^3$ Q-balls whereas the $\mathbb{C}P^5$ field takes the vacuum value. On the segment $R_1\le r\le R_2$ all three $\mathbb{C}P^N$ fields are nontrivial,  then on $R_2\le r\le R_3$ the $\mathbb{C}P^1$ field takes its vacuum value and on $R_3\le r\le R_4$ only the $\mathbb{C}P^5$ field is nontrivial. For $r>R_4$ all three $\mathbb{C}P^N$ fields vanish.

In Fig.\ref{hdensityBBS}, we give the 3D plot of the Hamitlonian density of the BBS solution (plotted in Fig.\ref{fig:fig12}(b)) 
for the $\mathbb{C}P^1-\mathbb{C}P^3-\mathbb{C}P^5$ model.

\section{Conclusions}\label{sec6}
 
We have presented several new compact solutions of the coupled $\mathbb{C}P^N$ nonlinear multi-component field sigma model.
Our model contains two or more  $\mathbb{C}P^N$  fields coupled through the multibody contact potentials. 
In our study we mostly focused  on the case of two coupled  $\mathbb{C}P^N$  models. 
We managed to get many different spherically symmetric geometrical shapes of compactons, characterized by numbers $(N_1,N_2)$
 that determine the number of complex fields in each $\mathbb{C}P^N$ component. The $(N_1,N_2)=(1,1)$ {\it i.e.}
$\mathbb{C}P^1-\mathbb{C}P^1$ model has a Q-ball $-$ Q-ball (BB) type solution. Another BB solutions are obtained for $(1,3)$ and $(3,3)$.
For $N_1$ or $N_2$ larger than 3, the solutions have the form of Q-ball $-$ Q-shell (BS) or Q-shell $-$ Q-shell (SS). 
Each solution depends on several constant parameters, including the coupling constants of the model. 
In our study we mostly looked at the dependence on two constants $(\omega_1,\omega_2)$ that appear due to the Q-ball ansatz and strongly 
influence the size of the compact Q-ball radii. 
One of the most interesting configurations that appears in the coupled model are the harbored solutions. 
They involve two radial compactons -- one surrounded by the other one. 
This is a new solution in comparison with other harbored solutions where in the center of the shell hides a black hole or some other massive body. 
Our harbored compactons have  the form of BS and SS structures, such that $f(r)>0$ overlap only with $g(r)=0$ and $g(r)>0$ overlap only with $f(r)=0$. 
In other words, each $\mathbb{C}P^N$ component does not interact with the other components.  
Such a solution can appear exclusively because of the compact nature of our solutions and the form of coupling  potential.
The stability of the solutions was evaluated indirectly by analyzing their energy. 
This energy is slightly larger than in the case of  non-coupled models. 
Changing the values of parameters $\omega_1$ and $\omega_2$ we change the size of compactons that form the solution. 
It leads to the change of position of energy density maxima which are associated with these compactons. 
When the energy density peaks of two compact components move apart the energy decreases. 
It means that the two-body potential plays the role of a repulsive force. 
Moreover, analyzing  the energy-charge scaling  $E\sim Q^\alpha,~~\alpha<1$ we can conclude that large balls or shells are energetically 
more favorable (more stable) than small ones. 

One would wonder if there is a way of obtaining  some approximate analytical solutions. The answer on this question is positive only in the sector of small amplitude fields. In such a case radial equations reduce to the system of coupled signum-Gordon type equations. If one consider the case of non overlapping compactons then each such compacton is the solution of the signum-Gordon equation. In such a case the approximation is exactly as that presented in \cite{Klimas:2017eft}. On the other hand, even in the sector of small amplitude fields the overlapping compactons are not so straightforward to be obtained as analytical solutions. Even having such a solution one would not get valuable information because the solution  is not generic. This is the reason why we do not present approximated analytical curves for the profile of radial curves in the case of overlapping compactons.

The repulsive nature of the two-body potential corresponds with the positive value of the coupling constant. 
We managed to find some solutions also for  the negative value of the coupling constant. They are 
quite different from the solutions obtained for the positive value of coupling constant and do exist only 
under certain restrictions on the values of the model parameters. These solutions would be reported in another paper.

The compact Q-ball/shell solutions containing only scalar fields allows for building more sophisticated configurations 
placing one spherical configuration next to another one. 
The only condition is that they do not touch each other. 
In principle there is no restriction for the number of such spherical structures. 
Clearly, the total configuration will not be  spherically symmetric any longer.  
This situation however can change after coupling the model to gravity or to electromagnetism. 
The infinite range of interaction requires the spherical symmetry of the solution in whole space. 
In such a case, only a single Q-ball/shell or a concentric structure of many shells can exist (in terms of energy density and not the individual ingredients of each $\mathbb{C}P^N$ model). 
Thinking on application to boson stars we see that coupling our model to electromagnetism and gravity is a 
natural and physically relevant extension. In this context, solutions  presented in this paper would have the role of limit solutions. 
This extension is under current analysis and we shall report on the results in further papers.    

\begin{center}
{\bf Acknowledgment}
\end{center}
Discussions during the YITP workshop YITP-W-20-03 on ``Strings and Fields 2020'' 
has been useful to complete this work. N.S. was supported in part by JSPS KAKENHI Grant Number JP B20K03278(1). L.C.K. was supported by FAPESC scholarship.

\bibliography{coupledQballs}

\begin{thebibliography}{45}%
\makeatletter
\providecommand \@ifxundefined [1]{%
 \@ifx{#1\undefined}
}%
\providecommand \@ifnum [1]{%
 \ifnum #1\expandafter \@firstoftwo
 \else \expandafter \@secondoftwo
 \fi
}%
\providecommand \@ifx [1]{%
 \ifx #1\expandafter \@firstoftwo
 \else \expandafter \@secondoftwo
 \fi
}%
\providecommand \natexlab [1]{#1}%
\providecommand \enquote  [1]{``#1''}%
\providecommand \bibnamefont  [1]{#1}%
\providecommand \bibfnamefont [1]{#1}%
\providecommand \citenamefont [1]{#1}%
\providecommand \href@noop [0]{\@secondoftwo}%
\providecommand \href [0]{\begingroup \@sanitize@url \@href}%
\providecommand \@href[1]{\@@startlink{#1}\@@href}%
\providecommand \@@href[1]{\endgroup#1\@@endlink}%
\providecommand \@sanitize@url [0]{\catcode `\\12\catcode `\$12\catcode
  `\&12\catcode `\#12\catcode `\^12\catcode `\_12\catcode `\%12\relax}%
\providecommand \@@startlink[1]{}%
\providecommand \@@endlink[0]{}%
\providecommand \url  [0]{\begingroup\@sanitize@url \@url }%
\providecommand \@url [1]{\endgroup\@href {#1}{\urlprefix }}%
\providecommand \urlprefix  [0]{URL }%
\providecommand \Eprint [0]{\href }%
\providecommand \doibase [0]{http://dx.doi.org/}%
\providecommand \selectlanguage [0]{\@gobble}%
\providecommand \bibinfo  [0]{\@secondoftwo}%
\providecommand \bibfield  [0]{\@secondoftwo}%
\providecommand \translation [1]{[#1]}%
\providecommand \BibitemOpen [0]{}%
\providecommand \bibitemStop [0]{}%
\providecommand \bibitemNoStop [0]{.\EOS\space}%
\providecommand \EOS [0]{\spacefactor3000\relax}%
\providecommand \BibitemShut  [1]{\csname bibitem#1\endcsname}%
\let\auto@bib@innerbib\@empty
\bibitem [{\citenamefont {Arodz}\ and\ \citenamefont
  {Lis}(2008)}]{Arodz:2008jk}%
  \BibitemOpen
  \bibfield  {author} {\bibinfo {author} {\bibfnamefont {H.}~\bibnamefont
  {Arodz}}\ and\ \bibinfo {author} {\bibfnamefont {J.}~\bibnamefont {Lis}},\
  }\bibfield  {title} {\enquote {\bibinfo {title} {{Compact Q-balls in the
  complex signum-Gordon model}},}\ }\href {\doibase 10.1103/PhysRevD.77.107702}
  {\bibfield  {journal} {\bibinfo  {journal} {Phys. Rev. D}\ }\textbf {\bibinfo
  {volume} {77}},\ \bibinfo {pages} {107702} (\bibinfo {year} {2008})},\
  \Eprint {http://arxiv.org/abs/0803.1566} {arXiv:0803.1566 [hep-th]}
  \BibitemShut {NoStop}%
\bibitem [{\citenamefont {Arodz}\ and\ \citenamefont
  {Lis}(2009)}]{Arodz:2008nm}%
  \BibitemOpen
  \bibfield  {author} {\bibinfo {author} {\bibfnamefont {H.}~\bibnamefont
  {Arodz}}\ and\ \bibinfo {author} {\bibfnamefont {J.}~\bibnamefont {Lis}},\
  }\bibfield  {title} {\enquote {\bibinfo {title} {{Compact Q-balls and
  Q-shells in a scalar electrodynamics}},}\ }\href {\doibase
  10.1103/PhysRevD.79.045002} {\bibfield  {journal} {\bibinfo  {journal} {Phys.
  Rev. D}\ }\textbf {\bibinfo {volume} {79}},\ \bibinfo {pages} {045002}
  (\bibinfo {year} {2009})},\ \Eprint {http://arxiv.org/abs/0812.3284}
  {arXiv:0812.3284 [hep-th]} \BibitemShut {NoStop}%
\bibitem [{\citenamefont {Friedberg}\ \emph {et~al.}(1976)\citenamefont
  {Friedberg}, \citenamefont {Lee},\ and\ \citenamefont
  {Sirlin}}]{Friedberg:1976me}%
  \BibitemOpen
  \bibfield  {author} {\bibinfo {author} {\bibfnamefont {R.}~\bibnamefont
  {Friedberg}}, \bibinfo {author} {\bibfnamefont {T.~D.}\ \bibnamefont {Lee}},
  \ and\ \bibinfo {author} {\bibfnamefont {A.}~\bibnamefont {Sirlin}},\
  }\bibfield  {title} {\enquote {\bibinfo {title} {{A Class of Scalar-Field
  Soliton Solutions in Three Space Dimensions}},}\ }\href {\doibase
  10.1103/PhysRevD.13.2739} {\bibfield  {journal} {\bibinfo  {journal} {Phys.
  Rev.}\ }\textbf {\bibinfo {volume} {D13}},\ \bibinfo {pages} {2739--2761}
  (\bibinfo {year} {1976})}\BibitemShut {NoStop}%
\bibitem [{\citenamefont {Coleman}(1985)}]{Coleman:1985ki}%
  \BibitemOpen
  \bibfield  {author} {\bibinfo {author} {\bibfnamefont {Sidney~R.}\
  \bibnamefont {Coleman}},\ }\bibfield  {title} {\enquote {\bibinfo {title} {{Q
  Balls}},}\ }\href {\doibase 10.1016/0550-3213(85)90286-X,
  10.1016/0550-3213(86)90520-1} {\bibfield  {journal} {\bibinfo  {journal}
  {Nucl. Phys.}\ }\textbf {\bibinfo {volume} {B262}},\ \bibinfo {pages} {263}
  (\bibinfo {year} {1985})},\ \bibinfo {note} {[Erratum: Nucl.
  Phys.B269,744(1986)]}\BibitemShut {NoStop}%
\bibitem [{\citenamefont {Friedberg}\ \emph {et~al.}(1987)\citenamefont
  {Friedberg}, \citenamefont {Lee},\ and\ \citenamefont
  {Pang}}]{Friedberg:1986tq}%
  \BibitemOpen
  \bibfield  {author} {\bibinfo {author} {\bibfnamefont {R.}~\bibnamefont
  {Friedberg}}, \bibinfo {author} {\bibfnamefont {T.~D.}\ \bibnamefont {Lee}},
  \ and\ \bibinfo {author} {\bibfnamefont {Y.}~\bibnamefont {Pang}},\
  }\bibfield  {title} {\enquote {\bibinfo {title} {{Scalar Soliton Stars and
  Black Holes}},}\ }\href {\doibase 10.1103/PhysRevD.35.3658} {\bibfield
  {journal} {\bibinfo  {journal} {Phys. Rev.}\ }\textbf {\bibinfo {volume}
  {D35}},\ \bibinfo {pages} {3658} (\bibinfo {year} {1987})}\BibitemShut
  {NoStop}%
\bibitem [{\citenamefont {Lee}(1987)}]{Lee:1986ts}%
  \BibitemOpen
  \bibfield  {author} {\bibinfo {author} {\bibfnamefont {T.~D.}\ \bibnamefont
  {Lee}},\ }\bibfield  {title} {\enquote {\bibinfo {title} {{Soliton Stars and
  the Critical Masses of Black Holes}},}\ }\href {\doibase
  10.1103/PhysRevD.35.3637} {\bibfield  {journal} {\bibinfo  {journal} {Phys.
  Rev.}\ }\textbf {\bibinfo {volume} {D35}},\ \bibinfo {pages} {3637} (\bibinfo
  {year} {1987})}\BibitemShut {NoStop}%
\bibitem [{\citenamefont {Kusenko}(1997)}]{Kusenko:1997zq}%
  \BibitemOpen
  \bibfield  {author} {\bibinfo {author} {\bibfnamefont {Alexander}\
  \bibnamefont {Kusenko}},\ }\bibfield  {title} {\enquote {\bibinfo {title}
  {{Solitons in the supersymmetric extensions of the standard model}},}\ }\href
  {\doibase 10.1016/S0370-2693(97)00584-4} {\bibfield  {journal} {\bibinfo
  {journal} {Phys. Lett.}\ }\textbf {\bibinfo {volume} {B405}},\ \bibinfo
  {pages} {108} (\bibinfo {year} {1997})},\ \Eprint
  {http://arxiv.org/abs/hep-ph/9704273} {arXiv:hep-ph/9704273 [hep-ph]}
  \BibitemShut {NoStop}%
\bibitem [{\citenamefont {Kusenko}\ and\ \citenamefont
  {Shaposhnikov}(1998)}]{Kusenko:1997si}%
  \BibitemOpen
  \bibfield  {author} {\bibinfo {author} {\bibfnamefont {Alexander}\
  \bibnamefont {Kusenko}}\ and\ \bibinfo {author} {\bibfnamefont {Mikhail~E.}\
  \bibnamefont {Shaposhnikov}},\ }\bibfield  {title} {\enquote {\bibinfo
  {title} {{Supersymmetric Q balls as dark matter}},}\ }\href {\doibase
  10.1016/S0370-2693(97)01375-0} {\bibfield  {journal} {\bibinfo  {journal}
  {Phys. Lett.}\ }\textbf {\bibinfo {volume} {B418}},\ \bibinfo {pages}
  {46--54} (\bibinfo {year} {1998})},\ \Eprint
  {http://arxiv.org/abs/hep-ph/9709492} {arXiv:hep-ph/9709492 [hep-ph]}
  \BibitemShut {NoStop}%
\bibitem [{\citenamefont {Kusenko}\ \emph {et~al.}(1998)\citenamefont
  {Kusenko}, \citenamefont {Kuzmin}, \citenamefont {Shaposhnikov},\ and\
  \citenamefont {Tinyakov}}]{Kusenko:1997vp}%
  \BibitemOpen
  \bibfield  {author} {\bibinfo {author} {\bibfnamefont {Alexander}\
  \bibnamefont {Kusenko}}, \bibinfo {author} {\bibfnamefont {Vadim}\
  \bibnamefont {Kuzmin}}, \bibinfo {author} {\bibfnamefont {Mikhail~E.}\
  \bibnamefont {Shaposhnikov}}, \ and\ \bibinfo {author} {\bibfnamefont
  {P.~G.}\ \bibnamefont {Tinyakov}},\ }\bibfield  {title} {\enquote {\bibinfo
  {title} {{Experimental signatures of supersymmetric dark matter Q balls}},}\
  }\href {\doibase 10.1103/PhysRevLett.80.3185} {\bibfield  {journal} {\bibinfo
   {journal} {Phys. Rev. Lett.}\ }\textbf {\bibinfo {volume} {80}},\ \bibinfo
  {pages} {3185--3188} (\bibinfo {year} {1998})},\ \Eprint
  {http://arxiv.org/abs/hep-ph/9712212} {arXiv:hep-ph/9712212 [hep-ph]}
  \BibitemShut {NoStop}%
\bibitem [{\citenamefont {Klimas}\ and\ \citenamefont
  {Livramento}(2017)}]{Klimas:2017eft}%
  \BibitemOpen
  \bibfield  {author} {\bibinfo {author} {\bibfnamefont {P.}~\bibnamefont
  {Klimas}}\ and\ \bibinfo {author} {\bibfnamefont {L.~R.}\ \bibnamefont
  {Livramento}},\ }\bibfield  {title} {\enquote {\bibinfo {title} {{Compact
  Q-balls and Q-shells in CPN type models}},}\ }\href {\doibase
  10.1103/PhysRevD.96.016001} {\bibfield  {journal} {\bibinfo  {journal} {Phys.
  Rev.}\ }\textbf {\bibinfo {volume} {D96}},\ \bibinfo {pages} {016001}
  (\bibinfo {year} {2017})},\ \Eprint {http://arxiv.org/abs/1704.01132}
  {arXiv:1704.01132 [hep-th]} \BibitemShut {NoStop}%
\bibitem [{\citenamefont {Klimas}\ \emph {et~al.}(2019)\citenamefont {Klimas},
  \citenamefont {Sawado},\ and\ \citenamefont {Yanai}}]{Klimas:2018ywv}%
  \BibitemOpen
  \bibfield  {author} {\bibinfo {author} {\bibfnamefont {Pawe\l{}}\
  \bibnamefont {Klimas}}, \bibinfo {author} {\bibfnamefont {Nobuyuki}\
  \bibnamefont {Sawado}}, \ and\ \bibinfo {author} {\bibfnamefont {Shota}\
  \bibnamefont {Yanai}},\ }\bibfield  {title} {\enquote {\bibinfo {title}
  {{Gravitating compact $Q$-ball and $Q$-shell solutions in the $\mathbb{C}P^N$
  nonlinear sigma model}},}\ }\href {\doibase 10.1103/PhysRevD.99.045015}
  {\bibfield  {journal} {\bibinfo  {journal} {Phys. Rev. D}\ }\textbf {\bibinfo
  {volume} {99}},\ \bibinfo {pages} {045015} (\bibinfo {year} {2019})},\
  \Eprint {http://arxiv.org/abs/1812.08363} {arXiv:1812.08363 [hep-th]}
  \BibitemShut {NoStop}%
\bibitem [{\citenamefont {Yanai}(2019)}]{Yanai:2019wpv}%
  \BibitemOpen
  \bibfield  {author} {\bibinfo {author} {\bibfnamefont {Shota}\ \bibnamefont
  {Yanai}},\ }\bibfield  {title} {\enquote {\bibinfo {title} {{Q-balls, -shells
  of a nonlinear sigma model with finite cosmological constants}},}\ }\href
  {\doibase 10.1088/1742-6596/1194/1/012114} {\bibfield  {journal} {\bibinfo
  {journal} {J. Phys. Conf. Ser.}\ }\textbf {\bibinfo {volume} {1194}},\
  \bibinfo {pages} {012114} (\bibinfo {year} {2019})}\BibitemShut {NoStop}%
\bibitem [{\citenamefont {Sawado}\ and\ \citenamefont
  {Yanai}(2020)}]{Sawado:2020ncc}%
  \BibitemOpen
  \bibfield  {author} {\bibinfo {author} {\bibfnamefont {Nobuyuki}\
  \bibnamefont {Sawado}}\ and\ \bibinfo {author} {\bibfnamefont {Shota}\
  \bibnamefont {Yanai}},\ }\bibfield  {title} {\enquote {\bibinfo {title}
  {{Compact, charged boson stars and shells in the $\mathbb{C}P^N$ gravitating
  nonlinear sigma model}},}\ }\href {\doibase 10.1103/PhysRevD.102.045007}
  {\bibfield  {journal} {\bibinfo  {journal} {Phys. Rev. D}\ }\textbf {\bibinfo
  {volume} {102}},\ \bibinfo {pages} {045007} (\bibinfo {year} {2020})},\
  \Eprint {http://arxiv.org/abs/2006.03424} {arXiv:2006.03424 [hep-th]}
  \BibitemShut {NoStop}%
\bibitem [{\citenamefont {Sawado}\ and\ \citenamefont
  {Yanai}(2021)}]{Sawado:2021rsc}%
  \BibitemOpen
  \bibfield  {author} {\bibinfo {author} {\bibfnamefont {Nobuyuki}\
  \bibnamefont {Sawado}}\ and\ \bibinfo {author} {\bibfnamefont {Shota}\
  \bibnamefont {Yanai}},\ }\bibfield  {title} {\enquote {\bibinfo {title}
  {{Phase analyses for compact, charged boson stars and shells harboring black
  holes in the $\mathbb{C}P^N$ nonlinear sigma model}},}\ }\href {\doibase
  10.1103/PhysRevD.103.125018} {\bibfield  {journal} {\bibinfo  {journal}
  {Phys. Rev. D}\ }\textbf {\bibinfo {volume} {103}},\ \bibinfo {pages}
  {125018} (\bibinfo {year} {2021})},\ \Eprint
  {http://arxiv.org/abs/2103.05877} {arXiv:2103.05877 [hep-th]} \BibitemShut
  {NoStop}%
\bibitem [{\citenamefont {Kleihaus}\ \emph {et~al.}(2010)\citenamefont
  {Kleihaus}, \citenamefont {Kunz}, \citenamefont {Lammerzahl},\ and\
  \citenamefont {List}}]{Kleihaus:2010ep}%
  \BibitemOpen
  \bibfield  {author} {\bibinfo {author} {\bibfnamefont {Burkhard}\
  \bibnamefont {Kleihaus}}, \bibinfo {author} {\bibfnamefont {Jutta}\
  \bibnamefont {Kunz}}, \bibinfo {author} {\bibfnamefont {Claus}\ \bibnamefont
  {Lammerzahl}}, \ and\ \bibinfo {author} {\bibfnamefont {Meike}\ \bibnamefont
  {List}},\ }\bibfield  {title} {\enquote {\bibinfo {title} {{Boson Shells
  Harbouring Charged Black Holes}},}\ }\href {\doibase
  10.1103/PhysRevD.82.104050} {\bibfield  {journal} {\bibinfo  {journal} {Phys.
  Rev.}\ }\textbf {\bibinfo {volume} {D82}},\ \bibinfo {pages} {104050}
  (\bibinfo {year} {2010})},\ \Eprint {http://arxiv.org/abs/1007.1630}
  {arXiv:1007.1630 [gr-qc]} \BibitemShut {NoStop}%
\bibitem [{\citenamefont {Witten}(1985)}]{Witten:1984eb}%
  \BibitemOpen
  \bibfield  {author} {\bibinfo {author} {\bibfnamefont {Edward}\ \bibnamefont
  {Witten}},\ }\bibfield  {title} {\enquote {\bibinfo {title} {{Superconducting
  Strings}},}\ }\href {\doibase 10.1016/0550-3213(85)90022-7} {\bibfield
  {journal} {\bibinfo  {journal} {Nucl. Phys. B}\ }\textbf {\bibinfo {volume}
  {249}},\ \bibinfo {pages} {557--592} (\bibinfo {year} {1985})}\BibitemShut
  {NoStop}%
\bibitem [{\citenamefont {Davis}\ and\ \citenamefont
  {Shellard}(1988)}]{Davis:1988jp}%
  \BibitemOpen
  \bibfield  {author} {\bibinfo {author} {\bibfnamefont {R.~L.}\ \bibnamefont
  {Davis}}\ and\ \bibinfo {author} {\bibfnamefont {E.~P.~S.}\ \bibnamefont
  {Shellard}},\ }\bibfield  {title} {\enquote {\bibinfo {title} {{The Physics
  of Vortex Superconductivity}},}\ }\href {\doibase
  10.1016/0370-2693(88)90673-9} {\bibfield  {journal} {\bibinfo  {journal}
  {Phys. Lett. B}\ }\textbf {\bibinfo {volume} {207}},\ \bibinfo {pages}
  {404--410} (\bibinfo {year} {1988})}\BibitemShut {NoStop}%
\bibitem [{\citenamefont {Lemperiere}\ and\ \citenamefont
  {Shellard}(2003{\natexlab{a}})}]{Lemperiere:2002en}%
  \BibitemOpen
  \bibfield  {author} {\bibinfo {author} {\bibfnamefont {Y.}~\bibnamefont
  {Lemperiere}}\ and\ \bibinfo {author} {\bibfnamefont {E.~P.~S.}\ \bibnamefont
  {Shellard}},\ }\bibfield  {title} {\enquote {\bibinfo {title} {{On the
  behavior and stability of superconducting currents}},}\ }\href {\doibase
  10.1016/S0550-3213(02)01028-3} {\bibfield  {journal} {\bibinfo  {journal}
  {Nucl. Phys. B}\ }\textbf {\bibinfo {volume} {649}},\ \bibinfo {pages}
  {511--525} (\bibinfo {year} {2003}{\natexlab{a}})},\ \Eprint
  {http://arxiv.org/abs/hep-ph/0207199} {arXiv:hep-ph/0207199} \BibitemShut
  {NoStop}%
\bibitem [{\citenamefont {Lemperiere}\ and\ \citenamefont
  {Shellard}(2003{\natexlab{b}})}]{Lemperiere:2003yt}%
  \BibitemOpen
  \bibfield  {author} {\bibinfo {author} {\bibfnamefont {Y.}~\bibnamefont
  {Lemperiere}}\ and\ \bibinfo {author} {\bibfnamefont {E.~P.~S.}\ \bibnamefont
  {Shellard}},\ }\bibfield  {title} {\enquote {\bibinfo {title} {{Vorton
  existence and stability}},}\ }\href {\doibase 10.1103/PhysRevLett.91.141601}
  {\bibfield  {journal} {\bibinfo  {journal} {Phys. Rev. Lett.}\ }\textbf
  {\bibinfo {volume} {91}},\ \bibinfo {pages} {141601} (\bibinfo {year}
  {2003}{\natexlab{b}})},\ \Eprint {http://arxiv.org/abs/hep-ph/0305156}
  {arXiv:hep-ph/0305156} \BibitemShut {NoStop}%
\bibitem [{\citenamefont {Itaya}\ \emph {et~al.}(2014)\citenamefont {Itaya},
  \citenamefont {Sawado},\ and\ \citenamefont {Suzuki}}]{Itaya:2014oza}%
  \BibitemOpen
  \bibfield  {author} {\bibinfo {author} {\bibfnamefont {Satoru}\ \bibnamefont
  {Itaya}}, \bibinfo {author} {\bibfnamefont {Nobuyuki}\ \bibnamefont
  {Sawado}}, \ and\ \bibinfo {author} {\bibfnamefont {Michitaka}\ \bibnamefont
  {Suzuki}},\ }\bibfield  {title} {\enquote {\bibinfo {title} {{Vortex
  solutions in a Witten-type model}},}\ }\href {\doibase
  10.1088/1742-6596/563/1/012014} {\bibfield  {journal} {\bibinfo  {journal}
  {J. Phys. Conf. Ser.}\ }\textbf {\bibinfo {volume} {563}},\ \bibinfo {pages}
  {012014} (\bibinfo {year} {2014})}\BibitemShut {NoStop}%
\bibitem [{\citenamefont {Bogolyubsky}\ and\ \citenamefont
  {Makhankov}(1976)}]{Bogolyubsky:1976yu}%
  \BibitemOpen
  \bibfield  {author} {\bibinfo {author} {\bibfnamefont {I.~L.}\ \bibnamefont
  {Bogolyubsky}}\ and\ \bibinfo {author} {\bibfnamefont {V.~G.}\ \bibnamefont
  {Makhankov}},\ }\bibfield  {title} {\enquote {\bibinfo {title} {{Lifetime of
  Pulsating Solitons in Some Classical Models}},}\ }\href@noop {} {\bibfield
  {journal} {\bibinfo  {journal} {Pisma Zh. Eksp. Teor. Fiz.}\ }\textbf
  {\bibinfo {volume} {24}},\ \bibinfo {pages} {15--18} (\bibinfo {year}
  {1976})}\BibitemShut {NoStop}%
\bibitem [{\citenamefont {Gleiser}(1994)}]{Gleiser:1993pt}%
  \BibitemOpen
  \bibfield  {author} {\bibinfo {author} {\bibfnamefont {Marcelo}\ \bibnamefont
  {Gleiser}},\ }\bibfield  {title} {\enquote {\bibinfo {title} {{Pseudostable
  bubbles}},}\ }\href {\doibase 10.1103/PhysRevD.49.2978} {\bibfield  {journal}
  {\bibinfo  {journal} {Phys. Rev. D}\ }\textbf {\bibinfo {volume} {49}},\
  \bibinfo {pages} {2978--2981} (\bibinfo {year} {1994})},\ \Eprint
  {http://arxiv.org/abs/hep-ph/9308279} {arXiv:hep-ph/9308279} \BibitemShut
  {NoStop}%
\bibitem [{\citenamefont {Copeland}\ \emph {et~al.}(1995)\citenamefont
  {Copeland}, \citenamefont {Gleiser},\ and\ \citenamefont
  {Muller}}]{Copeland:1995fq}%
  \BibitemOpen
  \bibfield  {author} {\bibinfo {author} {\bibfnamefont {Edmund~J.}\
  \bibnamefont {Copeland}}, \bibinfo {author} {\bibfnamefont {M.}~\bibnamefont
  {Gleiser}}, \ and\ \bibinfo {author} {\bibfnamefont {H.~R.}\ \bibnamefont
  {Muller}},\ }\bibfield  {title} {\enquote {\bibinfo {title} {{Oscillons:
  Resonant configurations during bubble collapse}},}\ }\href {\doibase
  10.1103/PhysRevD.52.1920} {\bibfield  {journal} {\bibinfo  {journal} {Phys.
  Rev. D}\ }\textbf {\bibinfo {volume} {52}},\ \bibinfo {pages} {1920--1933}
  (\bibinfo {year} {1995})},\ \Eprint {http://arxiv.org/abs/hep-ph/9503217}
  {arXiv:hep-ph/9503217} \BibitemShut {NoStop}%
\bibitem [{\citenamefont {Fodor}\ \emph {et~al.}(2008)\citenamefont {Fodor},
  \citenamefont {Forgacs}, \citenamefont {Horvath},\ and\ \citenamefont
  {Lukacs}}]{Fodor:2008es}%
  \BibitemOpen
  \bibfield  {author} {\bibinfo {author} {\bibfnamefont {Gyula}\ \bibnamefont
  {Fodor}}, \bibinfo {author} {\bibfnamefont {Peter}\ \bibnamefont {Forgacs}},
  \bibinfo {author} {\bibfnamefont {Zalan}\ \bibnamefont {Horvath}}, \ and\
  \bibinfo {author} {\bibfnamefont {Arpad}\ \bibnamefont {Lukacs}},\ }\bibfield
   {title} {\enquote {\bibinfo {title} {{Small amplitude quasi-breathers and
  oscillons}},}\ }\href {\doibase 10.1103/PhysRevD.78.025003} {\bibfield
  {journal} {\bibinfo  {journal} {Phys. Rev. D}\ }\textbf {\bibinfo {volume}
  {78}},\ \bibinfo {pages} {025003} (\bibinfo {year} {2008})},\ \Eprint
  {http://arxiv.org/abs/0802.3525} {arXiv:0802.3525 [hep-th]} \BibitemShut
  {NoStop}%
\bibitem [{\citenamefont {Fodor}\ \emph {et~al.}(2009)\citenamefont {Fodor},
  \citenamefont {Forgacs}, \citenamefont {Horvath},\ and\ \citenamefont
  {Mezei}}]{Fodor:2009xw}%
  \BibitemOpen
  \bibfield  {author} {\bibinfo {author} {\bibfnamefont {Gyula}\ \bibnamefont
  {Fodor}}, \bibinfo {author} {\bibfnamefont {Peter}\ \bibnamefont {Forgacs}},
  \bibinfo {author} {\bibfnamefont {Zalan}\ \bibnamefont {Horvath}}, \ and\
  \bibinfo {author} {\bibfnamefont {Mark}\ \bibnamefont {Mezei}},\ }\bibfield
  {title} {\enquote {\bibinfo {title} {{Oscillons in dilaton-scalar
  theories}},}\ }\href {\doibase 10.1088/1126-6708/2009/08/106} {\bibfield
  {journal} {\bibinfo  {journal} {JHEP}\ }\textbf {\bibinfo {volume} {08}},\
  \bibinfo {pages} {106} (\bibinfo {year} {2009})},\ \Eprint
  {http://arxiv.org/abs/0906.4160} {arXiv:0906.4160 [hep-th]} \BibitemShut
  {NoStop}%
\bibitem [{\citenamefont {Gleiser}\ \emph {et~al.}(2011)\citenamefont
  {Gleiser}, \citenamefont {Graham},\ and\ \citenamefont
  {Stamatopoulos}}]{Gleiser:2011xj}%
  \BibitemOpen
  \bibfield  {author} {\bibinfo {author} {\bibfnamefont {Marcelo}\ \bibnamefont
  {Gleiser}}, \bibinfo {author} {\bibfnamefont {Noah}\ \bibnamefont {Graham}},
  \ and\ \bibinfo {author} {\bibfnamefont {Nikitas}\ \bibnamefont
  {Stamatopoulos}},\ }\bibfield  {title} {\enquote {\bibinfo {title}
  {{Generation of Coherent Structures After Cosmic Inflation}},}\ }\href
  {\doibase 10.1103/PhysRevD.83.096010} {\bibfield  {journal} {\bibinfo
  {journal} {Phys. Rev. D}\ }\textbf {\bibinfo {volume} {83}},\ \bibinfo
  {pages} {096010} (\bibinfo {year} {2011})},\ \Eprint
  {http://arxiv.org/abs/1103.1911} {arXiv:1103.1911 [hep-th]} \BibitemShut
  {NoStop}%
\bibitem [{\citenamefont {Achilleos}\ \emph {et~al.}(2013)\citenamefont
  {Achilleos}, \citenamefont {Diakonos}, \citenamefont {Frantzeskakis},
  \citenamefont {Katsimiga}, \citenamefont {Maintas}, \citenamefont
  {Manousakis}, \citenamefont {Tsagkarakis},\ and\ \citenamefont
  {Tsapalis}}]{Achilleos:2013zpa}%
  \BibitemOpen
  \bibfield  {author} {\bibinfo {author} {\bibfnamefont {V.}~\bibnamefont
  {Achilleos}}, \bibinfo {author} {\bibfnamefont {F.~K.}\ \bibnamefont
  {Diakonos}}, \bibinfo {author} {\bibfnamefont {D.~J.}\ \bibnamefont
  {Frantzeskakis}}, \bibinfo {author} {\bibfnamefont {G.~C.}\ \bibnamefont
  {Katsimiga}}, \bibinfo {author} {\bibfnamefont {X.~N.}\ \bibnamefont
  {Maintas}}, \bibinfo {author} {\bibfnamefont {E.}~\bibnamefont {Manousakis}},
  \bibinfo {author} {\bibfnamefont {C.~E.}\ \bibnamefont {Tsagkarakis}}, \ and\
  \bibinfo {author} {\bibfnamefont {A.}~\bibnamefont {Tsapalis}},\ }\bibfield
  {title} {\enquote {\bibinfo {title} {{Oscillons and oscillating kinks in the
  Abelian-Higgs model}},}\ }\href {\doibase 10.1103/PhysRevD.88.045015}
  {\bibfield  {journal} {\bibinfo  {journal} {Phys. Rev. D}\ }\textbf {\bibinfo
  {volume} {88}},\ \bibinfo {pages} {045015} (\bibinfo {year} {2013})},\
  \Eprint {http://arxiv.org/abs/1306.3868} {arXiv:1306.3868 [hep-th]}
  \BibitemShut {NoStop}%
\bibitem [{\citenamefont {Correa}\ \emph {et~al.}(2019)\citenamefont {Correa},
  \citenamefont {de~Souza~Dutra}, \citenamefont {Frederico}, \citenamefont
  {Malomed}, \citenamefont {Oliveira},\ and\ \citenamefont
  {Sawado}}]{Correa:2019jaa}%
  \BibitemOpen
  \bibfield  {author} {\bibinfo {author} {\bibfnamefont {R.~A.~C.}\
  \bibnamefont {Correa}}, \bibinfo {author} {\bibfnamefont {A.}~\bibnamefont
  {de~Souza~Dutra}}, \bibinfo {author} {\bibfnamefont {T.}~\bibnamefont
  {Frederico}}, \bibinfo {author} {\bibfnamefont {Boris~A.}\ \bibnamefont
  {Malomed}}, \bibinfo {author} {\bibfnamefont {O.}~\bibnamefont {Oliveira}}, \
  and\ \bibinfo {author} {\bibfnamefont {N.}~\bibnamefont {Sawado}},\
  }\bibfield  {title} {\enquote {\bibinfo {title} {{Creating Oscillons and
  Oscillating Kinks in Two Scalar Field Theories}},}\ }\href {\doibase
  10.1063/1.5120500} {\bibfield  {journal} {\bibinfo  {journal} {Chaos}\
  }\textbf {\bibinfo {volume} {29}},\ \bibinfo {pages} {103124} (\bibinfo
  {year} {2019})},\ \Eprint {http://arxiv.org/abs/1907.07145} {arXiv:1907.07145
  [hep-th]} \BibitemShut {NoStop}%
\bibitem [{\citenamefont {Fadragas}\ and\ \citenamefont
  {Leon}(2014)}]{Fadragas:2014mra}%
  \BibitemOpen
  \bibfield  {author} {\bibinfo {author} {\bibfnamefont {Carlos~R.}\
  \bibnamefont {Fadragas}}\ and\ \bibinfo {author} {\bibfnamefont {Genly}\
  \bibnamefont {Leon}},\ }\bibfield  {title} {\enquote {\bibinfo {title} {{Some
  remarks about non-minimally coupled scalar field models}},}\ }\href {\doibase
  10.1088/0264-9381/31/19/195011} {\bibfield  {journal} {\bibinfo  {journal}
  {Class. Quant. Grav.}\ }\textbf {\bibinfo {volume} {31}},\ \bibinfo {pages}
  {195011} (\bibinfo {year} {2014})},\ \Eprint {http://arxiv.org/abs/1405.2465}
  {arXiv:1405.2465 [gr-qc]} \BibitemShut {NoStop}%
\bibitem [{\citenamefont {Ishihara}\ and\ \citenamefont
  {Ogawa}(2021)}]{Ishihara:2021iag}%
  \BibitemOpen
  \bibfield  {author} {\bibinfo {author} {\bibfnamefont {Hideki}\ \bibnamefont
  {Ishihara}}\ and\ \bibinfo {author} {\bibfnamefont {Tatsuya}\ \bibnamefont
  {Ogawa}},\ }\bibfield  {title} {\enquote {\bibinfo {title} {{Variety of
  nontopological solitons in a spontaneously broken U(1) gauge theory: Dust
  balls, shell balls, and potential balls}},}\ }\href {\doibase
  10.1103/PhysRevD.103.123029} {\bibfield  {journal} {\bibinfo  {journal}
  {Phys. Rev. D}\ }\textbf {\bibinfo {volume} {103}},\ \bibinfo {pages}
  {123029} (\bibinfo {year} {2021})},\ \Eprint
  {http://arxiv.org/abs/2103.13732} {arXiv:2103.13732 [hep-th]} \BibitemShut
  {NoStop}%
\bibitem [{\citenamefont {Arodz}(2002)}]{Arodz:2002yt}%
  \BibitemOpen
  \bibfield  {author} {\bibinfo {author} {\bibfnamefont {H.}~\bibnamefont
  {Arodz}},\ }\bibfield  {title} {\enquote {\bibinfo {title} {{Topological
  compactons}},}\ }\href@noop {} {\bibfield  {journal} {\bibinfo  {journal}
  {Acta Phys. Polon. B}\ }\textbf {\bibinfo {volume} {33}},\ \bibinfo {pages}
  {1241--1252} (\bibinfo {year} {2002})},\ \Eprint
  {http://arxiv.org/abs/nlin/0201001} {arXiv:nlin/0201001} \BibitemShut
  {NoStop}%
\bibitem [{\citenamefont {Arodz}(2004)}]{Arodz:2003mx}%
  \BibitemOpen
  \bibfield  {author} {\bibinfo {author} {\bibfnamefont {H.}~\bibnamefont
  {Arodz}},\ }\bibfield  {title} {\enquote {\bibinfo {title} {{Symmetry
  breaking transition and appearance of compactons in a mechanical system}},}\
  }\href@noop {} {\bibfield  {journal} {\bibinfo  {journal} {Acta Phys. Polon.
  B}\ }\textbf {\bibinfo {volume} {35}},\ \bibinfo {pages} {625--638} (\bibinfo
  {year} {2004})},\ \Eprint {http://arxiv.org/abs/hep-th/0312036}
  {arXiv:hep-th/0312036} \BibitemShut {NoStop}%
\bibitem [{\citenamefont {Liu}\ \emph {et~al.}(2021)\citenamefont {Liu},
  \citenamefont {Bo}, \citenamefont {Liu}, \citenamefont {Liu}, \citenamefont
  {Shi}, \citenamefont {Yuan}, \citenamefont {He},\ and\ \citenamefont
  {Wu}}]{Liu:2021}%
  \BibitemOpen
  \bibfield  {author} {\bibinfo {author} {\bibfnamefont {Bin}\ \bibnamefont
  {Liu}}, \bibinfo {author} {\bibfnamefont {Wan}\ \bibnamefont {Bo}}, \bibinfo
  {author} {\bibfnamefont {Jiandong}\ \bibnamefont {Liu}}, \bibinfo {author}
  {\bibfnamefont {Juan}\ \bibnamefont {Liu}}, \bibinfo {author} {\bibfnamefont
  {Jiu-lin}\ \bibnamefont {Shi}}, \bibinfo {author} {\bibfnamefont {Jinhui}\
  \bibnamefont {Yuan}}, \bibinfo {author} {\bibfnamefont {Xing-Dao}\
  \bibnamefont {He}}, \ and\ \bibinfo {author} {\bibfnamefont {Qiang}\
  \bibnamefont {Wu}},\ }\bibfield  {title} {\enquote {\bibinfo {title} {{Simple
  harmonic and damped motions of dissipative solitons in two-dimensional
  complex Ginzburg-Landau equation supported by an external V-shaped
  potential}},}\ }\href@noop {} {\bibfield  {journal} {\bibinfo  {journal}
  {Chaos, Solitons and Fractals}\ }\textbf {\bibinfo {volume} {150}},\ \bibinfo
  {pages} {111126} (\bibinfo {year} {2021})}\BibitemShut {NoStop}%
\bibitem [{\citenamefont {Thompson}\ and\ \citenamefont
  {Ghaffari}(1983)}]{PhysRevA.27.1741}%
  \BibitemOpen
  \bibfield  {author} {\bibinfo {author} {\bibfnamefont {J.~M.~T.}\
  \bibnamefont {Thompson}}\ and\ \bibinfo {author} {\bibfnamefont
  {R.}~\bibnamefont {Ghaffari}},\ }\bibfield  {title} {\enquote {\bibinfo
  {title} {Chaotic dynamics of an impact oscillator},}\ }\href {\doibase
  10.1103/PhysRevA.27.1741} {\bibfield  {journal} {\bibinfo  {journal} {Phys.
  Rev. A}\ }\textbf {\bibinfo {volume} {27}},\ \bibinfo {pages} {1741--1743}
  (\bibinfo {year} {1983})}\BibitemShut {NoStop}%
\bibitem [{\citenamefont {Nusse}\ \emph {et~al.}(1994)\citenamefont {Nusse},
  \citenamefont {Ott},\ and\ \citenamefont {Yorke}}]{PhysRevE.49.1073}%
  \BibitemOpen
  \bibfield  {author} {\bibinfo {author} {\bibfnamefont {Helena~E.}\
  \bibnamefont {Nusse}}, \bibinfo {author} {\bibfnamefont {Edward}\
  \bibnamefont {Ott}}, \ and\ \bibinfo {author} {\bibfnamefont {James~A.}\
  \bibnamefont {Yorke}},\ }\bibfield  {title} {\enquote {\bibinfo {title}
  {Border-collision bifurcations: An explanation for observed bifurcation
  phenomena},}\ }\href {\doibase 10.1103/PhysRevE.49.1073} {\bibfield
  {journal} {\bibinfo  {journal} {Phys. Rev. E}\ }\textbf {\bibinfo {volume}
  {49}},\ \bibinfo {pages} {1073--1076} (\bibinfo {year} {1994})}\BibitemShut
  {NoStop}%
\bibitem [{\citenamefont {Chin}\ \emph {et~al.}(1994)\citenamefont {Chin},
  \citenamefont {Ott}, \citenamefont {Nusse},\ and\ \citenamefont
  {Grebogi}}]{PhysRevE.50.4427}%
  \BibitemOpen
  \bibfield  {author} {\bibinfo {author} {\bibfnamefont {Wai}\ \bibnamefont
  {Chin}}, \bibinfo {author} {\bibfnamefont {Edward}\ \bibnamefont {Ott}},
  \bibinfo {author} {\bibfnamefont {Helena~E.}\ \bibnamefont {Nusse}}, \ and\
  \bibinfo {author} {\bibfnamefont {Celso}\ \bibnamefont {Grebogi}},\
  }\bibfield  {title} {\enquote {\bibinfo {title} {Grazing bifurcations in
  impact oscillators},}\ }\href {\doibase 10.1103/PhysRevE.50.4427} {\bibfield
  {journal} {\bibinfo  {journal} {Phys. Rev. E}\ }\textbf {\bibinfo {volume}
  {50}},\ \bibinfo {pages} {4427--4444} (\bibinfo {year} {1994})}\BibitemShut
  {NoStop}%
\bibitem [{\citenamefont {Adam}\ \emph {et~al.}(2017)\citenamefont {Adam},
  \citenamefont {Sanchez-Guillen},\ and\ \citenamefont
  {Wereszczynski}}]{Adam:2017pdh}%
  \BibitemOpen
  \bibfield  {author} {\bibinfo {author} {\bibfnamefont {C.}~\bibnamefont
  {Adam}}, \bibinfo {author} {\bibfnamefont {J.}~\bibnamefont
  {Sanchez-Guillen}}, \ and\ \bibinfo {author} {\bibfnamefont {A.}~\bibnamefont
  {Wereszczynski}},\ }\bibfield  {title} {\enquote {\bibinfo {title} {{BPS
  submodels of the Skyrme model}},}\ }\href {\doibase
  10.1016/j.physletb.2017.04.003} {\bibfield  {journal} {\bibinfo  {journal}
  {Phys. Lett. B}\ }\textbf {\bibinfo {volume} {769}},\ \bibinfo {pages}
  {362--367} (\bibinfo {year} {2017})},\ \Eprint
  {http://arxiv.org/abs/1703.05818} {arXiv:1703.05818 [hep-th]} \BibitemShut
  {NoStop}%
\bibitem [{\citenamefont {Adam}\ \emph {et~al.}(2018)\citenamefont {Adam},
  \citenamefont {Foster}, \citenamefont {Krusch},\ and\ \citenamefont
  {Wereszczynski}}]{Adam:2017srx}%
  \BibitemOpen
  \bibfield  {author} {\bibinfo {author} {\bibfnamefont {C.}~\bibnamefont
  {Adam}}, \bibinfo {author} {\bibfnamefont {D.}~\bibnamefont {Foster}},
  \bibinfo {author} {\bibfnamefont {S.}~\bibnamefont {Krusch}}, \ and\ \bibinfo
  {author} {\bibfnamefont {A.}~\bibnamefont {Wereszczynski}},\ }\bibfield
  {title} {\enquote {\bibinfo {title} {{BPS sectors of the Skyrme model and
  their non-BPS extensions}},}\ }\href {\doibase 10.1103/PhysRevD.97.036002}
  {\bibfield  {journal} {\bibinfo  {journal} {Phys. Rev. D}\ }\textbf {\bibinfo
  {volume} {97}},\ \bibinfo {pages} {036002} (\bibinfo {year} {2018})},\
  \Eprint {http://arxiv.org/abs/1709.06583} {arXiv:1709.06583 [hep-th]}
  \BibitemShut {NoStop}%
\bibitem [{\citenamefont {Klimas}\ \emph {et~al.}(2018)\citenamefont {Klimas},
  \citenamefont {Streibel}, \citenamefont {Wereszczynski},\ and\ \citenamefont
  {Zakrzewski}}]{Klimas:2018woi}%
  \BibitemOpen
  \bibfield  {author} {\bibinfo {author} {\bibfnamefont {P.}~\bibnamefont
  {Klimas}}, \bibinfo {author} {\bibfnamefont {J.~S.}\ \bibnamefont
  {Streibel}}, \bibinfo {author} {\bibfnamefont {A.}~\bibnamefont
  {Wereszczynski}}, \ and\ \bibinfo {author} {\bibfnamefont {W.~J.}\
  \bibnamefont {Zakrzewski}},\ }\bibfield  {title} {\enquote {\bibinfo {title}
  {{Oscillons in a perturbed signum-Gordon model}},}\ }\href {\doibase
  10.1007/JHEP04(2018)102} {\bibfield  {journal} {\bibinfo  {journal} {JHEP}\
  }\textbf {\bibinfo {volume} {04}},\ \bibinfo {pages} {102} (\bibinfo {year}
  {2018})},\ \Eprint {http://arxiv.org/abs/1801.05454} {arXiv:1801.05454
  [hep-th]} \BibitemShut {NoStop}%
\bibitem [{\citenamefont {Nelson}(2002)}]{nelson2002}%
  \BibitemOpen
  \bibfield  {author} {\bibinfo {author} {\bibfnamefont {David~R.}\
  \bibnamefont {Nelson}},\ }\href@noop {} {\emph {\bibinfo {title} {Defects and
  Geometry in Condensed Matter Physics}}}\ (\bibinfo  {publisher} {Cambridge
  University Press},\ \bibinfo {address} {Cambridge},\ \bibinfo {year}
  {2002})\BibitemShut {NoStop}%
\bibitem [{\citenamefont {Narayan}\ and\ \citenamefont
  {Fisher}(1993)}]{PhysRevB.48.7030}%
  \BibitemOpen
  \bibfield  {author} {\bibinfo {author} {\bibfnamefont {Onuttom}\ \bibnamefont
  {Narayan}}\ and\ \bibinfo {author} {\bibfnamefont {Daniel~S.}\ \bibnamefont
  {Fisher}},\ }\bibfield  {title} {\enquote {\bibinfo {title} {Threshold
  critical dynamics of driven interfaces in random media},}\ }\href {\doibase
  10.1103/PhysRevB.48.7030} {\bibfield  {journal} {\bibinfo  {journal} {Phys.
  Rev. B}\ }\textbf {\bibinfo {volume} {48}},\ \bibinfo {pages} {7030--7042}
  (\bibinfo {year} {1993})}\BibitemShut {NoStop}%
\bibitem [{\citenamefont {Ishiguro}\ \emph {et~al.}(1997)\citenamefont
  {Ishiguro}, \citenamefont {Sato},\ and\ \citenamefont
  {Takamaru}}]{PhysRevLett.78.4761}%
  \BibitemOpen
  \bibfield  {author} {\bibinfo {author} {\bibfnamefont {Seiji}\ \bibnamefont
  {Ishiguro}}, \bibinfo {author} {\bibfnamefont {Tetsuya}\ \bibnamefont
  {Sato}}, \ and\ \bibinfo {author} {\bibfnamefont {Hisanori}\ \bibnamefont
  {Takamaru}} (\bibinfo {collaboration} {The Complexity Simulation Group}),\
  }\bibfield  {title} {\enquote {\bibinfo {title} {V-shaped dc potential
  structure caused by current-driven electrostatic ion-cyclotron
  instability},}\ }\href {\doibase 10.1103/PhysRevLett.78.4761} {\bibfield
  {journal} {\bibinfo  {journal} {Phys. Rev. Lett.}\ }\textbf {\bibinfo
  {volume} {78}},\ \bibinfo {pages} {4761--4764} (\bibinfo {year}
  {1997})}\BibitemShut {NoStop}%
\bibitem [{\citenamefont {Rodriguez-Coppola}\ and\ \citenamefont
  {Perez-Alvarez}(1992)}]{Rodriguez_Coppola_1992}%
  \BibitemOpen
  \bibfield  {author} {\bibinfo {author} {\bibfnamefont {H}~\bibnamefont
  {Rodriguez-Coppola}}\ and\ \bibinfo {author} {\bibfnamefont {R}~\bibnamefont
  {Perez-Alvarez}},\ }\bibfield  {title} {\enquote {\bibinfo {title} {Exchange
  energy of a quasi-2d electron gas in a v-shaped potential},}\ }\href
  {\doibase 10.1088/0953-8984/4/50/013} {\bibfield  {journal} {\bibinfo
  {journal} {Journal of Physics: Condensed Matter}\ }\textbf {\bibinfo {volume}
  {4}},\ \bibinfo {pages} {10245--10256} (\bibinfo {year} {1992})}\BibitemShut
  {NoStop}%
\bibitem [{\citenamefont {Arodz}\ \emph {et~al.}(2005)\citenamefont {Arodz},
  \citenamefont {Klimas},\ and\ \citenamefont {Tyranowski}}]{Arodz:2005gz}%
  \BibitemOpen
  \bibfield  {author} {\bibinfo {author} {\bibfnamefont {H.}~\bibnamefont
  {Arodz}}, \bibinfo {author} {\bibfnamefont {P.}~\bibnamefont {Klimas}}, \
  and\ \bibinfo {author} {\bibfnamefont {T.}~\bibnamefont {Tyranowski}},\
  }\bibfield  {title} {\enquote {\bibinfo {title} {{Field-theoretic models with
  V-shaped potentials}},}\ }\href@noop {} {\bibfield  {journal} {\bibinfo
  {journal} {Acta Phys. Polon. B}\ }\textbf {\bibinfo {volume} {36}},\ \bibinfo
  {pages} {3861--3876} (\bibinfo {year} {2005})},\ \Eprint
  {http://arxiv.org/abs/hep-th/0510204} {arXiv:hep-th/0510204} \BibitemShut
  {NoStop}%
\bibitem [{\citenamefont {Kleihaus}\ \emph {et~al.}(2009)\citenamefont
  {Kleihaus}, \citenamefont {Kunz}, \citenamefont {Lammerzahl},\ and\
  \citenamefont {List}}]{Kleihaus:2009kr}%
  \BibitemOpen
  \bibfield  {author} {\bibinfo {author} {\bibfnamefont {Burkhard}\
  \bibnamefont {Kleihaus}}, \bibinfo {author} {\bibfnamefont {Jutta}\
  \bibnamefont {Kunz}}, \bibinfo {author} {\bibfnamefont {Claus}\ \bibnamefont
  {Lammerzahl}}, \ and\ \bibinfo {author} {\bibfnamefont {Meike}\ \bibnamefont
  {List}},\ }\bibfield  {title} {\enquote {\bibinfo {title} {{Charged Boson
  Stars and Black Holes}},}\ }\href {\doibase 10.1016/j.physletb.2009.03.066}
  {\bibfield  {journal} {\bibinfo  {journal} {Phys. Lett.}\ }\textbf {\bibinfo
  {volume} {B675}},\ \bibinfo {pages} {102--115} (\bibinfo {year} {2009})},\
  \Eprint {http://arxiv.org/abs/0902.4799} {arXiv:0902.4799 [gr-qc]}
  \BibitemShut {NoStop}%
\end{thebibliography}%

\end{document}